\documentclass[english,superscriptaddress,twocolumn,oneside, amsmath,amssymb,amsfonts,aps,pra,floatfix]{revtex4-2}
\usepackage[english]{babel}
\usepackage[utf8x]{inputenc}
\usepackage{microtype}
\usepackage{graphicx}
\usepackage{siunitx}
\usepackage{xspace}
\usepackage{xcolor}
\usepackage[hidelinks]{hyperref}
\usepackage{bm}
\usepackage{bbold}
\usepackage{csquotes}
\usepackage{booktabs}
\usepackage[notrig]{physics}

\usepackage[draft]{fixme}
\usepackage{xcolor}

\usepackage[capitalise]{cleveref}

\newcommand{\diag}[1]{\mathrm{diag}(#1)}
\newcommand{\mat}[1]{\begin{bmatrix} #1 \end{bmatrix}}
\renewcommand{\vec}[1]{\ensuremath{\bm{#1}}\xspace}
\renewcommand{\L}{\mathcal{L}}
\renewcommand{\H}{\mathcal{H}}
\newcommand{\U}{\hat{\mathcal{U}}}
\renewcommand{\O}{\hat{\mathcal{O}}}

\newcommand{\hc}{\text{H.c.}}

\newtheorem{definition}{Definition}

\newcommand{\affA}{Department of Physics and Astronomy, Aarhus University, DK-8000 Aarhus C, Denmark}
\newcommand{\affB}{Aarhus Institute of Advanced Studies, Aarhus University, DK-8000 Aarhus C, Denmark}

\date{\today}

\begin{document}

	\title{The superconducting circuit companion -- an introduction with worked examples}
	
	\author{S. E. Rasmussen}
	\email{stig@phys.au.dk}
	\affiliation{\affA}
	
	\author{K. S. Christensen}
	\affiliation{\affA}
	
	\author{S. P. Pedersen}
	\affiliation{\affA}
	
	\author{L. B. Kristensen}
	\affiliation{\affA}
	
	\author{T. B\ae{}kkegaard}
	\affiliation{\affA}
	
	\author{N. J. S. Loft}
	\affiliation{\affA}
	
	\author{N. T. Zinner}
	\email{zinner@phys.au.dk}
	\affiliation{\affA}
	\affiliation{\affB}

\begin{abstract}
    This tutorial aims at giving an introductory treatment of the circuit analysis of superconducting qubits, i.e., two-level systems in superconducting circuits. It also touches upon couplings between such qubits and how microwave driving and these couplings can be used for single- and two-qubit gates, as well as how to include noise when calculating the dynamics of the system. We also discuss higher-dimensional superconducting qudits.
    The tutorial is intended for new researchers with limited or no experience with the field but should be accessible to anyone with a bachelor's degree in physics. The tutorial introduces the basic methods used in quantum circuit analysis, starting from a circuit diagram and ending with a quantized Hamiltonian, that may be truncated to the lowest levels. We provide examples of all the basic techniques throughout the discussion, while in the last part of the tutorial we discuss several of the most commonly used circuits for quantum information applications. This includes both worked examples of single qubits and examples of how to analyze the coupling methods that allow multiqubit operations. In several detailed appendices, we provide the interested reader an introduction to more advanced techniques for handling larger circuit designs.
\end{abstract}

\maketitle

\tableofcontents

\section{Introduction}

Since Richard Feynman first proposed using quantum simulators to simulate physics \cite{Feynman1982,Lloyd1996}, an increasing amount of attention has been given to quantum processors and quantum technology, something which is only expected to increase further in the coming years \cite{DiVincenzo2000,Dowling2003,Ladd2010}. This increase in attention has led to swift progress within the field of quantum mechanics, taking it from basic science research to engineering of multiqubit quantum systems capable of performing actual calculations  \cite{Bernien2017,Harrow2017,Arute2019,Wang2019,Zhong2020}. During this evolution, a new discipline has emerged, coined quantum engineering, bridging the basic science of quantum mechanics with areas traditionally considered engineering fields \cite{Krantz2019}.
It is expected that the advent of quantum engineering will lead to computational speedups, making it possible to solve classically unsolvable problems \cite{Nielsen2010,Papageorgiou2013,Preskill2018}.

A particularly prominent platform for scalable quantum technology is superconducting circuits used for implementing qubits or even higher-dimensional qudits. 
Compared to other quantum technology schemes, such as trapped ions \cite{Cirac1995,Leibfried2003,Porras2004,Blatt2008,Haffner2008,Blatt2012}, ultracold atoms \cite{Jaksch2005,Lewenstein2007,Bloch2008,Gross2017,Schafer2020}, electron spins in silicon \cite{Loss1998,Kane1998,deSousa2004,Vrijen2000,Hollenberg2006,Morello2010} and quantum dots \cite{Imamoglu1999,Englund2005,Petta2005,Hanson2007,Zwanenburg2013}, nitrogen vacancies in diamonds \cite{Dutt2007,Hanson2006}, or polarized photons \cite{Knill2001,Pittman2001,Franson2002,Pittman2003}, which all encode quantum information in microscopic systems, such as ions, atoms, electrons, or photons, superconducting circuits are quite different. They are macroscopic in size and printed lithographically on wafers much similar to classical computer chips \cite{Yan2016,Barends2013,Oliver2013,Shcherbakova2015,Tsioutsios2020}. The fact that these systems exhibit microscopic behavior, i.e., quantum-mechanical effects, while being macroscopic in size has led to the notion of mesoscopic physics in order to describe this intermediate scale \cite{Caldeira1983,Leggett1987,Vool2017}.
A mesoscopic advantage of superconducting circuits is the fact that microscopic features such as energy spectra, coupling strengths, and coherence rates depend on macroscopic circuit parameters. This means that one can design circuits such that the properties of the resulting quantum-mechanical system, sometimes called an artificial atom \cite{You2005,Buluta2009,Buluta2011,You2011,Georgescu2014}, can be more or less tailormade to exhibit a particular behavior.

In this tutorial, we aim to give an introduction to circuit analysis of superconducting qubits intended for new researchers in the field. With this, we aim to give the tools needed for tailoring macroscopic circuits to a desired qubit behavior.
We refer to a (superconducting) qubit as the two lowest energy levels of a superconducting circuit or subcircuit, denoted by the Fock states $\ket 0$ and $\ket 1$.
There are, however, several examples of superconducting qubits which exploit higher-lying states for coupling \cite{Baekkegaard2019} or control \cite{Poletto2012,DiCarlo2010}.

The field of superconducting circuits is rapidly evolving, and new theoretical frameworks are emerging which make use of the larger Hilbert space of both harmonic and anharmonic resonator modes, e.g., bosonic qubits \cite{Vlastakis2013,Mirrahimi2014,Ofek2016,Rosenblum2018,Gao2019,Gertler2021,Joshi2021,Cai2021} or the Kerr-cat qubits \cite{Puri2017,Puri2019,Grimm2020}, which employs the entire circuit including drives to yield an effective potential where the two lowest levels are coherent cat states. 
Such continuous variable \cite{Braunstein2005,Lau2016} qubits are outside the scope of this tutorial, but an understanding of the fundamentals presented in this tutorial can act as a stepping stone towards an increased understanding of emerging superconducting circuit designs.

The present tutorial can be viewed as an introduction to more advanced reviews of the field, such as Refs. \cite{Devoret1997,Devoret2004,Burkard2004,You2005,Schoelkopf2008,Clarke2008,You2011,Girvin2014,Gambetta2017,Wendin2017,Gu2017,Vool2017,Krantz2019}, and is by no means a review of current state-of-the-art technology or practices, but rather a detailed introduction to the theoretical methods needed to analyze superconducting circuits in order to produce and manipulate qubits.
We do not discuss the actual experimental production of superconducting circuits, but limit the tutorial to theoretical analysis of such circuits.
The tutorial assumes knowledge of undergraduate-level quantum mechanics, electrodynamics, and analytical mechanics, meaning that the tutorial should be accessible to readers with a bachelor's degree in physics.

The tutorial is organized as follows: First, we present the basic circuit variables and components used in the analysis in \cref{sec:Analyzing lumped circuit}. Then we present the classical analysis used for finding the Hamiltonian of a given superconducting circuit in \cref{sec:EOM}, where we use the method of nodes. In \cref{sec:quantizeAndEffective} we quantize the Hamiltonian and in \cref{sec:recasting} we recast the Hamiltonian as interacting oscillators. In \cref{sec:rwa} we discuss time-averaged dynamics using the interaction picture. The truncation of anharmonic oscillators is discussed in \cref{sec:truncation}. The use of microwave driving for control and single-qubit gates is presented in \cref{sec:driving}, and the simple coupling of modes is presented in \cref{sec:coupling}, where two-qubit gates are discussed as well. In \cref{sec:noise} we introduce a method for treating noise in open two-level quantum systems, and finally in \cref{sec:examples} we present a variety of examples ranging from single qubit implementations to tunable couplers and multibody interactions.
In \cref{sec:summary} we present an overview of the methods and give a perspective on where to go from here.

To students and researchers entirely new to the field of superconducting qubits, who just want to start analyzing their first circuit, the amount of information in this tutorial might seem extensive at first. To distill this down to the essential information needed to get started we therefore recommend reading \cref{sec:Analyzing lumped circuit,sec:EOM,sec:quantizeAndEffective,sec:recasting,sec:truncation}, which should be sufficient for analyzing your first superconducting circuit.

\section{Lumped-element circuit diagrams}\label{sec:Analyzing lumped circuit}

In this section we start by introducing the dynamical variables used when analyzing superconducting circuits and then present the basic components of the circuits.

\begin{figure}
    \centering
    \includegraphics[width=.8\columnwidth]{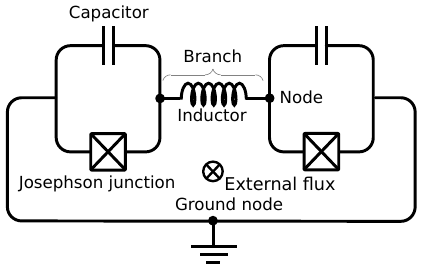}
    \caption{Example of a lumped-element circuit consisting of a Josephson junction and a capacitor in parallel connected by an inductor to another Josephson junction and capacitor pair. Such a Josephson junction and capacitor pair is considered a transmonlike qubit, see \cref{sec:transmon}. An external flux is threading the inductive loop of the circuit.}
    \label{fig:exampleCircuit}
\end{figure}

Our analysis takes its starting point in the lumped-element model. This model simplifies the description of a spatially distributed system (in our case a superconducting electrical circuit) into a topology of discrete entities. We assume that the attributes of the circuit (capacitance, inductance, and resistance) are idealized into electrical components (capacitors, inductors, and resistors) joined by a network of perfectly conducting wires. An example of a lumped circuit can be seen in \cref{fig:exampleCircuit}. We discuss the different components in \cref{sec:CircuitComponents}.

We assume all the circuits discussed in this tutorial to be superconducting, meaning that there is no electrical resistance in the circuit and all magnetic fields are expelled from the wires (the Meissner effect). We therefore ignore losses to the external environment in the following analysis. In other words, we will consider closed quantum systems for most of this tutorial.  
However, a realistic description of any quantum system should include some interactions with the environment, as these can never be completely ignored in an experiment. Notwithstanding, it is a good description to treat losses to the external environment as a correction to the dynamics of the system, something which we discuss in \cref{sec:noise}.

\subsection{Circuit variables}\label{subsec:Circuit variables and}

\begin{figure}
    \centering
    \includegraphics{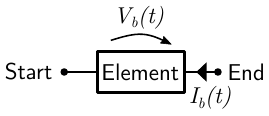}
    \caption{Arbitrary two-terminal component on a branch, $b$, between two nodes (dots). The voltage, $V_b(t)$, over the component is defined from the start of the branch to the end of the branch. The current, $I_b(t)$, through the branch is defined in the opposite direction.}
    \label{fig:2-terminal}
\end{figure}

Circuit analysis aims at finding the equations of motion of an electrical circuit.
Typically this means determining the current and voltage through all components of the circuit. For simplicity, we consider only circuit networks containing two-terminal components, i.e., components connected to two wires. Each such component is said to lie on a \emph{branch}, $b$, and is characterized by two variables at any given time $t$: The voltage, $V_b(t)$, across it and the current, $I_b(t)$, through it. We define the orientation of the voltage to be opposite to the direction of the current, see \cref{fig:2-terminal}. Thus these two are defined by the underlying electromagnetic field by
\begin{subequations}\label{eq:VandI}
\begin{align}
	V_b(t) &= \int_\text{start of $b$}^\text{end of $b$} \vec{E}(t)\cdot \dd\vec{\ell}, \label{eq:Vb}\\
	I_b(t) &= \frac{1}{\mu_0} \oint_b \vec{B}(t)\cdot \dd\vec{\ell}, \label{eq:Ib}
\end{align}
\end{subequations}
where $\mu_0$ is the vacuum permeability, and $\vec{E}$ and $\vec{B}$ are the electric field inside the wire and the magnetic field outside the wire, respectively. The closed loop in the second integral is done in vacuum encircling the given element. As we describe the circuits in the lumped-element model, the voltage and current are independent of the precise path the fields are integrated along in the following sense. For the line integral of the electric field in \cref{eq:Vb} we take the integration path to be well outside the wire of the inductors, meaning that the magnetic field is zero along the path. Similarly for the loop integral of the magnetic field in \cref{eq:Ib}, we take the integration path to be well outside the dielectric of the capacitors, meaning that the electric field is zero along the path. For more details on the integration of electromagnetic fields see, e.g., Ref. \cite{Griffiths1981}.

We define the branch flux and branch charge variables as
\begin{subequations}\label{eq:PhiandQ}
\begin{align}
\Phi_b(t) &= \int_{-\infty}^t V_b(t') \dd{t'}	\label{eq:flux_definition},\\ 
Q_b(t) &= \int_{-\infty}^t I_b(t')\dd{t'}	\label{eq:charge_definition},
\end{align}
\end{subequations}
where it is assumed that the system is at rest at $t'=-\infty$ with zero voltages and currents.
As there are less degrees of freedom in the circuit than there are branches in the circuit, these are, just as the currents and voltages, not completely independent but related through Kirchhoff's laws
\begin{subequations}\label{eq:KirchhoffsLaws1}
\begin{align}
&\sum\limits_{\substack{\text{all $b$ arriving}\\ \text{at $n$}}} Q_b = q_n, \label{eq:KCL} \\
&\sum\limits_\text{all $b$ around $l$}\Phi_b = \tilde\Phi_l, \label{eq:KVL}
\end{align}
\end{subequations}
where $q_n$ is the charge accumulated at node $n$ and $\tilde\Phi_l$ is the external magnetic flux through the loop $l$. A \emph{node} can be understood as a point where components, or branches, converge, see \cref{fig:exampleCircuit}, where we denote nodes with a dot. We can define any circuit as a set of nodes and a set of branches.

The notion of nodes and branches comes from graph theory, which is the natural mathematical language for analyzing circuits. The interested reader can find more details of fundamental graph theory and its application to electrical circuits in \cref{subsec:Fundamental graph theory}.

\subsection{Circuit components}\label{sec:CircuitComponents}

We consider primarily three different components of a superconducting circuit: linear capacitors, linear inductors, and nonlinear Josephson junctions. The two linear components should be well known to most readers, and we therefore introduce them only briefly. The Josephson junction, on the other hand, is a nonlinear component that is specific to superconducting circuits, and it is the main component when working with superconducting qubits.

As we are considering superconducting circuits we do not consider resistors or other losses. Such dissipative components are not easily included in the Hamiltonian formalism presented in this tutorial due to their irreversible nature. However, it can be done using, for instance, the Caldeira-Leggett model \cite{Caldeira1981,Burkard2004}.

\subsubsection{Capacitors}

The first component we consider is the capacitor. For a general capacitor, the charge on the capacitor is determined as a function of the voltage, $q(t) = f[V(t)]$. In this tutorial, we consider only linear capacitors where the voltage is proportional to the charge stored on the capacitor plates
\begin{equation}\label{eq:cap_charge_voltage}
	V(t) = \frac{q(t)}{C},
\end{equation}
where $C$ is the capacitance of the capacitor. This linear relationship is the defining property of the linear capacitor. In reality, this is merely an approximation, as there are small nonlinearities, which makes $C$ a function of $q$ and $V$. These effects are usually small and therefore it is standard to neglect them. \Cref{eq:cap_charge_voltage} can be rewritten to the flux-charge relation using \cref{eq:flux_definition} as
\begin{equation}\label{eq:cap_flux_charge}
	\dot{\Phi}(t) = V(t) = \frac{q(t)}{C},
\end{equation}
where the dot indicates differentiation with respect to $t$.
The charge $q(t)$ is equal to the branch charge, and using \cref{eq:charge_definition} we find the branch current 
\begin{equation}\label{eq:currentCapacitor}
    I(t) = C\ddot{\Phi}(t).
\end{equation}
The energy stored in the capacitor is found by integrating the power $P = V(t)I(t)$ from $t=-\infty$ to $t$
\begin{equation}\label{eq:cap_energy}
    E = \frac{1}{2} C \dot{\Phi} ^2(t).
\end{equation}
For superconducting circuits, typical values of the capacitances are of the order $\SI{10}{\femto\farad}$. In lumped-circuit diagrams we denote the capacitor as a pair of parallel lines, see \cref{fig:exampleCircuit}.

\subsubsection{Inductors}

The time-dependent current flowing through a general inductor is a function of the flux through it, $I(t) = f[\Phi(t)]$. 
For a linear inductor, the current is proportional to the magnetic flux,
\begin{equation}\label{eq:ind_flux_charge}
	I(t) = \dot{q}(t) = \frac{1}{L}\Phi(t),
\end{equation}  
where $L$ is the inductance of the inductor.
Integrating over the power as before, the energy stored in the inductor is then 
\begin{equation}\label{eq:ind_energy}
    E = \frac{1}{2L} \Phi^2(t).
\end{equation}
For superconducting qubits, typical values of linear inductances are of the order $\SI{1}{\nano\henry}$. In lumped-circuit diagrams, we denote the linear inductor as a coil, see \cref{fig:exampleCircuit}.

\begin{figure}
    \centering
    \includegraphics[width=.4\columnwidth]{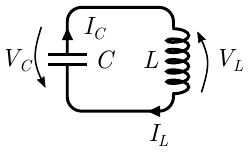}
    \caption{Simple $LC$-oscillator circuit. A capacitor with capacitance $C$ is connected in a closed circuit with an inductor of inductance $L$. The voltages over the two components are $V_C$ and $V_L$, respectively, while the currents are $I_C$ and $I_L$, respectively. The resulting equation of motion is a harmonic oscillator.}
    \label{fig:lc_oscillator}
\end{figure}

As a short clarifying example we consider the classical $LC$ oscillator shown in \cref{fig:lc_oscillator}. From Kirchhoff's current law in \cref{eq:KCL} we know that $I_C = I_L$, where $I_C$ and $I_L$ are the currents through the capacitor and inductor, respectively. Kirchhoff's voltage law gives us  $V_C = -V_L$, assuming no fluctuating external flux. Using \cref{eq:flux_definition,eq:charge_definition,eq:cap_flux_charge,eq:ind_flux_charge} we can set up the equations of motion for the system
\begin{equation}\label{eq:lc_eqom}
	\ddot{\Phi}(t) = -\frac{1}{LC}\Phi(t),
\end{equation}
where we introduce $\Phi(t) = \Phi_C(t) = -\Phi_L(t)$ to get rid of the subscripts. The system behaves as a simple harmonic oscillator in the flux. This is analogous to a spring, where the flux is the position, and the mass and spring constants are replaced by the capacitance and inverse inductance, respectively.

\subsubsection{Josephson junctions}\label{subsubsec:Josephson Junction}

So far we have considered only components with linear current-voltage relations. For reasons that will become clear when we quantize the lumped circuit, constructing a qubit from only linear components is by no means straightforward. We therefore need nonlinear components which come in the form of the Josephson junction. The Josephson junction plays a special role in superconducting circuits, as it has no simple analog in a nonsuperconducting circuit since it is related to charge quantization effects that occur in superconductors. We start with a short introduction to superconductivity (see Ref. \cite{Gallop1991} for more details).

When the temperature is decreased some materials undergo a phase transition where the resistivity drops to zero. Together with the Meissner effect, i.e., that the material perfectly expels all magnetic fields, the perfect conduction is the defining property of a superconductor.

The phase transition between the nonsuperconducting phase and the superconducting phase of a material happens because the conduction electrons condense into a so-called BCS ground state, which is characterized by an amplitude and a phase. \textit{A priori} it might seem impossible for electrons to condense into a single quantum state since the Pauli exclusion principle forbids this. However, as Cooper suggested, some attractive force between the electrons leads to the formation of electron pairs \cite{Cooper1956}, which have integer spin and thus behave like bosons. This makes it possible for these so-called Cooper pairs to condense into a single quantum ground state and in this state the solid becomes superconducting.

\begin{figure}
    \centering
    \includegraphics[width=\columnwidth]{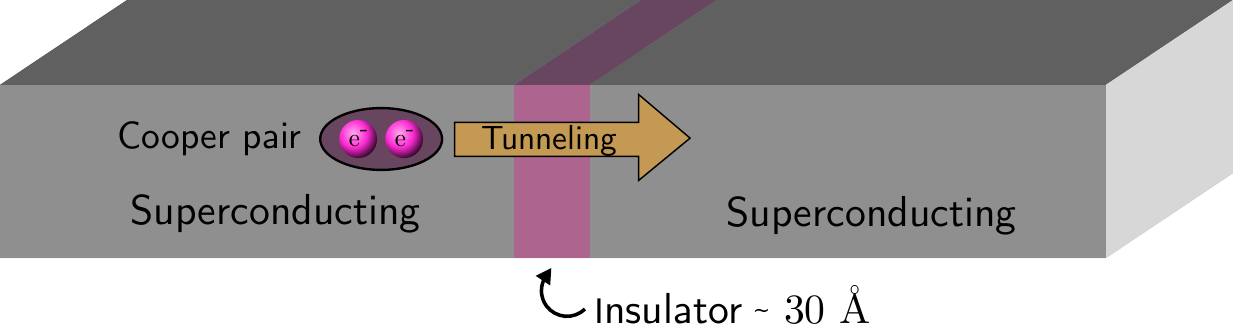}
    \caption{Sketch of a Josephson junction. Two superconducting materials are separated by a thin insulator, with a thickness of the order of $\SI{30}{\angstrom}$. If a nonsuperconducting metal is used as a separator, it can be several micrometers wide. Cooper pairs can tunnel back and forth between the two superconducting materials.}
    \label{fig:JJ}
\end{figure}

A Josephson junction consists of two superconducting islands separated by a thin insulator, a nonsuperconducting metal, or a narrow superconducting wire. Cooper pairs can then tunnel through the barrier from one island to the other, a phenomenon known as the Josephson effect \cite{Josephson1962,Josephson1974}, see \cref{fig:JJ}. The tunneling rate (current) and the voltage between the two islands depends on the superconducting phase difference, $\phi$, between the islands through \cite{Barone1982}
\begin{align}
	I(t) &= I_c \sin [\phi (t)]	\label{eq:josephson_current_phase},\\
	V(t) &= \frac{\hbar}{2e}\dot{\phi}	\label{eq:josephson_voltage_phase},
\end{align}
where $I_c$ is the critical current of the junction, which depends on the junction geometry. \Cref{eq:josephson_voltage_phase} allows us to relate the junction phase difference to the generalized flux through $\Phi = \hbar \phi / 2e$. The charge and flux are thus related through
\begin{equation}\label{eq:josephson_charge_flux}
	\dot{q}(t) = I_c\sin\left(2\pi\dfrac{\Phi(t)}{\Phi_0}\right),
\end{equation}
where we define the magnetic flux quantum $\Phi_0 = h/2e$. The Josephson junction works as a flux-dependent inductor with inductance given by \cite{Devoret2004}
\begin{equation}\label{eq:josephson_inductance}
	L(\Phi) = \left(\pdv{I}{\Phi}\right)^{-1} = \dfrac{L_J}{\cos\left(2\pi\dfrac{\Phi}{\Phi_0}\right)},
\end{equation}
where we define the Josephson inductance $L_J = \Phi_0/2\pi I_c$. Since the inductance is associated with the inertia of the Cooper pairs it is often referred to as kinetic inductance. 
See \cref{sec:fluxonium} for details on the use of large kinetic inductance. For superconducting qubits, typical values of Josephson inductances are of the order $\SI{100}{\nano\henry}$.
The energy of a Josephson junction is also nonlinear. We have
\begin{equation}\label{eq:JJ_energy}
    E = \frac{\Phi_0^2}{(2\pi)^2}\frac{1}{L_J} \left[ 1 - \cos \left( 2\pi \frac{\Phi}{\Phi_0} \right) \right],
\end{equation}
where we often neglect the constant term when dealing with the Lagrangian or Hamiltonian, as it is irrelevant for the dynamics of the system. We define the factor in front of the bracket to be the Josephson energy of the Josephson junction, $E_J = \Phi_0^2/(2\pi)^2L_J = \Phi_0I_c/2\pi $. In this tutorial we denote Josephson junctions as a boxed "x" in lumped-circuit diagrams, see \cref{fig:exampleCircuit}. In the literature sometimes an "x" without a box is used. 

It is conventional to simplify notation in a way such that charges and fluxes become dimensionless. This is done by using units where 
\begin{equation}\label{eq:dimensionless}
    \hbar = 2e = 1 \qquad \text{and thus} \qquad \frac{\Phi_0}{2\pi} = 1.
\end{equation}
This means that we get rid of the cumbersome factor of $2\pi/\Phi_0$ in the sinusoidal Josephson junction terms.
Note that in this convention the units of capacitance and inductance become inverse energy. Moreover, with this choice of units the junction phase differences are equal to the generalized flux $\phi = \Phi$, and the energy of a Josephson junction becomes equal to the critical current, $E_J = I_c$.

\subsubsection{dc SQUID}\label{sec:dcSquid}

\begin{figure}
    \centering
    \includegraphics[scale=.8]{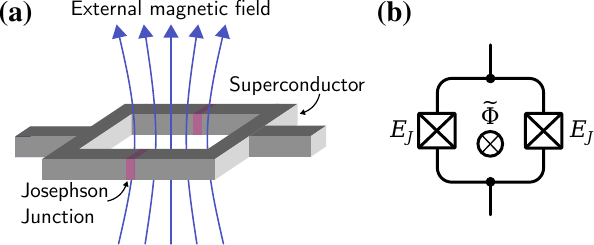}
    \caption{A dc superconducting quantum interference device (dc SQUID). \textbf{(a)} Implementation of a dc SQUID. \textbf{(b)} Corresponding circuit diagram.}
    \label{fig:squid}
\end{figure}

It is often desirable to be able to tune the parameters of the circuit externally. Therefore many circuits employ a direct current superconducting quantum interference device, or dc SQUID, instead of a single Josephson junction. A dc SQUID consists of two Josephson junctions on a ring, with an external magnetic field, $\tilde \Phi$, through the ring \cite{Jaklevic1964}, see \cref{fig:squid}(a). While this does not change the form of the energy of the Josephson junction, it has the advantage that it makes the front factor in \cref{eq:JJ_energy} tunable. To see this consider the circuit diagram in \cref{fig:squid}(b). The energy of this component must be the sum of two Josephson junctions
\begin{equation}\label{eq:squidInitial}
U = - E_J\cos \left(\Phi_L + \frac{\tilde\Phi}{2} \right) - E_J\cos \left(\Phi_R + \frac{\tilde\Phi}{2} \right),
\end{equation}
where $\Phi_{L/R}$ is the branch flux of the left and right branch, respectively, and we divide the external flux equally between the two arms of the dc SQUID following Kirchhoff's voltage law in \cref{eq:KVL}. Note that here we consider symmetrical junctions, but it is a neat exercise to extend it to asymmetrical junctions.

Since we are considering the arms of a loop, we can write $\Phi = \Phi_L = -\Phi_R$ in \cref{eq:squidInitial}.
Using the trigonometric identity $2\cos\alpha\cos\beta = \cos(\alpha - \beta) + \cos(\alpha+\beta)$ with $\alpha = \tilde\Phi/2$ and $\beta = \Phi$, we can rewrite \cref{eq:squidInitial} into the form
\begin{equation}
	U = -2E_J\cos \left(\frac{\tilde\Phi}{2}\right) \cos \Phi.	
\end{equation}
The so-called {\it fluxoid quantization condition} states that the algebraic sum of branch fluxes of all the inductive elements along the loop plus the externally applied flux must equal an integer number of superconducting flux quanta \cite{deaver1961,Doll1961,Krantz2019}, i.e.,
\begin{equation}
    \Phi + \tilde\Phi = 2\pi k,
\end{equation}
where $k$ is an integer. Together with Kirchhoff's voltage law in \cref{eq:KVL} this means that we can remove a degree of freedom. This explains how one goes from two branch fluxes, $\Phi_{L/R}$, to just one branch flux, $\Phi$, since the branch fluxes are the system degrees of freedom.
In other words we obtain, an effective Josephson energy of $E'_J(\tilde\Phi) = 2E_J|\cos(\tilde\Phi/2)|$, where the Josephson energy can be dynamically tuned through the external flux, $\tilde\Phi$. 
This idea is often implemented in superconducting circuits instead of a single Josephson junction so that the spacing of the energy levels can be tuned dynamically by tuning $\tilde\Phi$. However, we usually just place a single Josephson junction in a circuit diagram. Due to the sensitivity of the dc SQUID it has many uses especially in clinical applications such as magnetoencephalography \cite{Hamalainen1993,Gratta2001}, magnetocardiography, and magnetic resonance imaging (MRI), where they are used for detecting tiny magnetic fields  in living organisms \cite{Koelle1999,Kleiner2004}.

\subsubsection{Voltage and current sources}

We can treat constant voltage and current sources by representing them as capacitors or inductors. Consider a constant voltage source $V$. This can be represented by a very large but finite capacitor, in which an initially large charge $Q$ is stored such that $V = Q/C$ in the limit where $C \rightarrow \infty$.
Similarly, a constant current source can be represented by a very large but finite inductor, in which an initially large flux $\Phi$ is stored, such that $I=\Phi/L$ in the limit where $L \rightarrow \infty$.

\section{Equations of motion}\label{sec:EOM}

In order to describe the dynamics of the lumped-circuit diagrams we presented in the previous section, we now determine the equations of motion for the systems. The equations of motion depend on the circuit components and can be written in terms of the circuit variables using either the voltage and current in \cref{eq:VandI} or equivalently using the flux and charge in \cref{eq:PhiandQ}.
There are several ways of finding the equations of motion, and we start from the simplest approach; applying Kirchhoff's laws directly to the circuit. From this starting point, we then progress to the method of nodes and then to the Lagrangian and Hamiltonian.

\subsection{Applying Kirchhoff's laws directly}\label{sec:KirchhoffsDirectly}

The simplest way to find the equations of motion for a given circuit is to apply Kirchhoff's laws. We have already done this for the simple $LC$ oscillator example in \cref{fig:lc_oscillator}, which yielded the harmonic oscillator equation of motion in \cref{eq:lc_eqom}. To get a better feel for this procedure, let us consider a few additional examples.

\begin{figure}
    \centering
    \includegraphics[width=.4\columnwidth]{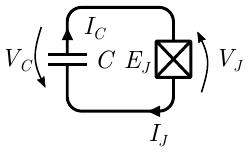}
    \caption{A Josephson junction and capacitor circuit. A capacitor with capacitance $C$ is connected in a closed circuit with a Josephson junction, $E_J$. The voltage over the two components is $V_C$ and $V_J$, respectively, while the currents are $I_C$ and $I_J$, respectively. The resulting equation of motion is a Duffing oscillator.}
    \label{fig:JJc_oscillator}
\end{figure}

The next natural step is to exchange the linear inductor in \cref{fig:lc_oscillator} with a nonlinear Josephson junction. This yields the circuit in \cref{fig:JJc_oscillator}. From Kirchhoff's current law in \cref{eq:KCL} we know that $I_C = I_J$, where $I_C$ and $I_J$ are the currents through the capacitor and Josephson junction, respectively. Kirchhoff's voltage law implies  $V_C = -V_J$. Using \cref{eq:flux_definition,eq:charge_definition,eq:cap_flux_charge,eq:ind_flux_charge} we can set up the equations of motion for the system,
\begin{equation}\label{eq:JJc_eqom}
	\ddot{\Phi}(t) = -\frac{I_c}{C}\sin\Phi(t),
\end{equation}
where we introduce $\Phi(t) = \Phi_C(t) = -\Phi_J(t)$. \Cref{eq:JJc_eqom} is identical to the equation of motion for a simple pendulum, with the critical current, $I_c$, playing the role of the gravitational constant and the capacitance, $C$, becoming the mass of the pendulum, similar to the case of the $LC$ circuit, see \cref{eq:lc_eqom}, which is the lowest order approximation to \cref{eq:JJc_eqom}. Contrary to \cref{eq:lc_eqom} this is not linear in $\Phi$, which is an effect of the introduction of the nonlinear Josephson junction.

\begin{figure}
    \centering
    \includegraphics[width=.8\columnwidth]{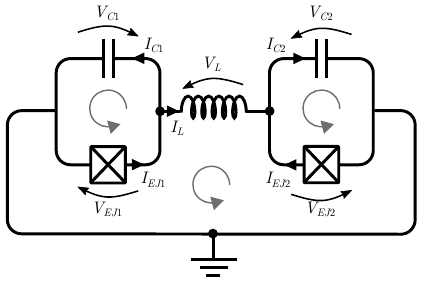}
    \caption{Example circuit of \cref{fig:exampleCircuit} with explicit directions of the voltages and currents shown. All loops are propagated counterclockwise.}
    \label{fig:exampleVI}
\end{figure}

We now continue to the more complicated example of \cref{fig:exampleCircuit}. This time Kirchhoff's voltage law gives us three equations, one for each loop of the circuit. We denote the left capacitor and Josephson junction $C_1$ and $E_{J,1}$, respectively. Similarly, we have to the right $C_2$, $E_{J,2}$. The connecting inductor is denoted by $L_{12}$. Defining the direction of the current and voltages as in \cref{fig:exampleVI}, we find the following equations from \cref{eq:KVL}
\begin{subequations}\label{eq:kvl_ex}
\begin{align}
    -\Phi_{EJ1} - \Phi_{C1} &= 0, \label{eq:kvl1_ex}\\
    \Phi_{EJ2} + \Phi_{C2} &= 0, \label{eq:kvl2_ex}\\
    \Phi_{EJ1} - \Phi_{EJ2} + \Phi_L &= \tilde \Phi,
\end{align}
\end{subequations}
where $\tilde \Phi$ is the external flux in the inductor loop. We propagate all loops counterclockwise, which yields negative signs on the terms in \cref{eq:kvl_ex} when the voltage of the given branch is in the opposite direction to the loop direction.
Note that we can also include external fluxes in the two capacitive loops. However, as we will see in \cref{sec:lagrangianApproach}, as long as we consider only time-independent fluxes, the external fluxes will only be relevant in purely inductive loops. 
From \cref{eq:kvl1_ex,eq:kvl1_ex} we define $\Phi_1 = \Phi_{EJ1} = - \Phi_{C1}$ and $\Phi_2 = \Phi_{EJ2} = - \Phi_{C2}$. Using this we can also express the flux through the inductor as $\Phi_L = \Phi_{2} - \Phi_{1} + \tilde\Phi$, which significantly reduces the number of variables.

From Kirchhoff's current law we find the following equations
\begin{equation}
    -I_{Cn} + I_{EJn} = \pm I_L,
\end{equation}
for $n = 1,2$, where the plus is for $n=1$ and the minus is for $n=2$. Inserting the current relations for the respective components, we find the following equations of motion
\begin{equation}\label{eq:eomExample}
    C_n\ddot \Phi_n = \pm \frac{1}{L_{12}}(\Phi_2 - \Phi_1 +\tilde\Phi) - E_{J,n} \sin\Phi_n,
\end{equation}
for $n=1,2$.

The end goal of our analysis is to quantize the circuit to treat it quantum mechanically. When doing quantum mechanics we are usually interested in the Hamiltonian of the system, as it is closely related to the energy spectrum and time evolution of the system. It is possible to infer the system Hamiltonian from the equations of motion. This is usually done by finding a Lagrangian that yields the equation of motion using Lagrange's equations [see \cref{eq:Lagrang}] and then performing a Legendre transformation. 

While the approach of applying Kirchhoff's law directly always yields the correct equations of motion, it quickly becomes cumbersome as the circuits increase in complexity. We, therefore, seek a method for determining the Lagrangian directly. This can be achieved using the method of nodes.

\subsection{Method of nodes}\label{sec:MethodofNodes}

In this section, we present the method of nodes which solves most practical problems involving Josephson junctions. The discussion follows the method proposed by Devoret \cite{Devoret1985,Vool2017}.

Our main obstacle when determining the Lagrangian of a given circuit is to remove superfluous degrees of freedom and determine how to include the external fluxes. As we saw above, we can solve these problems by manipulating Kirchhoff's law. Here we present an alternative approach.

We have already defined a node as a point where one or more components connect. We now further define a \emph{ground node} as a node connected to ground. These nodes are inactive since the flux through them is zero and thus they do not contribute to the dynamics of the system, and can thus be ignored. For the remaining nodes, we distinguish between active and passive nodes. An \emph{active node} is defined as a node where at least one capacitor and one inductor (either linear and Josephson junction) meet. A \emph{passive node} is defined as a node where only one type of component meet, either only capacitors or only inductors. It turns out that passive nodes represent superfluous degrees of freedom and therefore only yield constraints on the dynamics of the system. This is similar to how one may determine an effective capacitance for a serial or parallel collection of individual capacitances. 

Considering the example circuit in \cref{fig:exampleCircuit}, we can represent the circuit as a set of branches, $\mathcal{B}$, and a set of nodes, $\mathcal{N}$. The set of nodes consists of three nodes; two active nodes and a ground node. The set of branches is equal to the set of components in the circuit, i.e., the example circuit has five branches; two capacitor branches, two Josephson junction branches, and a single linear inductor branch.

\begin{figure}
    \centering
    \includegraphics[width=.6\columnwidth]{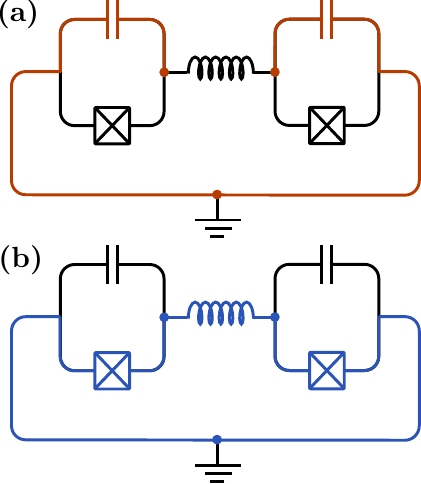}
    \caption{Highlight of \textbf{(a)} the capacitive subgraph and \textbf{(b)} the inductive subgraph of the example circuit in \cref{fig:exampleCircuit}.}
    \label{fig:subgraphs}
\end{figure}

We call such a representation consisting of a set of nodes and a set of branches a \emph{network graph} or simply a \emph{graph}. With this notation, we can divide the circuit into subgraphs. For a given circuit there are many possible subgraphs, but we focus on the capacitive subgraph and the inductive subgraph. The \emph{capacitive subgraph} contains only branches of capacitors and the nodes connected to such branches. The \emph{inductive subgraph} contains only branches of inductors and nodes connected to such branches. In the example circuit in \cref{fig:exampleCircuit} the two capacitor branches and all three nodes are in the capacitive subgraph, while the inductor branch and the two Josephson junction branches are in the inductive subgraph together with the three nodes. Notice how the nodes can be in both subgraphs at the same time, see \cref{fig:subgraphs}.

The capacitive subgraph consists only of linear capacitors, and thus we can express the energy of a capacitive branch in terms of the voltages, i.e., the derivative of the flux, using \cref{eq:cap_energy}. By doing this we have now broken the symmetry between the charge and flux, and the flux can now be viewed as the \enquote{position.} With this treatment, the capacitive energy becomes equivalent to the kinetic energy, while the inductive energy becomes equivalent to the potential energy.

This symmetry breaking also explains why passive nodes do not contribute to the dynamics of the system. A passive node in between two inductors does not have any kinetic energy and can therefore be considered stationary. On the other hand, a node in between two capacitors does have kinetic energy, but no potential energy, and can therefore be considered a free particle which does not interact with the rest of the system.

However, any realistic inductor (both linear and nonlinear) will always introduce some capacitance since a capacitance occurs whenever two conducting materials are in close proximity to each other. Consider the Josephson junction in \cref{fig:JJ}, it quite closely resembles a linear plate capacitor, thus it is expected that some parasitic capacitance will be present in parallel with the inductor. Nonetheless, we can often make this parasitic capacitance so small that it can be neglected in the lumped-element circuit. One should, however, be aware of these capacitances when designing superconducting circuits.

\subsubsection{Spanning tree}

We are now ready to consider the most important subgraph of the circuits: the spanning tree. The \emph{spanning tree} is constructed by connecting every node in the circuit to each of the other nodes by only one path. See \cref{def:spanning_tree} in \cref{subsec:Fundamental graph theory} for a more mathematical definition using graph theory. Note that there are often several choices for the spanning tree. This is not a problem for the analysis and can be seen analogous to the choice of a particular gauge in electromagnetic field theory or the choice of a coordinate system in classical mechanics.

Choosing a spanning tree for a given circuit partitions the branches into two sets: The set of branches on the spanning tree, $\mathcal{T}$, and its complementary set, $\bar{\mathcal{T}} = \mathcal{B} \setminus \mathcal{T}$, i.e., the branches \emph{not} on the spanning tree. We call the latter set the set of closure branches because its branches close the loop of the spanning tree. 

We use the spanning tree to determine where to include the external fluxes of the system. Following Kirchhoff's laws, the flux $\phi_n$ of node $n$ can be written as the sum of incoming and outgoing branch fluxes, with a suitable sign depending on the direction of the flux. With this in mind, we can write the branch fluxes in terms of the node fluxes
\begin{subequations}
\begin{align}
    \Phi_{b \in \mathcal{T}} &= \phi_n - \phi_{n'}, \\
    \Phi_{b \in \bar{\mathcal{T}}} &= \phi_n - \phi_{n'} + \tilde\Phi,
\end{align}
\end{subequations}
where $n$ and $n'$ are the nodes at the start and end of the given branch, respectively, and $\tilde\Phi$ is the external flux through the loop closed by the branch. Note that the external flux occurs only if the branch is a closure branch. The fact that external fluxes do not appear in every branch is due to Kirchhoff's law in \cref{eq:KVL}, which eliminates the external flux on some of the branches. One can therefore choose onto which branches these external fluxes should be included, as long as \cref{eq:KVL} is satisfied, which is exactly the choice we make by choosing the spanning tree. 

Substituting the node fluxes into the expressions for the energy of the different components, i.e., into \cref{eq:cap_energy,eq:ind_energy,eq:JJ_energy}, we can express the energies as a function of the node fluxes. The results can be seen in \cref{tab:energies}.

\begin{table}
	\centering
	\caption{Energies of different components on either the spanning tree or a closure branch of the circuit. The magnetic flux through the closure branch due to external fields is denoted $\tilde\Phi_b$. The time derivative of the magnetic flux is included for linear capacitors on closure branches for completeness. For the rest of this tutorial, we assume time-independent external fluxes, i.e., $\dot{\tilde\Phi}_b=0$. We refer to Refs. \cite{You2019,Riwar2021} for a discussion of time-dependent fluxes.}
	\label{tab:energies}
	\begin{tabular}{lcc}
		\toprule
		Element & Spanning tree & Closure branch \\
		\midrule
		& & \\[-1em]
		\parbox[l]{0cm}{Linear\\capacitor} & $\dfrac{C}{2}(\dot{\phi}_n - \dot{\phi}_{n'} )^2$ & $\dfrac{C}{2}(\dot{\phi}_n - \dot{\phi}_{n'} + \dot{\tilde\Phi}_b )^2$ \\[0.9em]
		\parbox[l]{0cm}{Linear\\inductor} & $\dfrac{1}{2L}(\phi_n - \phi_{n'} )^2$ & $\dfrac{1}{2L}(\phi_n - \phi_{n'} + \tilde\Phi_b )^2$ \\[0.9em]
		\parbox[l]{0cm}{Josephson\\junction} & $- E_J \cos\left(\phi_n - \phi_{n'} \right)$ & $ -E_J \cos(\phi_n - \phi_{n'} + \tilde\Phi_b )$ \\[0.5em]
		\bottomrule
	\end{tabular}
\end{table}

Note that if the circuit contains only time-independent external fluxes, it is often an advantage to choose a spanning tree containing as few capacitors as possible, such that the capacitors lie on the closure branches. The reason is that a time-independent external flux disappears from capacitive terms since $\dot{\tilde \Phi} = 0$. When working with time-independent external fluxes these are therefore only relevant in purely inductive loops. Time-dependent external fluxes are beyond the scope of this tutorial, see Refs. \cite{You2019,Riwar2021} for a treatment of this case. 

If we consider the example circuit in \cref{fig:exampleCircuit} we can choose the spanning tree in many different ways. Since we consider only time-independent external fluxes, a particularly nice choice of spanning would be over the two Josephson junction (JJ) branches, which means that any external flux will appear only in the linear inductor term. For this reason, we do not need to worry about any external fluxes through the two capacitive loops.

\subsection{Lagrangian approach}\label{sec:lagrangianApproach}

Having chosen a spanning tree for our circuit, we are now ready to determine its Lagrangian. The Lagrangian is found by subtracting the potential (inductive) energies from the kinetic (capacitive) energies
\begin{equation}\label{eq:defLagrangian}
    \L = T - U = T_\text{cap} - U_\text{ind} - U_\text{JJ},
\end{equation}
where $T$ is the kinetic energy and $U$ is the potential energy. The subscripts indicate the type of element each term refers to. 

With the definition of the Lagrangian and the energies of \cref{tab:energies}, we can write the Lagrangian for the example circuit in \cref{fig:exampleCircuit} as
\begin{equation}
\begin{aligned}\label{eq:exampleLagrangian}
\L &= \frac{C_1}{2}\dot\phi_1^2 + \frac{C_2}{2}\dot\phi_2^2 - \frac{1}{2L_{12}} (\phi_2 - \phi_1 + \tilde\Phi)^2 \\ 
&\phantom{=}+ E_{J,1}\cos\phi_1 + E_{J,2}\cos\phi_2,
\end{aligned}
\end{equation}
where $C_n$ and $E_{J,n}$ is the capacitance and Josephson energy of the capacitor and Josephson junction, respectively. The index $n=1,2$ corresponds to the left and right side, respectively. The inductance of the inductor is denoted $L_{12}$.
With the Lagrangian, one can obtain the equations of motion from Lagrange's equations
\begin{equation}\label{eq:Lagrang}
    \dv{t}\pdv{\L}{\dot \phi_n} = \pdv{\L}{\phi_n}.
\end{equation}
Applying this to the example circuit, we find the equations of motion
\begin{equation}
    \ddot \phi_n = \mp \frac{1}{L_{12}C_n}(\phi_2 - \phi_1 +\tilde\Phi) - \frac{E_{J,n}}{C_n} \sin\phi_n,
\end{equation}
where the minus is for $n=1$ and the plus is for $n=2$. This is identical to \cref{eq:eomExample} written up with node fluxes instead of branch fluxes.

\subsubsection{Using matrices}\label{sec:lagrangianMatrices}

Writing the Lagrangian as in \cref{eq:exampleLagrangian} can be rather tedious for larger circuits since it includes a lot of sums. We, therefore, seek a more elegant way to write the Lagrangian. This is achieved using matrix notation. First we list all the nodes 1 to $N$ and define a flux column vector $\vec{\phi}^T = (\phi_1, \dots, \phi_N)$, where $T$ indicates the transpose of the vector. Note that for a grounded circuit we do not include the ground node since its flux equals zero and it does not contribute to the true degrees of freedom in any case. We can always choose a ground node in our circuits as one mode will always decouple from the remaining modes for ungrounded circuits, see \cref{sec:changeOfBasis}.

We are now ready to set up the capacitive matrix $\vec C$ of the system. The nondiagonal matrix elements are minus the capacitance, $C_{jk}$, connecting nodes $j$ and $k$. The diagonal elements consist of the sum of the nondiagonal values in the corresponding row or column, multiplied by $-1$, i.e., $C_{jj} = \sum_{k\neq j} C_{jk}$. If a node is connected to ground via a capacitor, this capacitance must also be added to the diagonal element. With this $N\times N$ matrix we can write the kinetic energy term as
\begin{equation}\label{eq:E_kin}
    T = \frac{1}{2} \dot{\vec\phi}^T \vec C \dot{\vec\phi}.
\end{equation}

In the case of the example circuit in \cref{fig:exampleCircuit}, the flux column vector is $\vec{\phi}^T = (\phi_1, \phi_2)$, and the capacitive matrix becomes
\begin{equation}
    \vec C = \mat{C_1 & 0 \\ 0 & C_2}.
\end{equation}

We now consider the contribution from the linear inductors. We set up the inductive matrix $\vec {L^{-1}}$ in the same way as the capacitive matrix. The nondiagonal elements are $-1/L_{jk}$ if an inductance $L_{jk}$ connects nodes $j$ and $k$, and zero otherwise, while the diagonal elements consist of the sum of values in the corresponding row or column, multiplied by minus one, $1/L_{jj} = \sum_{k\neq j} 1/L_{jk}$. If a node is connected to the ground via an inductor, this inductance must also be added to the diagonal element. Of course, if no inductor is connecting two nodes, the element should be zero.
We must also include the external magnetic flux in this term. Thus the energy due to linear inductors becomes
\begin{equation}\label{eq:E_inductors}
    U_\text{ind} = \frac{1}{2} \vec\phi^T \vec {L^{-1}} \vec\phi + \sum_{b \in \bar{ \mathcal{T}}} \frac{1}{L_b} (\phi_n - \phi_{n'}) \tilde\Phi_b,
\end{equation}
where we remove all irrelevant constant terms. The second term sums over all the inductive closure branches of the circuit, where $n$ and $n'$ are the nodes connected by branch $b$. 

If we consider the example circuit again, the inductive matrix is
\begin{equation}
    \vec {L^{-1}}= \mat{1/L_{12} & -1/L_{12} \\ -1/L_{12} & 1/L_{12}},
\end{equation}
where $L_{12}$ is the inductance of the linear inductor.
With this the inductive energy of the example circuit becomes
\begin{equation}\label{eq:inductorEnergy}
    U_\text{ind} = \frac{1}{2} \vec\phi^T \vec {L^{-1}} \vec\phi + \frac{1}{L_{12}} (\phi_1 - \phi_2) \tilde\Phi,
\end{equation}
where $\tilde\Phi$ is the external flux through the inductive loop. When there are only a few linear inductors, as in the example circuit, it might be more straightforward to write the energy without the matrix notation.
We do not attempt to write the Josephson junction terms using matrix notation as they are nonlinear functions of the node flux variables.

\subsection{Hamiltonian approach}\label{sec:ObtainHamiltonian}

The Hamiltonian of the circuit can be found by a simple transformation of the Lagrangian through what is commonly referred to as a Legendre transformation. First, we define the conjugate momentum to each node flux by
\begin{equation}
    q_n = \pdv{\L}{\dot \phi_n},
\end{equation}
which in vector form becomes $\vec{q} = \vec C \dot{\vec \phi}$. If the capacitance matrix is invertible we can express $\dot{\vec \phi}$ as a function of $\vec{q}$. We denote the conjugate momenta as node charges since they correspond to the algebraic sum of the charges on the capacitances connected to node $n$.

The Hamiltonian can now be expressed in terms of the node charges, $q_n$, for the kinetic energy and node fluxes, $\phi_n$, for the potential energy through the Legendre transform
\begin{equation}
\begin{aligned}
    \H &= \dot{\vec{\phi}}^T \vec q - \L\\
    &= \frac{1}{2} \vec q^T \vec C^{-1} \vec q + U(\vec \phi),
\end{aligned}
\end{equation}
where the potential energy is a nonlinear function of the node fluxes. Note that the functional form of the Hamiltonian may differ depending on the choice of spanning tree. This is because the choice of flux-node coordinates is not unique, much like the electrodynamic potentials, which have a \enquote{gauge freedom} in which certain functions can be added to the potentials without any change to the physics, or more concretely; without changes to the electric and magnetic fields \cite{Griffiths1981}. Here a different choice of flux variables would correspond to a change of gauge as well and a physical quantity like the total energy should not change under such a transformation.

With the Hamiltonian, it is possible to find the equations of motion using Hamilton's equations
\begin{equation}\label{eq:Hamiltons}
    \dot \phi_n = \pdv{\H}{q_n}, \qquad \dot q_n = -\pdv{\H}{\phi_n},
\end{equation}
which yields results for the equations of motion that are equivalent to Lagrange's equations \cref{eq:Lagrang}.

\subsection{Normal modes}\label{sec:normalModes}

Lagrange's equations tell us that for all passive nodes $\dot q_n = 0$, since for a passive note we have $\pdv{\H}{\phi_n}=0$. This means that the circuit has at most the same number of true degrees of freedom as the number of active nodes except the ground node. The number of true degrees of freedom turn out to be identical to the number of normal modes of the system. If all inductors can be approximated as linear inductors (and external fluxes are ignored), the Lagrangian takes the form
\begin{equation}
    \L = \frac{1}{2} \vec{\dot{\phi}}^T \vec C \vec{\dot{\phi}} - \frac{1}{2} \vec\phi^T \vec {L^{-1}} \vec\phi.
\end{equation}
This simple form of the Lagrangian means that the equations of motion become
\begin{equation}
    \vec C \vec{\ddot \phi} = -\vec{L^{-1}} \vec \phi,
\end{equation}
which is essentially Hooke's law in matrix form where the capacitances play the role of the masses and inductances play the role of the spring constants \cite{Taylor2005}. The normal modes of the full systems can be found as the eigenvectors of the matrix product $\vec \Omega^2 = \vec C^{-1} \vec {L^{-1}}$ associated with nonzero eigenvalues. These nonzero eigenvalues correspond to the squared normal mode frequencies of the circuit. Note that $\vec{C}^{-1}$ and $\vec{L^{-1}}$ can always be diagonalized simultaneously since they are both positive definite matrices \cite{Taylor2005}.
It can be advantageous to find these eigenmodes and use them as a basis as it reduces the number of couplings between modes.

\subsection{Change of basis}\label{sec:changeOfBasis}

Here we present a method for changing into the normal mode basis of a circuit.
Given a circuit with $ N $ nodes and a matrix product $ \vec{\Omega}^2=\vec{C^{-1}}\vec{L^{-1}} $, let $ \vec{v}_{1},\vec{v}_{2},\dots,\vec{v}_{n} $ be the orthonormal eigenvectors of $\vec{\Omega}^2$, with eigenvalues $ \chi_{1},\chi_{2},\dots,\chi_{n} $. 
Let $ \vec{\phi}$ be the usual vector of the node fluxes of the circuit. We can then introduce the normal modes $ \vec{\psi} $ via
\begin{align}\label{eq:basischange}
\vec{\phi} = \vec{\mathcal{V}}\vec{\psi},
\end{align}
where 
\begin{align}\label{eq:vtransformation}
\vec{\mathcal{V}} = \mat{\mid & \mid & & \mid \\ \vec{v}_{1} & \vec{v}_{2} & \cdots & \vec{v}_{N} \\ \mid & \mid & & \mid},
\end{align}
is a matrix whose columns are the eigenvectors of $ \vec{\Omega}^2 $. The kinetic energy term in \cref{eq:E_kin} can now be written 
\begin{align}\label{eq:transformedEkin}
T = \frac{1}{2}\dot{\vec{\psi}}^{T}\vec{K}\dot{\vec{\psi}},
\end{align}
where we introduce the capacitance matrix in the transformed coordinates $ \vec{K} = \vec{\mathcal{V}}^{T}\vec{C}\vec{\mathcal{V}} $.

While we assume the columns of \cref{eq:vtransformation} to be the eigenvectors of $\vec \Omega^2$, this is not a requirement, and one can rotate to any frame using an orthonormal basis to construct $\vec{\mathcal{V}}$. However, only if one uses the eigenvectors of $\vec \Omega^2$ will the transformed capacitance matrix, $\vec{K}$, be diagonal with entries $ \lambda_{i} $. In terms of the canonical momenta $ \vec{p}$ conjugate to $ \vec{\psi} $, the kinetic energy takes the usual form
\begin{align}\label{eq:EkinTransformed}
T = \frac{1}{2}\vec{p}^{T}\vec{K}^{-1}\vec{p},
\end{align}
where the inverse of $ \vec{K} $ is trivial to find if it is  a diagonal matrix, yielding the entries $ 1/\lambda_{i} $. In the above, we have assumed that $\vec C$ is positive definite, which is usually the case. This means that $\lambda_{i} \neq0$. We comment on the case where $\vec C$ is not positive definite below. 

We must also consider how contributions from the higher-order terms of inductors behave under this coordinate transformation. Even though we have approximated all inductors as linear to find the normal modes, higher-order terms from Josephson junctions still contribute as corrections, often leading to couplings between the modes. Such terms transform the following way
\begin{align}
\phi_{k} - \phi_{l} \rightarrow \sum_{i}\left[(\vec{v}_{i})_{k} - (\vec{v}_{i})_{l}\right]\psi_{i}, \label{eq:modeTrans}
\end{align}
where $ (\vec{v}_{i})_{k} $ is the $ k $th entry of $ \vec{v}_{i} $. Considering for instance fourth-order terms in $\phi$, this can result in both two-body interactions as well as interactions beyond two body. These multibody interactions can complicate the equations of motion beyond what the change of basis adds in terms of simplification. Coordinate transformations are therefore often most useful in cases where the capacitors are symmetrically distributed, which results in simple normal modes.

The center-of-mass (CM) mode plays a special role in analytical mechanics, as it often decouples from the dynamics of the system. The same is the case for electrical circuits. The center-of-mass mode corresponds to $ \vec{v}_\text{CM} = (1,1,\dots,1)^{T}/\sqrt{N} $, which yields $ \psi_\text{CM} = (\phi_{1} + \phi_{2} + \cdots + \phi_{N})/\sqrt{N} $. This mode is always present and it corresponds to charge flowing equally into every node of the circuit from ground and oscillating back and forth between ground and the nodes. Furthermore, since all its entries are identical it always disappears in the linear combination of \cref{eq:modeTrans} [$(\vec{v}_\text{CM})_{k} - (\vec{v}_\text{CM})_{l} = 0$]. Hence, this mode is completely decoupled from the dynamics. 

The decoupling of this mode is related to how we can arbitrarily choose a node in our circuit as the ground node, whose node flux does not enter into our equations, or rather is identically set to zero. For an ungrounded circuit $\vec C$ is no longer positive definite and we have $ \lambda_\text{CM} = 0 $, making $ \vec{K} $ singular. We therefore always assume the circuit is grounded such that $ \lambda_\text{CM} \neq 0 $.

For an example of multibody interactions see \cref{sec:multibodyInteractions}. For other examples of changes of basis see \cref{sec:delftCoupler,sec:zeroPi}.

\section{Quantization and effective energies}\label{sec:quantizeAndEffective}

\subsection{Operators and commutators}\label{sec:quantize}

We now quantize the classical Hamiltonian to obtain a quantum-mechanical description of the circuit. This is done through canonical quantization, replacing all the variables and the Hamiltonian with operators
\begin{equation}
\begin{aligned}
\phi_n &\rightarrow \hat\phi_n, \\
q_n &\rightarrow \hat q_n, \\
\H &\rightarrow \hat \H,
\end{aligned}
\end{equation}
where $\hat\phi_n$ is the node flux operator corresponding to position coordinates, $\hat q_n$ is the conjugate momentum, and $\hat\H$ is the Hamiltonian operator. 
If the flux operator and the conjugate momentum operator are not constants of motion they obey the canonical commutator relation
\begin{equation}\label{eq:commutatorphiq}
    [\hat\phi_n, \hat q_m] = \hat \phi_n \hat q_m - \hat q_m \hat \phi_n = i\hbar \delta_{nm},
\end{equation}
where $\delta_{nm}$ is the Kronecker delta. The commutator relation in \cref{eq:commutatorphiq} does not hold if a given node, $n$, is not a true degree of freedom. This can happen in case the variable does not appear in $\hat\H$, 
and therefore the commutator between the variable and the Hamiltonian will be zero. This means that $\hat\phi_n$ or $\hat q_n$ will be constant of motion according to Heisenberg's equation of motion. This 
is of course also true for the classical variables as seen in Hamilton's equations in \cref{eq:Hamiltons}.

The commutator relation can be found using the value of the classical Poisson bracket, which determines the value of the corresponding commutator up to a factor of $i\hbar$, as Dirac argued \cite{Dirac1967}. Using this for the branch flux operators and the charge operators, both defined in \cref{eq:PhiandQ}, we find that the Poisson bracket is
\begin{equation}
    \{ \Phi_b, Q_b \} = \sum_n \left[\pdv{\Phi_b}{\phi_n}\pdv{Q_b}{q_n} - \pdv{Q_b}{\phi_n}\pdv{ \Phi_b}{q_n} \right] = \pm 1,
\end{equation}
where the sign is plus for a capacitive branch and minus for an inductive branch. Following Dirac's approach, we arrive at the following commutator relation
\begin{equation}\label{eq:commutatorPhiQ}
    [\hat \Phi_b, \hat Q_b] = \pm i\hbar,
\end{equation}
which is equivalent to the commutator in \cref{eq:commutatorphiq}.
Note that in general, these branch operators are not conjugate in the Hamiltonian. One must still find the true degrees of freedom before quantization is applied.

\subsection{Effective energies}\label{sec:effectiveEnergies}

Consider the generalized momentum $\hat{\vec q} = \vec C \dot{\hat{\vec \phi}}$. The time derivative of the generalized momentum is exactly the current through the capacitors, $\hat{\vec{I}} = \vec C \ddot{\hat{\vec \phi}}$. Note that in the limiting case of one node, this reduces to the current over a single parallel-plate capacitor, as it should. For this reason, it makes sense to think of the conjugate momentum as the sum of all charges on the capacitors attached to a given node. We therefore define
\begin{equation}
    \hat n_n = -\frac{\hat q_n}{2e}
\end{equation}
as the net number of Cooper pairs stored on the $n$th node. If we consider the kinetic energy of a circuit, we can write
\begin{equation}\label{eq:kineticEnergyEffective}
    \hat T = \frac{1}{2}  \hat {\vec q}^T \vec C^{-1} \hat{\vec q} = 4 \frac{e^2}{2} \hat{\vec n}^T \vec C^{-1} \hat{\vec n}.
\end{equation}
Now for each diagonal element, we have a contribution of $4E_{C,n}\hat n_n^2$, where we define the effective capacitive energy of the $n$th node as
\begin{equation}
    E_{C,n} = \frac{e^2}{2} (\vec C^{-1})_{(n,n)},  
\end{equation}
which is equivalent to the energy required to store a single charge on the capacitor. Note that in our dimensionless notation from \cref{eq:dimensionless} we have $\hat n_n = - \hat q_n$, while the effective energy becomes $E_{C,n} = (\vec C^{-1})_{(n,n)} / 8$. 

Similarly, we introduce the effective energies of the linear inductances and Josephson junctions, $E_{L,n}$ and $E_{J,n}$, of each node. The effective inductive energy is the diagonal elements of $\vec {L^{-1}}$, which is equivalent to the sum of the inverse inductances of the inductors connected to the given node. The effective Josephson energy is found as the sum of the Josephson energies of the junctions connected to the given node.

Returning to our example circuit in \cref{fig:exampleCircuit}, we can now write it using operators and effective energies. It becomes
\begin{equation}\label{eq:exampleOperators}
\begin{aligned}
    \hat \H &= 4 \left( E_{C,1}\hat n_1^2 + E_{C,2}\hat n_2^2\right) + E_{L,12} (\hat \phi_1 - \hat\phi_2 + \tilde\Phi)^2 \\ 
    &\phantom{=}- E_{J,1}\cos\hat\phi_1 - E_{J,2}\cos\hat\phi_2,
\end{aligned}
\end{equation}
where the coupling energy of the linear inductor is $E_{L,12} = 1/2L_{12}$. The effective energies of the Josephson junctions is the Josephson energies. Note that since our example does not include any coupling capacitors, we do not obtain any coupling term $\vec (\vec C^{-1})_{(1,2)}\hat n_1 \hat n_2$ since $\vec C$ is diagonal. In reality, this is rarely the case.

\section{Recasting to interacting harmonic oscillators}\label{sec:recasting}

We want to consider the low-energy limit of the superconducting circuit since we want to create a qubit using the two lowest-lying states of the nonlinear oscillator quantum system. This can be done by suppressing the kinetic energy of the system, such that the 'position' coordinate will be localized near the minimum of the potential. We consider a single anharmonic oscillator (AHO) as in \cref{fig:JJc_oscillator} but with a possible linear inductor in parallel, which means that we can omit subscripts in this section as there is only a single mode. The Hamiltonian we thus consider is
\begin{equation}
    \hat\H_\text{AHO} = 4E_C \hat n^2 + E_L\hat\phi^2 - E_J \cos \hat \phi.
\end{equation}
If the effective capacitive energy, $E_{C}$, of the mode is much smaller than the effective Josephson energy, $E_J$, the flux will be well localized near the bottom of the potential. This is equivalent to a heavy particle moving near its equilibrium position.
In this case, we can Taylor expand the potential part of the Hamiltonian up to fourth order in $\phi$ such that the Josephson-junction term takes the form
\begin{equation}\label{eq:cosExpansion}
    E_J\cos \phi = E_J - \frac{1}{2}E_J\phi^2 + \frac{1}{24}E_J \phi^4 + O(\phi^6).
\end{equation}
Throwing away the irrelevant constant term, we are left with a Hamiltonian consisting of second- and fourth-order terms. If we require the couplings between different parts of the superconducting circuit to be small, we can treat each mode individually as a harmonic oscillator perturbed by a quartic anharmonicity and possibly some couplings to other modes of the system. For each mode in our system, we have a simple harmonic oscillator (SHO) of the form 
\begin{equation}
    \hat \H_\text{SHO} = 4 E_C \hat n^2 + \left(E_L + \frac{1}{2}E_J\right) \hat \phi^2.
\end{equation}
The simple harmonic oscillator is well understood quantum mechanically, and using the algebraic approach \cite{Sakurai2011} we define the annihilation and creation operators
\begin{subequations}
\begin{align}
    \hat b &= \frac{1}{\sqrt{2}} \left( \frac{1}{\sqrt{\zeta}} \hat \phi - i \sqrt{\zeta} \hat n \right), \\
    \hat b^\dagger &= \frac{1}{\sqrt{2}} \left( \frac{1}{\sqrt{\zeta}} \hat \phi + i \sqrt{\zeta} \hat n \right),
\end{align}
\end{subequations}
where we define the impedance 
\begin{equation}\label{eq:impedance}
    \zeta = \sqrt{\frac{4 E_C}{E_L + E_J/2}}.
\end{equation}
When restoring dimensions and going away from the dimensionless notation defined in \cref{eq:dimensionless} the impedance in \cref{eq:impedance} must be multiplied with a factor of $R_Q/2\pi$, where $R_Q = h/(2e)^2 \simeq \SI{6.45}{\kilo\ohm}$ is the resistance quantum, which emerges in the quantum Hall effect.
The annihilation and creation operators fulfill the usual commutator relation
\begin{equation}
    [\hat b, \hat b^\dagger] = 1.
\end{equation}
Expressing the flux and conjugate momentum operators in terms of the annihilation and creation operators,
\begin{subequations}\label{eq:stepOperators}
  \begin{align}
      \hat \phi &= \sqrt{\frac{\zeta}{2}} (\hat b + \hat b^\dagger ), \\
      \hat n &= \frac{i}{\sqrt{2\zeta}} (\hat b - \hat b^\dagger ),
  \end{align} 
\end{subequations}
we can rewrite the oscillator part of the Hamiltonian as
\begin{equation}
    \hat \H_\text{SHO} = 4 \sqrt{E_C\left(E_L + \frac{1}{2}E_J\right)} \left(\hat N + \frac{1}{2} \right),
\end{equation}
where we introduce the usual number operator $\hat N = \hat b^\dagger \hat b$. 

Using the creation and annihilation operators, we can rewrite all quadratic and quartic interaction terms. The results are in \cref{tab:Overview} for the most commonly occurring terms.

\begin{table}
	\centering
	\caption{Overview of the different components and the operators they map to. Subscripts are included where appropriate and refer to different nodes. All constant terms are neglected. The impedance factor can be found in \cref{eq:impedance}.}
	\label{tab:Overview}
	\begin{tabular}{lcc}
		\hline \\[-1em]
		Component & Hamiltonian term & \begin{tabular}{@{}c@{}}Annihilation and\\ creation operators\end{tabular} \\[3pt]
		\hline & & \\[-1em]
		All terms & $\hat n^2 + \hat \phi ^2$ & $4\sqrt{E_{C}\left(E_{L} + \frac{1}{2}E_{J} \right)}\hat b^\dagger \hat b$ \\[3pt]
		\parbox[l]{0cm}{Linear\\capacitors} & $\hat{n}$ & $\frac{i}{\sqrt{2\zeta}}(\hat b - \hat b^\dagger)$ \\[3pt]
		& $\hat{n}^2$ & $-\frac{1}{2\zeta}(\hat b - \hat b^\dagger)^2$ \\[3pt]
		& $\hat{n}_i\hat{n}_j$ & $-\frac{1}{2\sqrt{\zeta_i\zeta_j}}(\hat b_i - \hat b_i^\dagger)(\hat b_j - \hat b_j^\dagger)$ \\[3pt]
		\parbox[l]{0cm}{Linear\\inductors} & $\hat{\phi}$ & $\sqrt{\frac{\zeta}{2}}(\hat b^\dagger + \hat b)$  \\[3pt]
		& $\hat{\phi}^2$ & $\frac{\zeta}{2}(\hat b^\dagger + \hat b)^2$\\[3pt]
		& $\hat{\phi}_i\hat{\phi}_j$ & $\frac{\sqrt{\zeta_i\zeta_j}}{2}(\hat b_i^\dagger + \hat b_i)(\hat b_j^\dagger + \hat b_j)$ \\[3pt]
		\parbox[l]{0cm}{Josephson\\junctions} & $\hat{\phi}^4$ & $\frac{\zeta^2}{4}(\hat b^\dagger + \hat b)^4$ \\[3pt]
		& $\hat{\phi}^3$ & $\frac{\zeta^{3/2}}{2^{3/2}}(\hat b^\dagger + \hat b)^3$ \\[3pt]
		& $\hat{\phi}_i^3\hat{\phi}_j$ & $\frac{\zeta_i^{3/2}\zeta_j^{1/2}}{4}(\hat b_i^\dagger + \hat b_i)^3(\hat b_j^\dagger + \hat b_j)$ \\[3pt]
		& $\hat{\phi}_i^2\hat{\phi}_j^2$ & $\frac{\zeta_i\zeta_j}{4}(\hat b_i^\dagger + \hat b_i)^2(\hat b_j^\dagger + \hat b_j)^2$ \\[3pt]
		\hline
	\end{tabular}
\end{table}

Returning to our example circuit in \cref{fig:exampleCircuit}, we can write the Hamiltonian in \cref{eq:exampleOperators} using annihilation and creation operators as
\begin{equation}\label{eq:HexampleStep}
\begin{aligned}
    \hat \H &=\omega_1 \hat b_1^\dagger \hat b_1 + \omega_2 \hat b_2^\dagger \hat b_2  + \frac{\alpha_1}{12} (\hat b_1^\dagger + \hat b_1)^4 \\
    &\phantom{=}+ \frac{\alpha_2}{12} (\hat b_2^\dagger + \hat b_2)^4 + g_{12} (\hat b_1^\dagger + \hat b_1)(\hat b_2^\dagger + \hat b_2)\\
    &\phantom{=} + \chi_1 (\hat b_1 + \hat b_1^\dagger) - \chi_2 (\hat b_2 + \hat b_2^\dagger),
\end{aligned}
\end{equation}
where we omit all constant terms. We further define
\begin{subequations}
\begin{align}
    \omega_n &= 4\sqrt{E_{C,n}\left(E_{L,12} + \frac{1}{2}E_{J,n} \right)},\\
    \alpha_n &= -\frac{\zeta_n^2}{8} E_{J,n}, \label{eq:anharmonicityExample}\\
    g_{12} &= -\sqrt{\zeta_1\zeta_2} E_{L,12},\label{eq:g12Example}\\
    \chi_n &= \sqrt{2\zeta_n}E_{L,12}\tilde\Phi,\label{eq:chi}\\
    \zeta_n &= \sqrt{\frac{4E_{C,n}}{E_{L,12} + E_{J,n}/2}},
\end{align}
\end{subequations}
where we refer to $\omega_n$ as the frequency, $\alpha_n$ as the anharmonicity, and $g_{12}$ the oscillator coupling strength.
Note that if the effective inductive energy is zero, $E_{L,n} = 0$, then the anharmonicity in \cref{eq:anharmonicityExample} becomes $\alpha_n = - E_{C,n}$, which is often the case.

Note that in the presence of an external flux, one should be careful in identifying the minimum of the potential around which one can then perform the expansion, as in \cref{eq:cosExpansion}.

\section{Time-averaged dynamics}\label{sec:rwa}

When analyzing the Hamiltonian of the circuit, it is often advantageous to consider which terms dominate the time evolution and which terms only give rise to minor corrections. The latter can often be neglected without changing the overall behavior of the system. It can often be difficult to determine which terms dominate, as different scales influence the dynamics of the system. This stems from the fact that the frequencies, $\omega_n$, of the oscillators are usually of the order $\si{\GHz}$ while the interactions between the different oscillators are usually much smaller, on the order of $\si{\MHz}$. We therefore employ separation of scales to remove the large energy differences of the modes from the Hamiltonian. This makes it possible to see the details of the interactions. In order to do this, we first introduce the concept of the interaction picture, where the interacting part of the Hamiltonian is in focus.

To summarize which terms we consider in the Hamiltonian, we divide the terms into three categories.
\begin{itemize}
    \item Large trivial terms: Well understood energy difference terms, such as the qubit frequencies, which we remove using separation of scales by transforming into the interaction picture. Usually of the order $\si{\GHz}$.
    \item Smaller but interesting terms: The dominant part of the interaction we are interested in. Usually of the order $\si{\MHz}$
    \item Small negligible terms: The suppressed part of the interaction, which does not contribute significantly to the time evolution. These can be removed using the rotating-wave approximation (RWA).
\end{itemize}
Note, however, that the above categorization is only a guide, and one should always consider each term in relation to the concrete system at hand.

\subsection{Interaction picture}\label{sec:interactionPicture}

Consider the state $\ket{\psi, t}_S$ at time $t$. This state satisfies the Schr\"o{}dinger equation,
\begin{equation}\label{eq:schrodinger}
    i\pdv{t}\ket{\psi, t}_S = \hat \H \ket{\psi, t}_S,
\end{equation}
where $\hat \H$ is the Hamiltonian. The subscript $S$ refers to the Schr\"o{}dinger picture. In the Schr\"o{}dinger picture operators are time independent and states are time dependent. We wish to change into the interaction picture by splitting the Hamiltonian in a way such that the dynamics are separated from the noninteracting part, $\hat \H = \hat \H_0 + \hat \H_{I,S}$. There are often several ways to make this splitting depending on what interaction we want to highlight. This separation comes at the cost that both the operators and states become time dependent. 
The advantage of using a specific splitting of the full Hamiltonian is that we can highlight some desired physics while ignoring other parts that are well understood. This is analogous to choosing a reference frame rotating with the Earth when doing classical physics in a reference frame fixed on the surface of the Earth.

States in the interaction picture are defined as
\begin{equation}
    |\psi, t\rangle_I = e^{i\hat \H_0t}|\psi, t\rangle_S,
\end{equation}
where the subscript $I$ refers to the interaction picture. The operators in the interaction picture are defined as
\begin{equation}\label{eq:OI}
    \O_I = e^{i\hat \H_0t} \O_S e^{-i\hat \H_0t},
\end{equation}
where $\O_S$ is an operator in the Schr\"o{}dinger picture.

It is then possible to show that the state satisfies the following Schr\"o{}dinger equation
\begin{equation}
    i\pdv{t}|\psi, t\rangle_I = \hat \H_I|\psi, t\rangle_I,
\end{equation}
where $\hat \H_I = e^{i\hat \H_0 t}\hat \H_{I,S}e^{-i\hat \H_0t}$ is the interaction part of the Hamiltonian in the interaction picture. 

In general, one can transform a Hamiltonian to any so-called rotating frame using the transformation rule
\begin{equation} \label{eq:H_transformation_rotating}
    \hat \H \rightarrow \hat \H_R = \U(t)^\dagger \hat\H\, \U(t) + i \dv{ \U(t)^\dagger}{t} \U(t),
\end{equation}
where $ \U(t) $ is a unitary transformation. This transformation rule holds for any unitary transformation and is quite useful to keep in mind. 
Note that \cref{eq:H_transformation_rotating} is equivalent to transforming the Hamiltonian into the interaction picture $\hat \H \rightarrow \hat \H_I$ when $\U(t) = \exp(-i\hat \H_0 t)$, and $\hat \H_0$ is the noninteracting part of the Hamiltonian, as the second term removes the noninteracting part of the Hamiltonian.

One can also show that the time evolution of the operators in the interaction picture is governed by a Heisenberg equation of motion
\begin{equation}\label{eq:HeisenbergEOM}
    \dv{t} \O_I = i [\hat \H_0, \O_I],
\end{equation}
where we assume no explicit time dependence in $\O_S$. 
Note that this implies that the voltage of the $b$th branch can be calculated as $\hat V_b = i [\hat \H, \hat \Phi_b]$.
For more information about the interaction picture see e.g., Ref. \cite{Sakurai2011}.

\subsection{Rotating-wave approximation}

\begin{figure}
    \centering
    \includegraphics[width=\columnwidth]{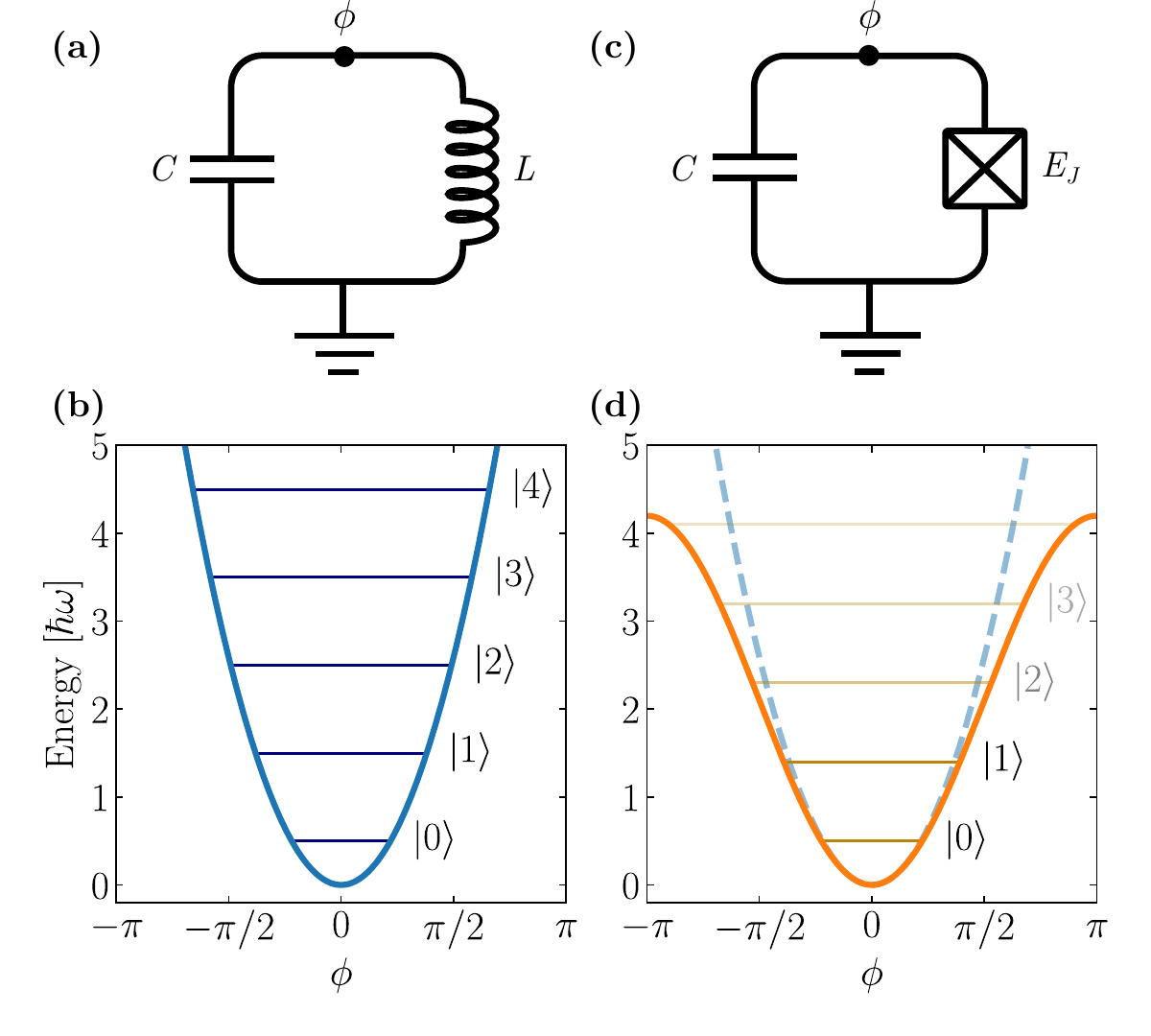}
    \caption{\textbf{(a)} Circuit of an $LC$ oscillator with inductance $L$ and capacitance $C$. We denote the phase on the superconducting island $\phi$, while the ground node has phase zero. \textbf{(b)} Energy potential of a quantum harmonic oscillator, as can be obtained by an LC-circuit. Here the energy levels are equidistantly spaced $\hbar\omega$ apart, where $\omega=\sqrt{1/LC}$. \textbf{(c)} Josephson junction qubit circuit, where the linear inductor is replaced by a nonlinear Josephson junction of energy $E_J$. \textbf{(d)} The Josephson junction changes the harmonic potential (blue dashed) into a sinusoidal potential (orange solid), yielding nonequidistant energy levels.}
    \label{fig:HOvsTransmon}
\end{figure}

Consider now the weakly anharmonic oscillator as seen in \cref{fig:HOvsTransmon}(c), which has the quantized Hamiltonian 
\begin{equation}\label{eq:quantizedAHO}
    \hat \H = \omega \hat b^\dagger \hat b + \frac{\alpha}{12}(\hat b^\dagger + \hat b)^4,
\end{equation}
where we remove all constant terms. The frequency is $\omega = \sqrt{8E_CE_J}$ and the anharmonicity is $\alpha = - E_C$ where $E_C$ and $E_J$ are the effective capacitive energy and Josephson energy, respectively. Now we choose the first term as the noninteracting Hamiltonian, $\hat \H_0 = \omega \hat b^\dagger \hat b$. We want to figure out how the annihilation and creation operators behave in the interaction picture, i.e., we want to calculate \cref{eq:OI} for the annihilation and creation operators. First, we notice that $\hat \H_{0}^{n} \hat b^\dagger  = \hat b^\dagger  (\hat \H_0 + \omega)^n$. Using this and expanding the exponential functions, we can prove that  
\begin{equation}
    e^{i\hat \H_0t} \hat b^\dagger e^{-i\hat \H_0t} = \hat b^\dagger e^{i\omega t}.
\end{equation}
By taking the complex conjugate, we find a similar expression for $\hat b$, but with a minus in the exponential factor on the right-hand side. 

We now wish to consider how different combinations of the annihilation and creation operators transform in the interaction picture. Starting with the number operator $\hat N = \hat b^\dagger \hat b$, we see the exponential factor from $\hat b^\dagger$ cancels the exponential factor from $\hat b$, meaning that the number operator is unaffected by the transformation. This is not surprising as the noninteracting Hamiltonian is chosen exactly as the number operator. However, if we consider terms like $J\hat b^\dagger \hat b^\dagger$ we find that in the interaction picture they take the form $J\hat b^\dagger \hat b^\dagger e^{2i\omega t}$. If $\omega$ is sufficiently large compared to the factor, $J$, in front of the term (which is often the case in superconducting circuit Hamiltonians, where $\omega \sim \si{\GHz}$, while other terms are usually of the order $J\sim \si{\MHz}$), these terms will oscillate very rapidly on the timescale induced by $J$. The time average over such terms on a timescale of $\tau \sim 1/J$ is zero, and we can therefore neglect them as they only give rise to minor corrections. This is the rotating-wave approximation which is widely used in atomic physics \cite{Allen1987,Bransden2003}. The story is the same for $\hat b \hat b$ terms. 
All terms that do not conserve the number of excitations (or quanta) of the system, i.e., terms where the number of annihilation operators is not equal to the number of creation operators, will rotate rapidly and can therefore safely be neglected. Note that while these individual terms are nonconserving, they always appear in conjugate pairs in the Hamiltonian such that the full Hamiltonian conserves the excitations, as it should.

It is important to point out that despite the naming, the 'conservation' is not related to a conservation law resulting from a symmetry, i.e., like in Noether's theorem. Rather, the statement here means that the excitation conserving terms are much more important than the nonconserving terms as long as the conditions for using the rotating-wave approximation are satisfied.

Now consider the anharmonicity term of \cref{eq:quantizedAHO}. When only including excitation conserving terms and removing irrelevant constants, the anharmonicity term takes the form 
\begin{equation}
\begin{aligned}
    \frac{\alpha}{12}(\hat b^\dagger + \hat b)^4 &= \alpha\left(\frac{1}{2} \hat b^\dagger \hat b^\dagger \hat b \hat b + \hat b^\dagger \hat b\right)\\ 
    &\phantom{=}+ \text{Nonconserving terms}.
\end{aligned}
\end{equation}
The last term, $\hat b^\dagger \hat b$, is the number operator, and we can therefore consider it a correction to the frequency, such that the dressed frequency becomes $\tilde \omega =  \omega + \alpha = \sqrt{8E_CE_J} - E_C$. The remaining $(\hat b^\dagger \hat b^\dagger \hat b \hat b)$ term makes the oscillator anharmonic. For this reason, we call $\alpha$ the anharmonicity of the anharmonic oscillator. 
If we remove terms that do not conserve the number of excitation, the Hamiltonian takes the form (in the Schr\"o{}dinger picture)
\begin{equation}\label{eq:Hanharmonicity}
    \hat \H = \tilde \omega \hat b^\dagger \hat b + \frac{\alpha}{2}\hat b^\dagger \hat b^\dagger \hat b \hat b.
\end{equation}

Next, consider an interaction term like the one in \cref{eq:HexampleStep}
\begin{equation}
    (\hat b_i^\dagger + \hat b_i^)(\hat b_j^\dagger + \hat b_j) = \hat b_i^\dagger\hat b_j + \hat b_i\hat b_j^\dagger + \hat b_i^\dagger\hat b_j^\dagger + \hat b_i\hat b_j.
\end{equation}
Changing into the interaction picture, we realize that the two last terms obtain a phase of $\exp{[\pm i (\omega_i + \omega_j) t]}$, which can be considered a fast oscillating term if the frequencies $\omega_i + \omega_j$ are much larger than the interaction strength, which is usually the case. We can therefore safely neglect these nonconserving terms. The two first terms on the other hand obtain a phase of $\exp({\pm i \delta t})$, where $\delta=\omega_i - \omega_j$ is called the detuning of the two oscillators. It is therefore tempting to say that these terms only contribute if $\omega_i \approx \omega_j$. This is, however, not the whole story. More precisely, we find that 
\begin{equation}
\begin{aligned}
    \hat b^\dagger_i \hat b_j e^{i\delta t} + \hat b_i \hat b_j^\dagger e^{-i\delta t} &= (\hat b^\dagger_i \hat b_j + \hat b_i \hat b_j^\dagger) \cos \delta t\\
    &+ i(\hat b^\dagger_i \hat b_j - \hat b_i \hat b_j^\dagger) \sin \delta t,
\end{aligned}
\end{equation}
which can be useful in some situations, e.g., when driving qubits, see \cref{sec:driving}. However, as a general rule of thumb, one can neglect these terms unless $\omega_i \simeq \omega_j$, i.e., $\delta=0$.
For a more general discussion on the validity of the time averaging dynamics see Ref. \cite{Gamel2010}.

If we consider the example circuit in \cref{fig:exampleCircuit}, under the assumption that $|\alpha_n| \ll \omega_n$, we can time average its Hamiltonian in \cref{eq:HexampleStep}. We choose the noninteracting Hamiltonian as $\H_0 = \omega_1 \hat b_1^\dagger \hat b_1 + \omega_2 \hat b_2^\dagger \hat b_2$, which means that the interacting part of the Hamiltonian in \cref{eq:HexampleStep} becomes
\begin{equation}\label{eq:InteractonHamiltonian}
\begin{aligned}
    \hat \H_I &= \frac{\alpha_1}{2} (\hat b_1^\dagger\hat b_1^\dagger\hat b_1\hat b_1 + 2\hat b_1^\dagger \hat b_1)    + \frac{\alpha_2}{2} (\hat b_2^\dagger\hat b_2^\dagger\hat b_2\hat b_2 + 2\hat b_2^\dagger \hat b_2) \\ 
    &\phantom{=}+ g_{12} \left( \hat b_1^\dagger\hat b_2e^{i\delta t} + \hat b_1\hat b_2^\dagger e^{-i\delta t} \right.\\
    &\phantom{=+g_{12}}+ \left.\hat b_1\hat b_2e^{-i(\omega_1+\omega_2) t} + \hat b_1^\dagger\hat b_2^\dagger e^{i(\omega_1+\omega_2) t}\right)  \\
    &\phantom{=} - \sum_{n=1}^2(-1)^n\chi_n (\hat b_ne^{-i\omega_n t} + \hat b_n^\dagger e^{i\omega_n t}),
\end{aligned}
\end{equation}
where we define the detuning $\delta = \omega_1 - \omega_2$. Assuming $\omega_1 \simeq \omega_2$, i.e., $\delta \simeq 0$, and if we further assume that $\omega_n \gg g_{12}$, then the coupling terms $\hat b_1 \hat b_2$ and $\hat b_1^\dagger \hat b_2^\dagger$ are fast oscillating and can thus be neglected. 

We can also write the Hamiltonian in the Schr\"o{}dinger picture, removing terms that do not conserve excitations. This yields
\begin{equation}\label{eq:HexampleTimeAveraged}
\begin{aligned}
    \hat \H &= \tilde\omega_1 \hat b_1^\dagger \hat b_1 + \tilde\omega_2 \hat b_2^\dagger \hat b_2  + \frac{\alpha_1}{2} \hat b_1^\dagger\hat b_1^\dagger\hat b_1\hat b_1 \\
    &\phantom{=}+ \frac{\alpha_2}{2} \hat b_2^\dagger\hat b_2^\dagger\hat b_2\hat b_2 + g_{12} (\hat b_1^\dagger\hat b_2 + \hat b_1\hat b_2^\dagger),
\end{aligned}
\end{equation}
where we introduce the revised frequencies $\tilde \omega_n = \omega_n + \alpha_n$. Writing the Hamiltonian in this frame without nonconserving terms reveals the effect of the anharmonicity.
In \cref{eq:HexampleTimeAveraged} we also assume that $\omega_n \gg \chi_n$, meaning that all terms related to the external flux are neglected. However, since $\chi_n$ depends on $\tilde\Phi$, which can be controlled externally, it is possible to tune $\chi_n$ such that the terms involving $\chi_n$ are not suppressed. This can be used to drive the modes, i.e., to add excitations to the two degrees of freedom.

\section{Truncation}\label{sec:truncation}

A harmonic oscillator, as one gets from a regular LC-circuit, has a spectrum consisting of an infinite number of equally spaced energy eigenstates [see \cref{fig:HOvsTransmon}(b)]. This is not desirable, as we wish to consider only the lowest states of the system in order to realize a qubit. However, when we introduce a Josephson junction instead of a linear inductor, we introduce an anharmonicity, compare \cref{fig:HOvsTransmon}(a) and (c).  The anharmonicity stems from the $(\hat b^\dagger \hat b^\dagger \hat b \hat b)$ terms [see \cref{eq:Hanharmonicity}] and can be viewed as perturbations to the harmonic oscillator Hamiltonian if $|\alpha| \ll \omega$. This anharmonicity changes the spacing between the energy levels of the harmonic oscillator, making it an anharmonic oscillator [see \cref{fig:HOvsTransmon}(d)]. Formally, the anharmonicity is defined as the difference between the first and second energy gap, while we define the relative anharmonicity as the anharmonicity divided by the first energy gap
\begin{equation}\label{eq:anharmonicity}
    \alpha = E_{12} - E_{01}, \qquad  \alpha_r = \frac{\alpha}{E_{01}}.
\end{equation}
Note that this anharmonicity is the same anharmonicity factor in front of the $(\hat b^\dagger \hat b^\dagger \hat b \hat b)$ terms mentioned in previous sections.

To operate only on the two lowest levels of the oscillator, the anharmonicity must be larger than the bandwidth of operations on the qubit. That is, if we want to drive excitation between the two lowest levels of the anharmonic oscillator, the anharmonicity must be larger than the amplitude of the driving field (also known as the Rabi frequency, see \cref{sec:driving}). If the anharmonicity is smaller than the amplitude of the driving field, we cannot distinguish between the energy gaps of the oscillator, and we end up driving multiple transitions in the spectrum instead of just the lowest one.

Taking this into account we find that as a rule-of-thumb, the relative anharmonicity should be at least a couple of percent for the system to make an effective qubit. In actual numbers, this converts to an anharmonicity around 100-$\SI{300}{\MHz}$ for a qubit frequency around 3-$\SI{6}{\GHz}$ \cite{Koch2007,Krantz2019}. It does not matter whether the anharmonicity is positive or negative. For transmon-type qubits, it will be negative, while it can be either positive or negative for flux-type qubits. The relative anharmonicity is proportional to $\sqrt{E_J/E_C}$, which means that this ratio must be of a certain size for the anharmonicity to have an effect. This is in contrast to what was discussed at the beginning of \cref{sec:recasting}, where we argued that we required this ratio to be as low as possible to allow for the expansion of cosines. Thus we need to find a suitable regime for the ratio, $E_J/E_C$. This regime is usually called the transmon regime and is around 50-100.

In the following section we assume that we have a sufficiently large anharmonicity to truncate the system into a two-level system. However, nothing is stopping us from keeping more levels, as we do in \cref{sec:qutrit}.

As an alternative to the methods for truncation presented in this tutorial, black-box quantization can be useful for determining the effective low-energy spectrum of a weakly anharmonic Hamiltonian \cite{Nigg2012,Solgun2014,Solgun2015}. This approach is especially useful when dealing with impedances in the circuit, but is beyond the scope of this tutorial.

\subsection{Two-level model (qubit)}\label{sec:twoLevelTruncation}

\begin{table}
	\centering
	\caption{Overview of the different combinations of the annihilation and creation operators and their truncation to two-dimensional Pauli operators. Subscripts are included for the interaction terms and refer to different nodes. All constant terms are ignored.}
	\label{tab:Overview2}
	\begin{tabular}{cc}
		\hline
		\begin{tabular}{@{}c@{}}Annihilation and\\ creation operators\end{tabular}  & Pauli operators \\
		\hline\\[-8pt]
	    $\hat b^\dagger - \hat b$ & $-i\sigma^y$\\[3pt]
	    $\hat b^\dagger + \hat b$ & $\sigma^x$ \\[3pt]
		$(\hat b^\dagger - \hat b)^2$ & $-\sigma^z$ \\[3pt]
		$(\hat b^\dagger + \hat b)^2$ & $-\sigma^z$ \\[3pt]
		$(\hat b^\dagger + \hat b)^3$ & $3\sigma^x$ \\[3pt]
		$(\hat b^\dagger + \hat b)^4$ & $-6\sigma^z$  \\[3pt]
		$(\hat b_i^\dagger - \hat b_i)(\hat b_j^\dagger - \hat b_j)$ & $-\sigma^y_i\sigma^y_j$  \\[3pt]
		$(\hat b_i^\dagger + \hat b_i)(\hat b_j^\dagger + \hat b_j)$ & $\sigma^x_i\sigma^x_j$ \\[3pt]
		$(\hat b_i^\dagger + \hat b_i)^3(\hat b_j^\dagger + \hat b_j)$ & $3\sigma^x_i\sigma^x_j$ \\[3pt]
		$(\hat b_i^\dagger + \hat b_i)^2(\hat b_j^\dagger + \hat b_j)^2$ & $\sigma^z_i\sigma^z_j - 2\sigma^z_i - 2\sigma^z_j$ \\[3pt]
		\hline
	\end{tabular}
\end{table}

\begin{figure}
    \centering
    \includegraphics[width=.5\columnwidth]{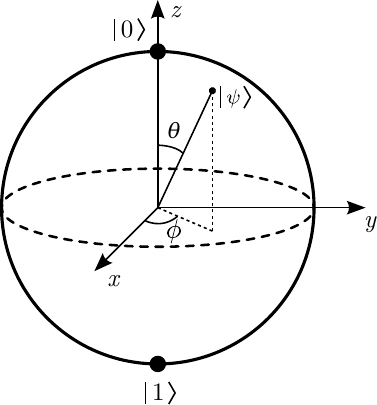}
    \caption{The Bloch sphere. Each point on the Bloch sphere corresponds to a quantum state. Rotations around the sphere correspond to transformations of the state.}
    \label{fig:BlochSphere}
\end{figure}

In a two-level system, which is equivalent to a qubit, we can represent the state of the system with two-dimensional vectors
\begin{equation}
    |0\rangle \sim \mat{1 \\ 0 }, \qquad |1\rangle \sim \mat{ 0 \\ 1 }.
\end{equation}
In this reduced Hilbert space all operators can be expressed by the Pauli matrices,
\begin{equation}
    \sigma^x = \mat{ 0 & 1 \\ 1 & 0 }, \quad \sigma^y = \mat{ 0 & -i \\ i & 0 }, \quad \sigma^z = \mat{ 1 & 0 \\ 0 & -1 },
\end{equation}
and the identity, since these four matrices span all $2\times 2$ Hermitian matrices. If we view the unitary operations as rotations in the Hilbert space, we can parameterize the superposition of the two states using a complex phase, $\phi$, and a mixing angle, $\theta$
\begin{equation}\label{eq:arbitraryState}
	|\psi\rangle =\alpha \ket{0} + \beta\ket{1} =  \cos\left(\frac{\theta}{2}\right) |0\rangle + e^{i\phi}\sin\left(\frac{\theta}{2}\right) |1\rangle,
\end{equation}
where $0\leq \theta \leq \pi$ and $0\leq \phi < 2\pi$ and $|\alpha|^2 + |\beta|^2 =1 $. With this, we can illustrate the qubit as a unit vector on the Bloch sphere, see \cref{fig:BlochSphere}. It is conventional to let the north pole represent the $\ket{0}$ state, while the south pole represents the $\ket{1}$ state. These lie on the $z$ axis, which is called the longitudinal axis as it represents the quantization axis for the states in the qubit. The $x$ and $y$ axes are called the transverse axes. 

Solving the Schr\"{o}dinger equation in \cref{eq:schrodinger} for the state in \cref{eq:arbitraryState} shows that it precesses around the $z$ axis at the qubit frequency. However, changing into a frame rotating with the frequency of the qubit, following the approach in \cref{sec:interactionPicture}. makes the Bloch vector stationary.

Unitary operations can be seen as rotations on the Bloch sphere and the Pauli matrices are thus the generators of rotations. Linear operators will then be represented by $2\times 2$ matrices as
\begin{align}
M_{2}[\hat{O}] = \mat{\mel{0}{\O}{0} & \mel{0}{\O}{1} \\ \mel{1}{\O}{0} & \mel{1}{\O}{1}}.
\end{align}
In general we denote the $ n\times n $ matrix representation of an operator $ \O$ with $ M_{n}[\O] $.

In order to apply this mapping to the Hamiltonian, we must map each operator in each term. As an example, we truncate the $(\hat b^\dagger + \hat b)^3$ term from \cref{tab:Overview2}:
\begin{align*}
    (\hat b^\dagger + \hat b )^3|0\rangle &= (\hat b^\dagger + \hat b )^2 |1\rangle\\
	&= (\hat b^\dagger + \hat b )\left(\sqrt{2}|2\rangle+|0\rangle\right) \\
	&= \sqrt{6}|3\rangle+ 3|1\rangle,\\
    (\hat b^\dagger + \hat b )^3 |1\rangle &= (\hat b^\dagger + \hat b )^2\left(\sqrt{2}|2\rangle+|0\rangle\right) \\
    &= (\hat b^\dagger + \hat b )\left(\sqrt{6}|3\rangle+3|1\rangle\right) \\
    &= \sqrt{24}|4\rangle +  6\sqrt{2}|2\rangle + 3|0\rangle .
\end{align*}
Using the orthonormality of the states we obtain the representation of the operator
\begin{align*}
    M_{2}[(\hat b^\dagger + \hat b)^3] &= \mat{ \langle0|(\hat b^\dagger + \hat b)^3|0\rangle & \langle0|(\hat b^\dagger + \hat b)^3|1\rangle \\
	\langle1|(\hat b^\dagger + \hat b)^3|0\rangle & \langle1|(\hat b^\dagger + \hat b)^3|1\rangle } \\
	&= \mat{ 0 & 3 \\
    	3 & 0 } = 3\sigma^x.
\end{align*}
Truncation of the remaining terms is presented in \cref{tab:Overview2}.

If we consider the example circuit in \cref{fig:exampleCircuit}, after we remove nonconserving terms as in \cref{eq:HexampleTimeAveraged} and assume an anharmonicity large enough for truncation to a two-level system, we obtain the following Hamiltonian:
\begin{equation}\label{eq:linearCouplingHamiltonian}
    \hat \H = -\frac{\tilde \omega_1}{2} \sigma_1^z -\frac{\tilde \omega_2}{2} \sigma_2^z + g_{12}(\sigma_1^+ \sigma_2^- + \sigma_1^- \sigma_2^+),
\end{equation}
where we define $\sigma^{\pm}_n = (\sigma^x_n \mp i\sigma^y_n)/2$. This Hamiltonian represents two qubits that can interact by swapping excitation between them, i.e., interacting via a swap coupling.

\subsection{Three-level model (qutrit)}\label{sec:qutrit}

It can be desirable to truncate to the three lowest levels of the anharmonic oscillator, i.e., the three lowest states of \cref{fig:HOvsTransmon}(d). This can, e.g., be useful if one wants to study qutrit systems \cite{Lanyon2009,Fedorov2012,Baekkegaard2019}, or the leakage from the qubit states to higher states \cite{Loft2020,Rasmussen2019}. In this case, the operators will be represented as $3 \times 3$ matrices. The matrix representation of the annihilation and creation operators become
\begin{subequations}\label{eq:3LevelStepMatrixRep}
\begin{equation}
M_3[\hat b^\dagger] = \mat{
0 & 0 & 0 \\
1 & 0 & 0 \\
0 & \sqrt{2} & 0
}, \quad M_3[\hat b] = \mat{
0 & 1 & 0 \\
0 & 0 & \sqrt{2} \\
0 & 0 & 0
},
\end{equation}
while the number operator is
\begin{equation}
 M_3[\hat b^\dagger \hat b] = \mat{
0 & 0 & 0 \\
0 & 1 & 0 \\
0 & 0 & 2
},
\end{equation}
and powers of $\hat b^\dagger + \hat b$ become
\begin{align}
M_3[(\hat b^\dagger + \hat b)^2] &= \mat{
1 & 0 & \sqrt{2} \\
0 & 3 & 0 \\
\sqrt{2} & 0 & 5
}, \\  M_3[(\hat b^\dagger + \hat b)^3] &= \mat{
0 & 3 & 0 \\
3 & 0 & 6\sqrt{2} \\
0 & 6\sqrt{2} & 0 
}, \\ M_3[(\hat b^\dagger + \hat b)^4] = & \mat{
3 & 0 & 6\sqrt{2} \\
0 & 15 & 0 \\
6\sqrt{2} & 0 & 39
}.\label{eq:M3fourthOrder}
\end{align}
\end{subequations}
From \cref{eq:M3fourthOrder} it is clear to see the varying size of the anharmonicity, as the differences $15-3 = 12$ and $39-15 = 24$ between the levels changes. This pattern continues for higher levels and means that we can distinguish between all the levels in principle. 

As we are dealing with $3\times 3$ matrices we can no longer use the Pauli spin-1/2 matrices as a basis for the operators. In this case one can use the Gell-Mann matrices as a basis. However, often it is more convenient to leave the annihilation and creation operators as above.
We are not limited by three levels, and it is possible to truncate the system to an arbitrary number of levels, thus creating a so-called qudit.

It is also possible to truncate the system before expanding the cosine functions of the Josephson junctions. This approach is discussed in \cref{sec:exactTrunc} where we also truncate an anharmonic oscillator to the four lowest levels.

\section{Microwave driving}\label{sec:driving}

\begin{figure}
    \centering
    \includegraphics[width=.5\columnwidth]{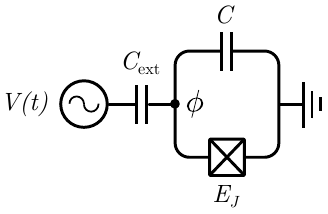}
    \caption{Circuit diagram of a single transmonlike superconducting qubit capacitively coupled to a microwave drive line.}
    \label{fig:driving}
\end{figure}

Single-qubit rotations in superconducting circuits can be achieved by capacitive microwave driving. In this section, we go through the steps of analyzing a microwave-controlled transmonlike qubit and then generalize to a $d$-level qudit. To this end, we consider the superconducting qubit seen in \cref{fig:driving}, which is capacitively coupled to a microwave source. Using the approach presented in \cref{sec:MethodofNodes} the Lagrangian of this circuit becomes
\begin{equation}\label{eq:drivingLagrangian}
    \L = \frac{C}{2}\dot \phi^2 + E_J \cos \phi + \frac{C_\text{ext}}{2}\left(V(t) - \dot \phi\right)^2,
\end{equation}
where $\phi$ is the node flux. Expanding the last term, we obtain
\begin{equation}
    \L = \L_0 + \frac{C_\text{ext}}{2}\left(V(t)^2 + \dot \phi^2 - 2V(t)\dot\phi\right),
\end{equation}
where $\L_0$ is the static part of the Lagrangian, i.e., the two first terms of \cref{eq:drivingLagrangian}.
The first term in the parenthesis is an irrelevant offset term, the second term is a change of the capacitance of the node, while the last term is our driving term. We throw away the offset term and rewrite
\begin{equation}
    \L = \frac{C+C_\text{ext}}{2} \dot \phi^2 + E_J\cos\phi - C_\text{ext} V(t) \dot \phi.
\end{equation}
The conjugate momentum of the node flux, $\phi$, is then 
\begin{equation}
    q = (C + C_\text{ext}) \dot\phi - C_\text{ext} V(t).
\end{equation}
Doing the usual Legendre transformation, our Hamiltonian takes the form
\begin{equation}\label{eq:H_full_driving}
    \H = \underbrace{\frac{1}{2}\frac{1}{C+C_\text{ext}}q^2 - E_J\cos \phi}_{\H_\text{AHO}} + \underbrace{\frac{C_\text{ext}}{C+C_\text{ext}}  V(t) q}_{\H_\text{ext}},
\end{equation}
where we denote the anharmonic oscillator part of the Hamiltonian $\H_\text{AHO}$ and the external driving part $\H_\text{ext}$. We are now ready to perform the quantization and the driving part becomes
\begin{equation} \label{eq:driving_Hamiltonian_term}
    \hat \H_\text{ext} = \frac{i}{\sqrt{2\zeta}}  \frac{C_\text{ext}}{C + C_\text{ext}} V(t) (\hat b^\dagger - \hat b).
\end{equation}
Assuming a large enough anharmonicity, we can truncate the Hamiltonian into the two lowest levels
\begin{equation}
    \hat \H = -\frac{1}{2} \omega\sigma^z + \Omega V(t) \sigma^y,
\end{equation}
where $\omega$ is the qubit frequency and $\Omega = C_\text{ext} /[\sqrt{2\zeta}(C+C_\text{ext})]$ is the Rabi frequency of the transition between the ground state and the excited state. Note that the size of the Rabi frequency is limited by the size of the anharmonicity, as discussed in \cref{sec:truncation}. The name Rabi frequency may cause a bit of confusion at first as it is not the frequency of the driving microwave but rather the amplitude. However, the Rabi frequency is named so since it is equal to the frequency of oscillation between the two states in a qubit when the driving frequency, $\omega_\textup{ext}$, is equal to the qubit frequency, $\omega$, i.e. when we drive the qubit 'on resonance' \cite{Bransden2003}.

We now change into a frame rotating with the frequency of the qubit, also known as the interaction frame as discussed in \cref{sec:rwa}. In particular we use $\H_0 = -\omega\sigma_z/2$ for the transformation in \cref{eq:H_transformation_rotating}. In this frame the Hamiltonian becomes
\begin{equation}\label{eq:drivingHamiltonian}
\hat\H_R=\Omega V(t)\left( \cos(\omega t) \sigma^y - \sin(\omega t) \sigma^x \right),
\end{equation}
which is equivalent to the external driving part of the Hamiltonian in the interaction picture, i.e., $\H_R = \H_\text{ext}^I$.
We assume that the driving voltage is sinusoidal
\begin{equation}\label{eq:voltage}
\begin{aligned}
    V(t) &= V_0 \eta(t) \sin (\omega_\text{ext} t + \varphi) \\
     &= V_0 \eta(t) \left[ \cos (\varphi) \sin(\omega_\text{ext} t) + \sin(\varphi) \cos(\omega_\text{ext} t)\right],
\end{aligned}
\end{equation}
where $V_0$ is the amplitude of the voltage, $\eta(t)$ is a dimensionless envelope function, $\omega_\text{ext}$ is the external driving frequency, and $\varphi$ is the phase of the driving. One usually defines the in-phase component $I=\cos(\varphi)$ and the out-of-phase component $Q = \sin(\varphi)$ \cite{Krantz2019}. Inserting the voltage in \cref{eq:voltage} into the Hamiltonian in \cref{eq:drivingHamiltonian} and rewriting we obtain 
\begin{equation}
\begin{aligned}
\hat \H_R = \frac{1}{2} \Omega V_0 \eta(t) \big\lbrace- & \left[ Q \sin(\delta t) +I \cos(\delta t)\right]\sigma^x \\  +& \left[Q \cos(\delta t) -I \sin(\delta t)\right]\sigma^y \big\rbrace,
\end{aligned}
\end{equation}
where $\delta  = \omega - \omega_\text{ext}$ is the difference between the qubit frequency and the driving frequency and we neglect fast oscillating terms, i.e., terms with $\omega + \omega_\text{ext}$, following the rotating-wave approximation. This Hamiltonian can be written very simple in matrix form
\begin{equation}
    \hat \H_R = -\frac{1}{2} \Omega V_0 \eta(t) \mat{ 0 & e^{-i(\delta t - \varphi)} \\ e^{i(\delta t - \varphi)} & 0 }.
\end{equation}
From this, we conclude that if we apply a pulse at the qubit frequency, i.e., $\omega_\text{ext} = \omega$, we can rotate the state of the qubit around the Bloch sphere in \cref{fig:BlochSphere}. By setting $\varphi=0$, i.e., using only the $I$ component we rotate about the $x$ axis. By setting $\varphi = \pi/2$, i.e., using only the $Q$ component, we rotate about the $y$ axis.

\subsection{Single-qubit gates}\label{sec:onequbitGates}

One of the objectives of using superconducting circuits is to be able to perform high-quality gate operations on qubit degrees of freedom \cite{Nielsen2010}. Microwave driving of the qubits can be used to perform single-qubit rotation gates. To see how this works we consider the unitary time-evolution operator of the driving Hamiltonian. At qubit frequency, i.e., $\delta = 0$, it takes the form
\begin{equation}\label{eq:timeEvolutionOperator}
\begin{aligned}
    \U(t) &=\exp \left[-i\int_0^t \hat \H_R(t') \dd{t'}\right] \\
    &= \exp \left[ \frac{i}{2}  \Theta (t) (I\sigma^x - Q\sigma^y) \right],
\end{aligned}
\end{equation}
where we take the Pauli operators outside the integral as there is no time dependence other than on the envelope $\eta(t)$. Note that this holds only for $\delta=0$, as here the Hamiltonian commutes with itself at different times. For nonzero $\delta$, one needs to solve the full Dyson's series in principle \cite{Sakurai2011}.
\Cref{eq:timeEvolutionOperator} is known as Rabi driving and can be used for engineering efficient single-qubit gate operations. The angle of rotation is defined as
\begin{equation}
    \Theta (t) = \Omega V_0 \int_0^t \eta(t')\dd{t'},
\end{equation}
which depends on the macroscopic design parameters of the circuit, via the coupling $\Omega$, the envelope of the pulse, $\eta(t)$, and the amplitude of the pulse, $V_0$. The latter two can be controlled using arbitrary wave generators (AWGs). In case one wishes to implement a $\pi$ pulse one must adjust these parameters such that $\Theta (\tau) = \pi$, where $\tau$ is the length of the driving pulse. 

Consider a $\pi$ pulse. For the in-phase case, i.e., $\varphi = 0$, the time-evolution operator takes the form
\begin{equation}
    \U_\textsc{x}(\tau) = \exp\left[ \frac{i}{2}\pi \sigma^x\right] = \mat{0 & 1 \\ 1 & 0},
\end{equation}
which is a Pauli-\textsc{x} gate, also known as a \textsc{not}-gate, which maps $\ket{0}$ to $\ket{1}$ and vice versa \cite{Lucero2008,Chow2009,Motzoi2009,Rasmussen2020a}. This corresponds to a rotation by $\pi$ radians around the $x$ axis of the Bloch sphere. By changing the value of $\Theta(\tau)$ it is possible to change the angle of the rotation.
Had we instead considered the out-of-phase case, i.e., $\varphi = \pi/2$ then we would have obtained a Pauli-\textsc{y} gate which maps $\ket{0}$ to $i\ket{1}$ and $\ket{1}$ to $-i\ket{0}$, corresponding to a rotation around the $y$ axis of the Bloch sphere.

A Pauli-\textsc{z} gate can be implemented in one of three ways:
\begin{itemize}
    \item By detuning the qubit frequency with respect to the driving field for some finite amount of time. This introduces an amplified phase error, which can be modeled as effective qubit rotations around the $z$ axis \cite{Lucero2010}.
    \item Driving with an off-resonance microwave pulse. This introduces a temporary Stark shift, which causes a phase change, corresponding to a rotation around the $z$ axis.
    \item Virtual \textsc{z} gates where a composition of \textsc{x} and \textsc{y} gates rotates the qubit state around the $x$ and $y$ axes, which is equivalent to a rotation around the $z$ axis \cite{Johnson2015}. This can be achieved very effectively simply by adjusting the phases of subsequent microwave gates \cite{McKay2017}.
\end{itemize}

Finally, we note that the Hadamard gate can be performed as a combination of two rotations: a $\pi$ rotation around the $z$ axis and a $\pi/2$ rotation around the $y$ axis.

\subsection{Generalization to qudit driving}\label{sec:quditDriving}

Now let us generalize the discussion to a $d$-dimensional qudit. Quantizing and truncating the anharmonic oscillator part of the Hamiltonian in \cref{eq:H_full_driving} to $d$ levels, the qudit Hamiltonian becomes
\begin{equation}\label{eq:quditHamilton}
    \hat \H_\text{osc} = \sum_{n=0}^{d-1} \omega_n \ketbra{n},
\end{equation}
where $\omega_n$ is the energy of qudit state $\ket{n}$. This is a rewriting of the $\omega \hat b^\dagger \hat b$ term and the anharmonicity term, where the anharmonicity has been absorbed into the set of $\omega_n$. Starting from \cref{eq:driving_Hamiltonian_term} and for simplicity setting the phase in \cref{eq:voltage} to $\pi/2$ such that $ V(t) = V_0 \cos(\omega_\text{ext} t) $, we can move to the rotating frame as was also done above for the qubit using \cref{eq:H_transformation_rotating}.
We choose the frame rotating with the external driving frequency
\begin{align}
    \H_0 = \sum_{n=0}^{d-1} n\,\omega_\text{ext} \ketbra{n},
\end{align}
which is contrary to what we did for the qubit, where we rotated into a frame equal to the qubit frequency.
We see that for a qubit ($d=2$) we get $ \H_0 = - \omega_\text{ext} \sigma_z / 2 $ up to a global constant, which we could have also chosen to use above, instead of the qubit frame.

Applying \cref{eq:H_transformation_rotating} to the qudit Hamiltonian in \cref{eq:quditHamilton}, we get
\begin{align}
    \hat\H_{\text{osc},R} = \sum_{n=0}^{d-1} \left( \omega_n - n \omega_\text{ext} \right) \ketbra{n}.
\end{align}
The same transformation is performed on $ \H_\text{ext} $ by using the standard expansion of the bosonic operators. By expanding the cosine in the voltage drive using Euler's formula, the total Hamiltonian in the rotating frame can be found. It becomes
\begin{align}\label{eq:quditDriving}
    \hat\H_R = \sum_{n=0}^{d-1} \delta_n \ketbra{n} + i\Omega_n \left( \ketbra{n+1}{n} - \ketbra{n}{n+1}\right),
\end{align}
where $ \delta_n = \omega_n - n\omega_\text{ext} $ is the detuning of the $n$th state relative to the ground state driven by the external field and
\begin{align} \label{eq:coupling_to_ext_general}
    \Omega_n = \sqrt{n+1} \Omega = \sqrt{n+1} \frac{C_\text{ext}}{C + C_\text{ext}} \frac{V_\text{0}}{\sqrt{2\zeta}}
\end{align}
is the Rabi frequency of the $n$th transition. Thus, by using a single drive, we achieve great control over this specific qudit transition. Transitions between other neighboring qudit states can be performed simultaneously by using a multimode driving field.
Note that the $i$ in the second term of \cref{eq:quditDriving} comes from the choice of $\varphi=\pi/2$, which can of course be changed if desired.

The external field enables transitions between two states in the qudit if the effective detuning, $\Delta_{n,n+1}$, is small compared to the size of the Rabi frequency, $\Omega_n$. The effective detuning between the $n$th and $(n+1)$th states is given as the difference between the detuning of the two states:
\begin{equation}\label{eq:effectiveDetuning}
    \Delta_{n,n+1} = \delta_{n+1} - \delta_n = \omega_{n+1} - \omega_n - \omega_\text{ext},
\end{equation}
from which we see that the frequency of the external field, $\omega_\text{ext}$, has to match the energy difference between the two states, $\omega_{n+1} - \omega_n$, for the driving to be efficient.

As mentioned in \cref{sec:truncation}, leakage to other states when driving between two states depends on the size of the anharmonicity. This can be understood from \cref{eq:effectiveDetuning}. For a small anharmonicity, $\Delta_{n,n+1}$ is approximately the same for all $n$ since $\omega_n$ will be approximately the same for all $n$, thus it becomes difficult to single out the desired transition we want to drive since the driving frequency, $\omega_\text{ext}$, will overlap with multiple transition frequencies. Luckily, tailored control pulse methods such as derivative removal by adiabatic gate (DRAG) and its improvements \cite{Motzoi2009,Gambetta2011} can reduce this leakage significantly, which allows for relative anharmonicities of just a couple of percent. The topic of tailored control pulses is beyond the scope of this discussion, and we refer to the cited works.

\section{Coupling of modes}\label{sec:coupling}

In our central example of \cref{fig:exampleCircuit}, we considered direct inductive coupling. While this coupling is rather straightforward theoretically it is rather difficult to implement experimentally. We therefore now consider simpler ways to couple qubits. By coupling qubits we also open up the possibility of implementing two-qubit gates. Examples of more sophisticated approaches to coupling qubits are discussed in \cref{sec:dynamicalCouplers}.

\subsection{Capacitive coupling}

\begin{figure}
    \centering
    \includegraphics[scale=.8]{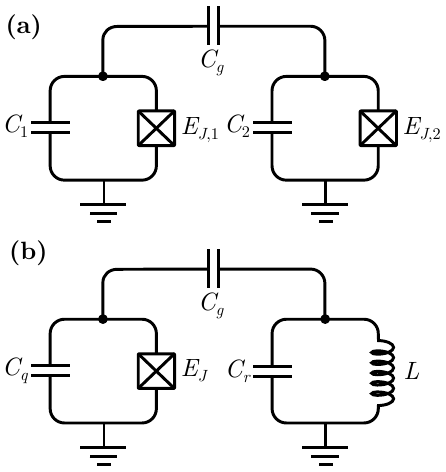}
    \caption{\textbf{(a)} Two transmonlike qubits coupled by a single capacitor with capacitance $C_g$, resulting in a static coupling between the modes. \textbf{(b)} A transmonlike qubit coupled to a linear resonator via a capacitor of capacitance $C_g$.}
    \label{fig:capacitiveCoupling}
\end{figure}

The simplest form of coupling both experimentally and theoretically is arguably capacitive coupling. Consider two transmonlike qubits coupled by a single capacitor with capacitance $C_g$, as seen in \cref{fig:capacitiveCoupling}(a). Note the similarities between this coupling and the circuit in \cref{fig:exampleCircuit}. As we see, the resulting Hamiltonian of \cref{fig:capacitiveCoupling}(a) is close to the Hamiltonian in \cref{eq:HexampleTimeAveraged}. However, capacitive coupling are much simpler to achieve experimentally.

The Hamiltonian is easily found following the approach in \cref{sec:MethodofNodes} 
\begin{equation}\label{eq:capacitiveCoupledH}
    \H = \frac{1}{2}\vec q^T \vec C^{-1} \vec q - E_{J,1} \cos \phi_1 - E_{J,2} \cos \phi_2,
\end{equation}
where $\vec q = (q_1, q_2)^T$ is the vector of conjugate momentum and the capacitance matrix is
\begin{equation}
    \vec C = \mat{C_1 + C_g & -C_g \\ -C_g & C_2 + C_g},
\end{equation}
which is invertible
\begin{equation}
    \vec C^{-1} = \frac{1}{C_\Sigma}\mat{C_2 + C_g & C_g \\ C_g & C_1 + C_g} \simeq \mat{\frac{1}{C_1} & \frac{C_g}{C_1C_2} \\ \frac{C_g}{C_1C_2} & \frac{1}{C_2}},
\end{equation}
where $C_\Sigma = \det (C)= C_1C_2 + C_1C_g + C_2C_g$. In the approximation of the second step above, we assume that the shunting capacitances are larger than the coupling capacitance, $C_n \gg C_g$, as is usually the case.
After rewriting to interacting harmonic oscillators the diagonal elements of $\vec C^{-1}$ contribute to the respective modes with the frequencies
\begin{equation}\label{eq:capacitivelyCoupledFreq}
    \omega_n = \sqrt{E_{C,n}E_{J,n}} + \alpha_n,
\end{equation}
where the effective capacitive energy is $E_{C,n} = (C_n+C_g)/C_\Sigma$ and the anharmonicity is $\alpha_n = -E_{C,n}$.
The off-diagonal elements on the other hand contribute to the interaction. The interaction term of the Hamiltonian is
\begin{equation}
    \H_\text{int} = \frac{C_g}{C_\Sigma}q_1q_2.
\end{equation}
Quantizing the Hamiltonian and changing into annihilation and creation operators the interaction part takes the form
\begin{equation}\label{eq:capacitivelyCoupledHint}
    \H_\text{int} = g_{12} \left( \hat b_1^\dagger \hat b_2 + \hat b_2\hat b_1^\dagger  \right),
\end{equation}
where we remove terms that do not conserve the total number of excitations by using the RWA. The coupling strength is 
\begin{equation}\label{eq:capacitivelyCoupledStrength}
    g_{12} = \frac{C_g}{2C_\Sigma \sqrt{\zeta_1\zeta_2}},
\end{equation}
where $\zeta_n$ is the impedance in \cref{eq:impedance}. Note the similarity with \cref{eq:g12Example} if one defines $E_{C,12} = C_g/2C_\Sigma$.
Such a coupling is called a transverse coupling since the interaction Hamiltonian only has nonzero matrix elements in off-diagonal entries. This is contrary to the longitudinal coupling discussed in \cref{sec:inductiveCoupling}.

\subsection{Two-qubit gates}

As with the single-qubit gates in \cref{sec:onequbitGates}, we can calculate the time-evolution operator, as in \cref{eq:timeEvolutionOperator} of the interacting Hamiltonian in order to determine the gate operation. However, contrary to microwave driving we cannot turn the interaction on and off directly. Luckily there are several approaches to this problem, the simplest being tuning the two qubits in and out of resonance such that the interaction terms time average to zero due to the RWA discussed in \cref{sec:rwa}. 
Examples of more complex and tunable coupling schemes are discussed \cref{sec:dynamicalCouplers}.

Consider the interaction part of the Hamiltonian in \cref{eq:capacitivelyCoupledHint}, we calculate the time-evolution operator of the two-level truncation of this
\begin{equation}\label{eq:2qubitGate}
\begin{aligned}
    \U(t) &= \exp \left[ i \int_0^t \eta(t') \hat \H_\text{int} \dd t' \right] \\
    &= \exp \left[ i \Theta(t) (\sigma_1^+ \sigma_2^- + \sigma_1^-\sigma_2^+ )\right] \\
    &= \mat{1 & 0 & 0 & 0\\ 0 & \cos \Theta(t) & -i\sin \Theta(t) & 0 \\ 0 & -i\sin \Theta(t) & \cos \Theta(t) & 0 \\ 0 & 0 & 0 & 1},
\end{aligned}
\end{equation}
where $\eta(t)$ is the envelope constructed so that it correspond to tuning the two qubits in and out of resonance, and we assume that this is the only part of the integral with time dependence. We also assume that the Hamiltonian commutes with itself at different times.
The angle of the coupling is given as
\begin{equation}
    \Theta(t) = g \int_0^t \eta(t') \dd t',
\end{equation}
which depends on the coupling strength, $g$, and the envelope $\eta(t)$. By setting $\Theta(\tau) = \pi/2$ we obtain the $i$\textsc{swap} gate from \cref{eq:2qubitGate} and taking $\Theta(\tau) = \pi/4$ we find the $\sqrt{i\textsc{swap}}$ gate.

Note that a similar procedure to the $i$\textsc{swap} gate can be used to create a $\textsc{cz}$ gate \cite{Strauch2003}.

\subsection{Linear resonators: control and measurement}\label{sec:resonators}

So far we have considered how to engineer anharmonic oscillators and truncate them into qubits as well as how to drive the qubits. However, for a qubit to be useful we must also be able to control it and perform measurements on it \cite{DiVincenzo2000}. These two things can be accomplished by coupling the qubit to a linear resonator, which is a simple harmonic oscillator \cite{Blais2003}.

Consider therefore the circuit presented in \cref{fig:capacitiveCoupling}(b) consisting of a transmonlike qubit capacitively coupled to an $LC$ oscillator or linear resonator. This circuit is similar to the example circuit presented in \cref{fig:capacitiveCoupling}(a) and the analysis up until truncation is identical with $1 \rightarrow q$, $2 \rightarrow r$, and only one anharmonicity meaning that we must change $-E_{J,2} \cos\phi_2 $ to $ \phi_r^2/2L$ in \cref{eq:capacitiveCoupledH}. Thus, we can truncate only the mode with the anharmonicity which results in the following Hamiltonian
\begin{equation}\label{eq:JanyesCummings}
    \hat \H_\text{JC} = \omega_r \hat b^\dagger \hat b + \frac{1}{2}\omega_q \sigma^z + g\left( \sigma^+\hat b + \sigma^- \hat b^\dagger \right),
\end{equation}
where $\hat b^\dagger$ and $\hat b$ are the creation and annihilation operators for the linear resonator, $\sigma^z$ is the $z$ Pauli operator of the qubit, and $\sigma^\pm$ represents the process of exciting and de-exciting the qubit. The qubit frequency is given as in \cref{eq:capacitivelyCoupledFreq}, the resonator frequency is given by
\begin{equation}
    \omega_r = \sqrt{E_{C,r}/L} ,
\end{equation}
and the coupling strength is given as in \cref{eq:capacitivelyCoupledStrength}

The Hamiltonian in \cref{eq:JanyesCummings} is known as the Jaynes-Cummings (JC) Hamiltonian, which was initially used in quantum optics to describe a two-level atom in a cavity \cite{Jaynes1963,Shore1993,Gerry2005}. Since then, the model has found application in many areas of physics, including superconducting electronic circuits, where a qubit is typically coupled to a transmission line resonator \cite{Wallraff2004,Frunzio2005,Wallraff2007,Leek2007,Fink2008,Reed2010,Kerman2013,Sank2016,Blais2020}. Because the Jaynes-Cummings Hamiltonian comes from quantum optics and cavity quantum electrodynamics (cavity QED), coupling between superconducting circuits and linear resonators is often denoted circuit QED.

Consider the limit where the qubit frequency is far detuned from the resonator frequency compared to the coupling rate and resonator linewidth $\kappa = \omega_r/Q$, where $Q$ is the quality factor of the resonator, i.e., $\Delta = |\omega_r - \omega_q| \gg g, \kappa$.
This is known as the dispersive limit since there is no direct exchange of energy between the two systems, i.e., only dispersive interactions between the resonator and the qubit occur. Using second-order perturbation theory we see that the qubit and the resonator change each other's frequencies \cite{Blais2004,Wallraff2005,Boissonneault2009}.

In the dispersive regime the Jaynes-Cummings Hamiltonian can be approximately diagonalized using the unitary transformation $ e^{\hat S}$ where $\hat S = \lambda ( \sigma^+\hat b - \sigma^- \hat b^\dagger )$ and $\lambda = g/\Delta$ is a small parameter. This transformation is called a Schrieffer-Wolff transformation \cite{Bravyi2011}. Using the Baker-Campbell-Hausdorff formula \cite{Sakurai2011} to second order in $\lambda$ we find the Hamiltonian in the dispersive regime becomes
\begin{equation}\label{eq:Hdisp}
    \hat \H_\text{disp} = e^{\hat S}\hat \H_\text{JC} e^{-\hat S} = (\omega_r + \chi \sigma^z)\hat b^\dagger \hat b + \frac{1}{2}\tilde \omega_q \sigma^z,
\end{equation}
where we define $\chi = g^2/\Delta$ as the qubit dependent frequency shift or dispersive shift. The qubit frequency is Lamb shifted to $\tilde \omega_q = \omega_q + \chi$, induced by the vacuum fluctuations in the resonator. Note that \cref{eq:Hdisp} is derived for a two-level atom/qubit. Taking the second excited state into account modifies the expression for the shift into
\begin{equation}
    \chi = - \frac{g_{01}^2}{\Delta}\left( \frac{1}{1+\Delta/\alpha} \right),
\end{equation}
where $g_{01}$ is the coupling rate between the 0 and 1 state of the qubit and $\alpha$ is the anharmonicity of the qubit, \cref{eq:anharmonicity}.

One can interpret the dispersive qubit-resonator interaction in two ways. Either as a shift of the qubit frequency by a quantity proportional to the photon population of the resonator $2\chi \langle \hat b^\dagger \hat b \rangle$ or as a qubit-dependent pull of the resonator frequency, $\omega_r \rightarrow \omega_r \pm \chi$. 

In the first interpretation the bare qubit frequency is modified by a Lamb shift and by an additional amount proportional to the number of photons populating the resonator. This is known as the ac Stark shift. It has the consequence that fluctuations in the photon number of the resonator induce small shifts in the qubit frequency, which brings it slightly out of its rotating frame and cause dephasing \cite{Schuster2005,Gambetta2006,Schuster2007,Rigetti2012,Sears2012,Zhang2017,Yan2018b}. In an experiment, it is therefore important to reduce photon-number fluctuations of the resonator, e.g., by keeping the process properly thermalized.

In the second interpretation, the resonator frequency is dependent on the state of the qubit. This means that it is possible to make a quantum nondemolition (QND) measurement of the qubit by shinning microwaves into the resonator at a frequency close to $\omega_r$ and then measuring the transmitted signal using standard homodyne techniques \cite{Wiseman2009}. However, the approximation in \cref{eq:Hdisp} is only valid in the small-photon limit, i.e., when when the resonator photon number, $\hat N = \hat b^\dagger \hat b$ is less than the critical photon number $N_c = \Delta^2 /4g^2$. This sets an upper limit to the power of the resonator as a probe while maintaining the conditions for a QND measurement. However, this is not the whole story; Ref. \cite{Sank2016} has shown that level crossings with other states of the qubit-resonator system induce state transitions which can be explained by the Jaynes-Cummings model. This is beyond the scope of our discussion, and we refer to the cited work for more information.

In the other limit, when the detuning between the qubit and the resonator frequency is small compared to the coupling rate, i.e., $\Delta \ll g$ we obtain a hybridization of the energy levels of the two systems. This opens up for a Rabi mode splitting, where each transition between the qubit and the resonator splits into two states with distance $\sqrt{N}g/\pi$ where $N$ denotes the resonator mode, i.e., the photon number. Thus excitation is coherently swapped between the two systems. While this cannot be used to perform measurements on the qubit it can be used to mediate couplings between two qubits by coupling another qubit to the resonator \cite{Blais2007,Majer2007}. We do not dive deeper into the details of measurements and couplings to linear resonators. For an experimental-minded review see, e.g., Ref. \cite{Krantz2019}.

\subsection{Inductive coupling}\label{sec:inductiveCoupling}

\begin{figure}
    \centering
    \includegraphics[scale=.8]{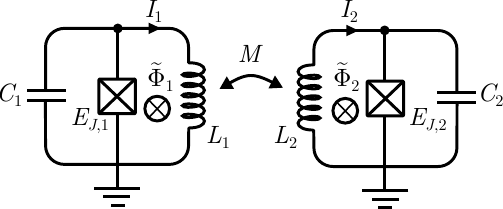}
    \caption{Mutual inductive coupling between two modes.}
    \label{fig:inductiveCoupling}
\end{figure}

So far we have considered only direct inductance as a way of coupling two qubits.  In this section we consider the mutual inductance of two modes as a means of coupling the modes. Consider therefore the two circuits in \cref{fig:inductiveCoupling}, consisting of a Josephson junction, a capacitor, and a linear inductor. Such circuits are known as rf SQUIDs \cite{Friedman2002}. Each of the circuits have the following Hamiltonian
\begin{equation}
    \hat \H_j = 4E_{C,j} \hat n_j^2 + \frac{1}{2L_j} \hat\phi_j^2 - E_{J,j} \cos (\hat \phi_j + \tilde \Phi_j).
\end{equation}
If two such circuits are brought into proximity of each other they will share a mutual inductance, yielding an interaction Hamiltonian
\begin{equation}\label{eq:HintMutualInd}
\begin{aligned}
\hat \H_\text{int} &= M_{12} \hat I_1 \hat I_2 \\ 
&= M_{12} I_{c1} \sin(\hat \phi_1+ \tilde \Phi_j) I_{2c} \sin (\hat \phi_2 + \tilde \Phi_j),
\end{aligned}
\end{equation}
where $\hat I_j$ is the current operator of the Josephson junction, see \cref{eq:josephson_current_phase}. The mutual inductance $M_{12}$ between the two circuits depends on the relative geometrical placement of the circuits. This can be increased, e.g., by overlapping the circuits \cite{Johnson2011} or by letting them share the same wire or Josephson-junction inductor \cite{Nori2005,Grajcar2006,Niskanen2007,Ashhab2008}. 

Consider now the case of no external flux, i.e., $\tilde \Phi = 0$. If we expand the potential to fourth order the interaction Hamiltonian takes the form
\begin{equation}
    \hat \H = M_{12} \left[\hat \phi_1 \hat \phi_2 - \frac{1}{36}(\hat\phi_1 \hat \phi_2^3 + \hat\phi_1^3 \hat \phi_2)\right].
\end{equation}
Truncating into a two-level model using \cref{tab:Overview,tab:Overview2} we find that the coupling becomes transverse
\begin{equation}
    \hat \H_\text{int} = g_{x} \sigma_1^x\sigma_2^x,
\end{equation}
where the coupling constant is
\begin{equation}
    g_x =  \frac{1}{2}M_{12}\sqrt{\zeta_1\zeta_2} \left[ 1 - \frac{1}{24} (\zeta_1 + \zeta_2) \right],
\end{equation}
with impedances given by \cref{eq:impedance}.

Consider now an external flux of $\tilde \Phi = \pi/2$. In this case, the sine terms obtain a phase effectively changing the terms into cosines. Expanding these to second-order yields
\begin{equation}
    \hat \H_\text{int} = \frac{M_{12}}{4}\left[-(\hat \phi_1^2 + \hat \phi_2^2) + \frac{1}{144}(\hat \phi_1^4 + \hat \phi_2^4) + \hat \phi_1^2\hat \phi_2^2 \right],
\end{equation}
where we recognize the two first terms as corrections to the qubit frequencies, the next two terms as corrections to the anharmonicities, and finally the last term is the interaction term. Considering only the last term and truncating into a two-level model we find (see \cref{tab:Overview,tab:Overview2})
\begin{equation}\label{eq:Hintgz}
    \hat \H '_\text{int} = g_z (\sigma_1^z \sigma_2^z - 2\sigma_1^z - 2\sigma_2^z).
\end{equation}
The first term is a longitudinal coupling between the two qubits 
with coupling constant
\begin{equation}
    g_z = \frac{1}{16}M_{12}\zeta_1\zeta_2.
\end{equation}
It is called longitudinal because all off-diagonal matrix elements are zero, contrary to transverse coupling. Longitudinal coupling can be used to create entanglement without exchanging energy between the modes by enabling a so-called phase gate \cite{Blais2007,Paik2016,Krantz2019}.
From the last two terms in \cref{eq:Hintgz}, we see that we obtain further corrections to the qubit frequencies.

\section{Noise and decoherence}\label{sec:noise}

So far we have considered only closed quantum systems, i.e., systems without interaction with the environment. This is usually a good approximation as we are dealing with cryogenic and thus isolated superconducting circuits. However, even in the best experimental setups, random and uncontrollable processes in the environment surrounding the system do occur. These are sources of noise and lead to decoherence of the quantum system. It is, therefore, necessary to develop a formalism to treat this theoretically as well. We assume that the Hamiltonian of the system and the environment is separable and has the form
\begin{equation}\label{eq:HsysCoupledToEnv}
    \hat\H = \hat\H_\text{sys} + \hat\H_\text{env} + \nu \hat S \cdot \hat \lambda, 
\end{equation}
where $\hat\H_\text{sys}$ is the Hamiltonian of the system, $\hat\H_\text{env}$ is the Hamiltonian of the environment, and the interaction strength between the system and the environment is given by $\nu$, while $\hat S$ is an operator within the system Hamiltonian $\hat \H_\text{sys}$ and $\hat \lambda$ represents the noisy environment which produces fluctuations $\delta \lambda$.

The treatment of open quantum systems is a whole subject on its own and a complete treatment is beyond the scope of this tutorial. We therefore present only a method for modeling noise in qubit systems. For a more extensive treatment of open quantum systems see e.g. Ref. \cite{Breuer2002}, and for an introduction on how to treat noise in an experiment see, e.g., Ref. \cite{Krantz2019}.

\subsection{Bloch-Redfield model}\label{sec:blochRedfield}

Consider an arbitrary state on the Bloch sphere as in \cref{eq:arbitraryState}. The density matrix for such a pure state is \cite{Nielsen2010}
\begin{equation}\label{eq:densityMatrix}
    \rho = \ketbra{\psi} = \frac{1}{2} (I + \vec a \cdot \vec \sigma) = \mat{|\alpha|^2 & \alpha\beta^* \\ \alpha^*\beta & |\beta|^2},
\end{equation}
where $I$ is the identity matrix, $\vec a$ is the Bloch vector, and $\vec \sigma = (\sigma^x, \sigma^y, \sigma^z)$ is the vector of Pauli spin matrices. If $\rho$ represents a pure state, $\psi$, then $\tr \rho^2=1$ and the Block vector becomes a unit vector, $\vec a = (\sin\theta\cos\phi, \sin\theta\sin\phi, \cos\theta)$, where $\theta$ and $\phi$ are the angles of the Bloch vector. If, on the other hand, $|\vec a| < 1$ the density matrix $\rho$ represents a mixed state with $\tr \rho^2<1$. In this case the Bloch vector terminates at points inside the unit sphere. 

In the Bloch-Redfield formulation of two-level systems, sources of noise are weakly coupled to the system with short correlation times compared to the system dynamics \cite{Wangsness1953,Bloch1957,Redfield1965,Breuer2002}. The noise in this formulation is determined by two rates: The longitudinal relaxation rate and the transverse relaxation rate.

\begin{figure}
    \centering
    \includegraphics[width=\columnwidth]{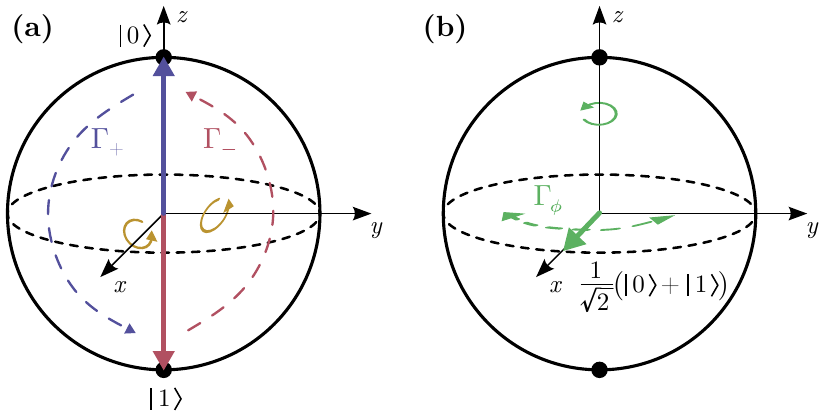}
    \caption{Bloch-sphere representation of noise. \textbf{(a)} \textbf{Longitudinal relaxation} is the result of energy exchange between the qubit and the environment. Transverse noise couples to the qubit and drives a rotation (transition) around an axis in the $x$-$y$ plane. Longitudinal relaxation is driven both by emission of energy to the environment, $\Gamma_-$ and absorption of energy from the environment, $\Gamma_+$. For a typical superconducting qubit, the temperature is much lower than the frequency of the qubit, $k_B T \ll \hbar \omega$, which suppresses the absorption rate, such that $\Gamma_1 \simeq \Gamma_-$. \textbf{(b)} \textbf{Pure dephasing} is the result of longitudinal noise that drives a rotation around the $z$ axis. Due to stochastic frequency fluctuations, a Bloch vector will diffuse both clockwise and counterclockwise around the $z$ axis parallel to the equator.}
    \label{fig:noise}
\end{figure}

\subsubsection{Longitudinal relaxation}

The longitudinal relaxation rate, $\Gamma_1 = 1/T_1$, describes depolarization along the qubit quantization axis, often referred to as “energy decay” or “energy relaxation”, which is why it is often referred to as the relaxation time. Longitudinal relaxation is caused by transverse noise, via the $x$ or $y$ axis on the Bloch sphere, see \cref{fig:noise}(a). Depolarization of the superconducting circuit occurs due to exchange of energy with the environment, leading both to excitation and relaxation of the qubits, meaning that one can write
\begin{equation}\label{eq:longitudinalNoise}
   \Gamma_1 = \Gamma_+ + \Gamma_-.
\end{equation}
Due to Boltzmann statistics and the fact that superconducting qubits are operated at low temperatures ($T \lesssim \SI{20}{\milli\kelvin}$) and with a qubit frequency in the $\si{\GHz}$ regime, the qubits generally lose energy to the environment, meaning that the excitation rate $\Gamma_+$ is suppressed exponentially as
\begin{equation}
    \frac{\Gamma_+}{\Gamma_-} = e^{-\beta\omega}, 
\end{equation}
where $\beta = 1/k_B T$ is the inverse of the Boltzmann constant multiplied with the temperature. Note, however, that empirically we often see stray population of the excited state much higher than we would expect from this theory. From Maxwell-Boltzmann statistics, we would expect a thermal population of the excited state of $P_{\ket{1}}\sim 10^{-5}$\%, but the measured excited-state population is often orders of magnitudes higher at around 1\% \cite{Jin2015}.

The longitudinal relaxation rate can be determined using Fermi's golden rule
\begin{equation}
    \Gamma_1 = \frac{1}{\hbar^2}|\mel{0}{\hat S}{1}|^2 S_\lambda (\omega_q),
\end{equation}
where $\hat S$ is the transverse coupling of the qubit to the environment, i.e., a coupling of the type $\sigma^x$ or $\sigma^y$. The qubit frequency is denoted $\omega_q$. The noise power spectral density
\begin{equation}
    S_\lambda (\omega) = \int_{\infty}^{-\infty} \dd t \langle \hat\lambda(t) \hat\lambda(0) \rangle e^{-i\omega t},
\end{equation}
characterizes the frequency distribution of the noise power for a stationary noise process $\hat\lambda$. Note that the Wiener-Khintchine theorem states that $S_\lambda(\omega)$ is the Fourier transform of the autocorrelation function $c_\lambda(t) = \langle \hat\lambda (t) \hat\lambda (0) \rangle$ of the noise source $\lambda$ \cite{Wiener1930,Champeney1987}.

The longitudinal relaxation rate can be measured by preparing the qubit in state $\ket{1}$ (e.g. using a $\pi$ pulse as in \cref{sec:driving}) and then making multiple measurements of the qubit excited-state population at a set of subsequent times \cite{Krantz2019}.

\subsubsection{Transverse relaxation}
The transverse relaxation time
\begin{equation}\label{eq:transverseNoise}
    \Gamma_2 = \frac{1}{T_2} = \frac{\Gamma_1}{2} + \Gamma_\phi
\end{equation}
describes the loss of coherence of a superposition. As seen in \cref{eq:transverseNoise} it is caused both by transverse noise, which leads to energy (longitudinal) relaxation of the excited-state component of the superposition state, and by longitudinal noise, which cause fluctuations of the qubit frequency and leads to pure dephasing, see \cref{sec:pureDephasing} below. Note that the sum in \cref{eq:transverseNoise} is only valid for weak noise that also is only correlated at short times \cite{Slichter1990}.

We introduce transverse relaxation as it is a measurable quantity, contrary to pure dephasing which can only be inferred using \cref{eq:transverseNoise}. Transverse relaxation can be measured using Ramsey interferometry \cite{Ramsey1950,Krantz2019}.
In Ramsey interferometry, a $\pi/2$ pulse rotates the Bloch vector from $\ket{0}$ to the equator of the Bloch sphere. If we know the qubit frequency perfectly, it should remain stationary at the equator, and if we apply another $\pi/2$ pulse at some time later, we should measure $\ket{1}$. However, if our knowledge of the qubit frequency and our assumed frame does not match the qubit's actual rotation frame, then the state will not remain stationary at the equator of the Bloch sphere after the first $\pi/2$ pulse is applied. Instead, it will precess around the equator at a frequency equal to the difference between the assumed frame and the actual qubit frequency. This means that if we perform two $\pi/2$ pulses with variable delay in between, we should observe oscillations in the measured state of the qubit. In reality, one often chooses a frame that is intentionally detuned from the qubit frequency so that these oscillations are observed even for the perfectly calibrated qubit. This means that an error in qubit frequency will result in a difference from the expected oscillation frequency \cite{Chen2018}.
For simple Markovian noise, these oscillations are exponentially damped with characteristic time $T_2$ \cite{Krantz2019}.

\subsubsection{Pure dephasing}\label{sec:pureDephasing}

The pure dephasing rate $\Gamma_\phi$ describes depolarization in the $x$-$y$ plane of the Bloch sphere. It is referred to as “pure dephasing,” to distinguish it from other phase-breaking processes such as energy excitation or relaxation. Pure dephasing is caused by longitudinal noise that couples to the qubit via the $z$ axis. This longitudinal noise causes the qubit frequency, $\omega$, to fluctuate such that it is no longer equal to the interaction frame frequency, causing the Bloch vector to precess forward or backward in the interacting frame as seen in \cref{fig:noise}(b).

To lowest order, the pure dephasing rate is orthogonal to the difference between the two diagonal matrix elements \cite{Breuer2002,Burkard2004}
\begin{equation}
    \Gamma_\phi = \frac{1}{\hbar^2}\left(|\mel{0}{\hat S}{0} - \mel{1}{\hat S}{1}|^2 \right)S_\lambda (0),   
\end{equation}
where $\hat S$ is the longitudinal coupling of the qubit to the environment, i.e., a coupling of the type $\sigma^z$.
This means that pure dephasing disappears if $\mel{0}{\hat\H_\text{env}}{0} = \mel{1}{\hat\H_\text{env}}{1}$. For superconducting circuits, this can often be realized by tuning the system to the so-called \enquote{sweet spot} using external flux biasing. This means that the transverse relaxation becomes approximately half the longitudinal relaxation rate as in \cref{eq:transverseNoise}. Thus decreasing the longitudinal relaxation rate becomes the main focus when developing qubits with the goal of increasing the lifetime. Note, however, in reality, pure dephasing will never disappear entirely due to effects beyond the linear theory, such as higher-order corrections, other noise sources, or nonmarkovian effects. Nevertheless these effects are small at the \enquote{sweet spot}, and therefore relaxation noise will often be the dominant source of noise.

Note that pure dephasing is in principle reversible as there is no energy exchange with the environment, which means that it can be undone without destroying any quantum information \cite{Bylander2011}. It is also worth noting that qubit dephasing is subject to broadband noise since noise at any frequency can modify the qubit frequency and cause dephasing.

The impact of noise alters the density matrix of \cref{eq:densityMatrix} giving us the Bloch-Redfield density matrix \cite{Ithier2005}
\begin{equation}
    \rho_\text{BR} = \mat{1+(|\alpha|^2-1)e^{-\Gamma_1t} & \alpha\beta^* e^{-i\delta t} e^{-\Gamma_2 t} \\ \alpha^*\beta e^{-i\delta t} e^{-\Gamma_2 t} & |\beta|^2 e^{-\Gamma_1t}}.
\end{equation}
Note that the longitudinal relaxation rate influences the diagonal, while the transverse influences only the off diagonal. We also include the phase difference $\delta = \omega_q - \omega_\text{ext}$ between the qubit frequency, $\omega_q$, and the rotating frame frequency, $\omega_\text{ext}$, which is needed in order to perform Ramsey interferometry.

\subsection{Master equation}

As we are interested in the effect of noise on the dynamics of the system we consider a so-called {\it master equation} for the system. A master equation describes the time evolution of a system (in our case an electrical circuit) where we model the system as an ensemble of states described by a density matrix $\rho$, and where we can determine the transition between the states by a transition matrix \cite{Breuer2002}.

From the time-dependent Schr\"{o}dinger equation in \cref{eq:schrodinger}, we can derive a master equation for the closed system called the Liouville-von Neumann equation
\begin{equation}
    \dot \rho (t) = -i [\hat \H, \rho(t)],
\end{equation}
where $\hat \H $ is the Hamiltonian of the system and $\rho$ is the density matrix in \cref{eq:densityMatrix}. Note how it resembles Heisenberg's equations of motion in \cref{eq:HeisenbergEOM}, but with a different sign since the density matrix is a dynamical variable, i.e. it is an operator in contrast to a quantum state or wave function.

As we are interested in the effect of noise we must add other terms to the Liouville-von Neumann equation. For a system that is weakly coupled to the environment, the evolution is described by the Lindblad master equation \cite{Breuer2002,Burkard2004,Manzano2020}
\begin{equation}\label{eq:lindblad}
    \dot \rho (t) = -i [\hat \H, \rho(t)] + \sum_{i} \Gamma_i \left( \hat L_i \rho \hat L_i^\dagger - \frac{1}{2} \{\hat L_i \hat L_i^\dagger, \rho \} \right),
\end{equation}
where $\{\cdot, \cdot \}$ is the anticommutator, and $\hat L_i$ are the so-called jump operators representing the interaction between the system and the environment.

For the case of a two-level model with both longitudinal and transverse relaxation weakly coupled to the environment, the Lindblad master equation takes the form
\begin{equation}\label{eq:lindblad2}
\begin{split}
    \dot \rho (t) &= -i [\hat \H, \rho(t)] \\
    &\phantom{=}+ \Gamma_{-} \left( \sigma_- \rho \sigma_+ - \frac{1}{2} \{\hat \sigma_- \hat \sigma_+, \rho \} \right)\\
    &\phantom{=}+ \Gamma_{+} \left( \sigma_+ \rho \sigma_- - \frac{1}{2} \{\hat \sigma_+ \hat \sigma_-, \rho \} \right)\\
    &\phantom{=}+ \Gamma_{\phi} \left( \sigma_z \rho \sigma_z-\rho \right)
    ,
\end{split}
\end{equation}
where the decoherence rates, $\Gamma_i$, can be found in \cref{eq:longitudinalNoise,eq:transverseNoise}. \Cref{eq:lindblad} can be used to simulate a system including noise and is usually solved numerically using, e.g., QuTiP \cite{Johansson2013}.

\section{Examples}\label{sec:examples}

\begin{figure}
    \centering
    \includegraphics[width=\columnwidth]{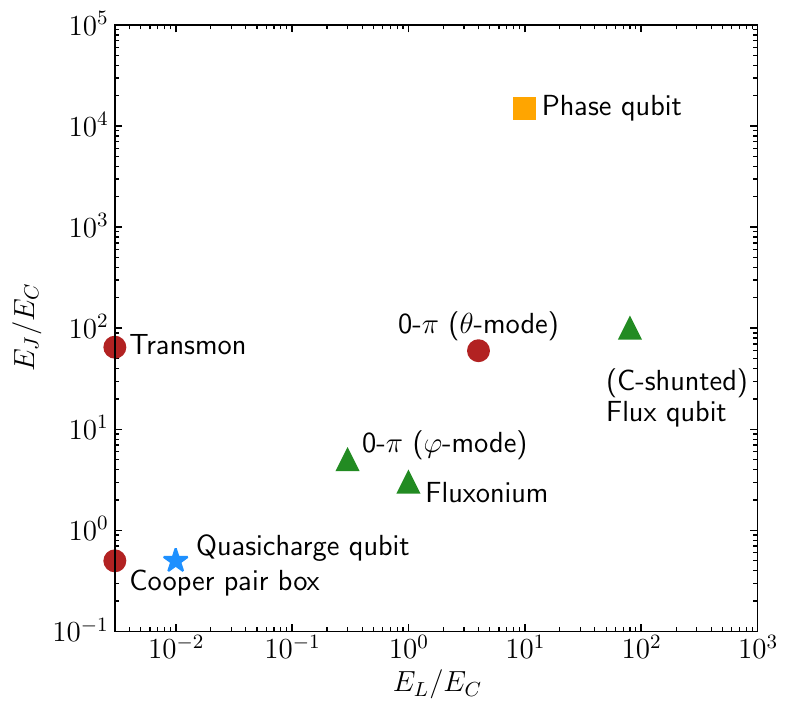}
    \caption{Parameter space of the \enquote{qubit zoo.} The qubits are plotted according to their effective Josephson energy, $E_J$, and inductive energy, $E_L$, both normalized by their effective capacitive energy, $E_C$. The marker indicates the type of qubits, with a yellow square indicating the phase qubit, red dots indicating charge qubits, green triangles indicating flux qubits, and a blue star for the quasicharge qubit. Note that the placement of the qubits is only approximate as the effective energies are not definitive. Also, note that the 0-$\pi$ qubit is plotted twice, once for each of its modes, where the $\varphi$ mode works similar to a fluxonium qubit, while the $\theta$ mode works similarly to the transmon qubit.}
    \label{fig:qubitZoo}
\end{figure}

In this section, we present some examples of more or less well known superconducting qubits. We start from some simple early single-qubit designs, then move to the transmon and flux qubit, and finally, we discuss couplings between qubits. In \cref{fig:qubitZoo} we present an overview of the qubits discussed in the examples.

There are four fundamental types of qubits: Phase qubits, charge qubits, flux qubits, and quasicharge qubits. These qubits can be ordered in pairs according to the behavior of quantum fluctuation in the Cooper pair condensate. Charge qubits with their single-charge tunneling are dual to flux qubits with single-flux tunneling, while phase qubits with phase oscillation are dual to the quasicharge qubits with quasicharge oscillations. These fundamental qubits can be seen in \cref{fig:qubitZoo}.

The simplest realization of a superconducting qubit is a phase qubit. It is a current-biased Josephson junction, which essentially is just a Josephson junction with a current applied across it. It operates in the so-called phase regime where $E_{C}\ll E_{J}$. In this regime the Josephson tunneling dominates over the charging of the capacitor, making the anharmonicity quite small. This can be interpreted as a low kinetic (capacitive) energy compared to the potential (inductive) energy of the system. The bias current introduces the anharmonicity and adjusting the bias current closer to the critical current of the  Josephson junction increases the anharmonicity. The fact that one can tune the anharmonicity dynamically is a strength of this qubit. However, the phase qubit has a rather large decoherence noise and we do not go into further details with the phase qubit as it is rarely used in modern circuit designs. For more details see Refs. \cite{Martinis1985,Devoret1985}.

\subsection{Charge qubits}

Central types of qubits are the so-called charge qubits. These have their name from the fact that the basis states of the qubit are charge eigenstates, meaning that they are only dependent on the number of excess Cooper pairs in a disconnected superconducting island, and mostly independent of the node fluxes. We start from the single Cooper pair box and move on to the transmon qubit, which is based on the Cooper pair box.

\subsubsection{Single Cooper pair box}\label{sec:SCPB}

In 1997 the first charge qubit, known as the single Cooper pair box (SCPB), was invented \cite{Shnirman1997,CPB1998,Nakamura1999}. As with the phase qubit, it is not used in modern qubit implementations due to bad coherence times. However, we go into detail with this qubit as it forms the basis for the renowned transmon qubit as well as being a nice simple example of how to analyze a circuit.

\begin{figure}
	\centering
	\includegraphics[scale=0.75]{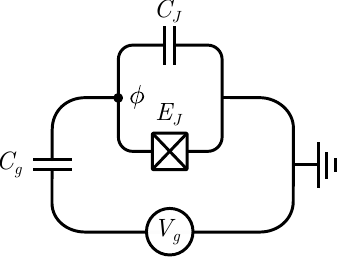}
	\caption{Circuit diagram of the single Cooper pair box, consisting of a Josephson junction, with energy $E_J$ and parasitic capacitance $C_J$, in series with a gate capacitor with capacitance $C_g$. The gate voltage is denoted $V_g$, and the system is connected to ground in the right corner. There is only one active node denoted by a dot.}
	\label{fig:scpb}
\end{figure}

The SCPB consists of a Josephson junction with energy $E_J$ and a capacitor with capacitance $C_g$ in series, with a superconducting island in between them. A parasitic capacitance $C_J$ is included in the Josephson junction. This is a lumped-circuit element representation of the natural capacitance that the junction will have by way of construction. The circuit is biased with a gate voltage $V_g$ over the capacitor, which makes it possible to transfer electrons from the reservoir to the superconducting island via the gate capacitance $C_g$. The circuit is connected to ground and thus there is only one active node with flux $\phi$ through it. The corresponding circuit diagram can be seen in \cref{fig:scpb}.

We follow the method presented in \cref{sec:MethodofNodes}. In order to write the Lagrangian, we must consider the fixed gate voltage. We model this as an external node with a well defined flux $\phi_V = V_gt$, meaning $\dot{\phi}_V = V_g$. Setting $\vec{\phi}^T = (\phi,\phi_V)$ we write the Lagrangian
\begin{equation}
    \L = \frac{1}{2}\dot{\vec{\phi}}^T \vec C \dot{\vec{\phi}} + E_J \cos \phi,
\end{equation}
where the capacitance matrix is 
\begin{equation}
	\vec C = \mat{
	C_J + C_g & -C_g \\
	-C_g & C_g
	}.
\end{equation}
Since we know that $\dot{\phi}_V = V_g$ is a classical externally controlled variable, it should not be quantized. Therefore, we only calculate one conjugate momentum
\begin{equation}
	q  = (C_g + C_j)\dot{\phi} - C_g V_g.
\end{equation}
Solving for $\dot \phi $ we perform a Legendre transformation and find the Hamiltonian
\begin{equation}
    \H = \frac{1}{2(C_g+C_J)}(q+C_g V_g)^2 - \frac{C_g V_g^2}{2} - E_J \cos \phi.
\end{equation}
We now change into conventional notation and define the effective capacitive energy
\begin{equation}\label{eq:ECSCB}
    E_C = \frac{e^2}{2(C_g + C_J)},
\end{equation}
which means that we can write the Hamiltonian as
\begin{equation}\label{eq:chargQubitH}
    \hat \H = 4E_C (\hat n - n_g )^2 - E_J \cos \hat \phi,
\end{equation}
where we quantize the dynamic variables and remove constant terms. We  further define the offset charge $n_g = C_g V_g/2e$.

\begin{figure}
	\centering
	\includegraphics[width=\columnwidth]{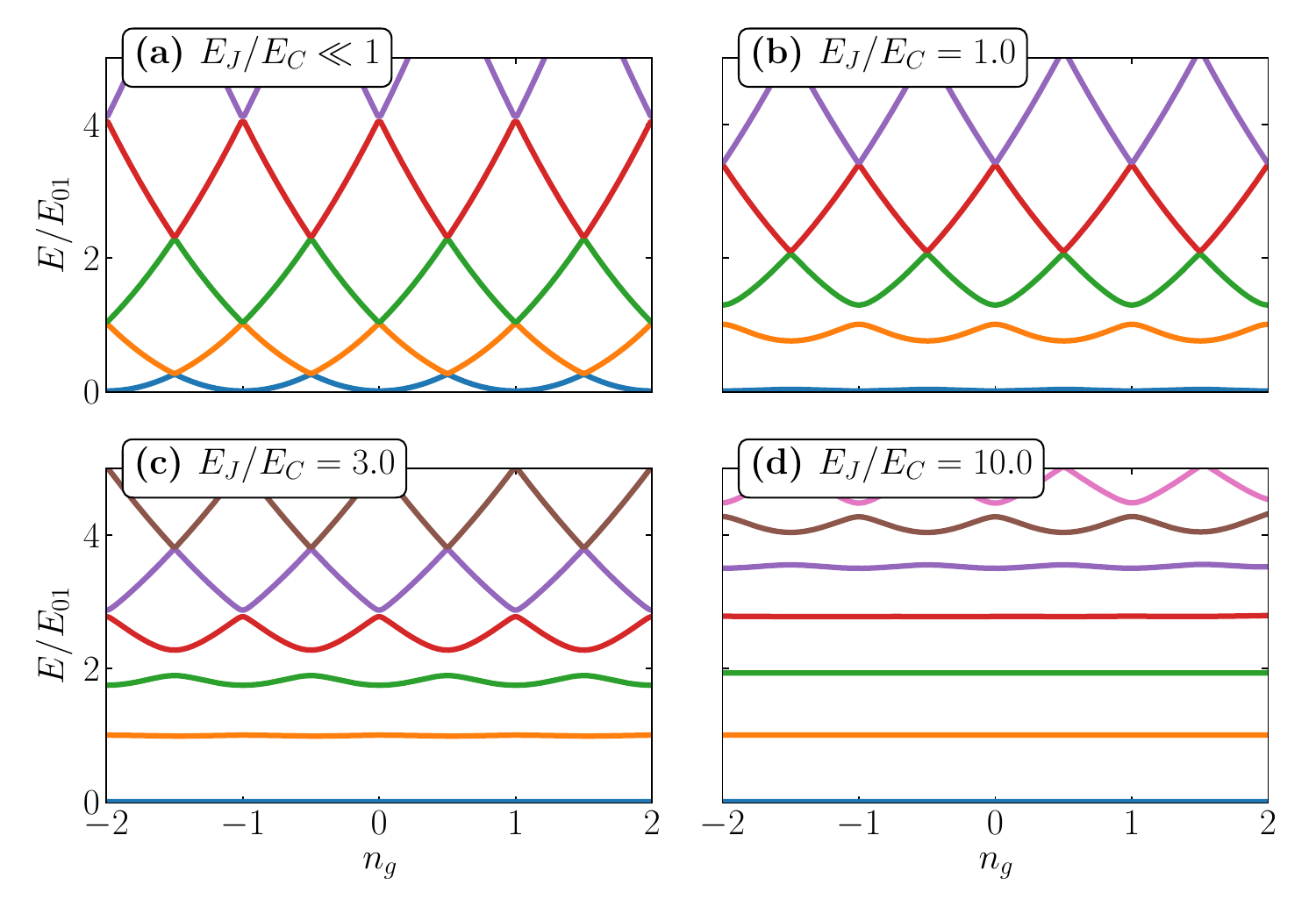}
	\caption{The energies of the lowest-lying states of the single Cooper pair box and transmon qubit as a function of the bias charge $n_g$. The difference between the two lowest bands is approximately equal to $E_J$ at the avoided crossing.}
	\label{fig:cpbEnergies}
\end{figure}

We can now discuss the operational regime of the Cooper pair box. When the Josephson energy is much smaller than the capacitive energy ($E_J/E_C \ll 1$), the energy spectrum of the system becomes a set of parabolas when plotted against $n_g$, one for each eigenvalue of $\hat{n}$. The parabolas cross at $n_g = n+1/2$, where $n\in\mathbb{Z}$, see \cref{fig:cpbEnergies}(a). If we consider the eigenstates of $\hat n$ we find that the states $\ket{n}$ and $\ket{n+1}$ are degenerate at $n_g = n+1/2$. These states are essentially charge states of the capacitor. In this picture, the Hamiltonian of the capacitor becomes
\begin{equation}
	\hat{\H}_C = 4E_C \sum_{n=-\infty}^\infty (n-n_g)^2 \ketbra{n}{n},
\end{equation}
which in matrix representation is just a diagonal matrix with $(n-n_g)^2$ on the diagonal.

Introducing the Josephson junction lifts the degeneracy and introduces an avoided crossing at $n_g = n+1/2$. The matrix representation now becomes a tridiagonal matrix with $E_J/2$ on the diagonals below and above the main diagonal, which consists of the entries from the capacitor discussed above. 

To show this we must relate the phase states of the Josephson junction $\ket{\phi}$ to the charge states $\ket{n}$. This can be done through a Fourier transform (this treatment is analogous to the treatment of a one-dimensional solid, see, e.g., Ref. \cite{Sakurai2011})
\begin{equation}\label{eq:FTphi2n}
    \ket{\phi} = \frac{1}{\sqrt{2\pi}} \sum_{n=-\infty}^\infty e^{-i n\phi} \ket{n}.
\end{equation}
Note that, since $n$ is a discrete variable, the phase must be $2\pi$ periodic. This is in agreement with the fact that we consider $\phi$ as the phase of the Josephson junction. The commutator between the two corresponding operators is
\begin{equation}
    [\hat \phi, \hat n] \sim i,
\end{equation}
where the '$\sim$' indicates that this is only true up to the association $\phi \sim \phi + 2\pi$. Since the phase is continuous, the inverse transformation of \cref{eq:FTphi2n} is
\begin{equation}\label{eq:FTn2phi}
    \ket{n} = \frac{1}{\sqrt{2\pi}} \int_0^{2\pi}\dd{\phi}e^{i\phi n} \ket{\phi}.
\end{equation}
Now writing the last term of \cref{eq:chargQubitH} as the sum of exponentials and inserting the identity relation we find
\begin{align*}
    \hat \H_J &= -E_J \cos \hat \phi \\
    &= -\frac{E_J}{2} \int_{0}^{2\pi} \dd{\phi} \ketbra{\phi} \left(e^{i\phi} + e^{-i\phi} \right) \\
    &= -\frac{E_J}{4\pi} \iint_{0}^{2\pi} \dd{\phi} \dd{\phi'} \sum_{n=-\infty}^\infty \ketbra{\phi}{\phi'} \left(e^{-in\phi}e^{i(n+1)\phi'} \right. \\
    &\phantom{\frac{E_J}{4\pi} \iint_{0}^{2\pi} \dd{\phi} \dd{\phi'} \sum_{n=-\infty}^\infty \ketbra{\phi}{\phi'}}+\left. e^{in\phi'}e^{-i(n+1)\phi} \right) \\
    &=- \frac{E_J}{2} \sum_{n=-\infty}^\infty \left( \ketbra{n}{n+1} + \ketbra{n+1}{n}\right),
\end{align*}
where we apply both \cref{eq:FTphi2n,eq:FTn2phi} in the second to last step and use the definition of the delta function as an integral in the last step.

Solving the full system, $\hat \H = \hat \H_C + \hat \H_J$, using either Mathieu functions \cite {Meixner1980} or numerically, yields the avoided crossings seen in \cref{fig:cpbEnergies}(b). The distance between these avoided crossings is approximately equal to the Josephson-junction energy $E_J$ for the lowest states in the spectrum.
To realize a qubit we set $n_g$ equal to some half-integer, which yields two states close to each other but with a large gap to higher states \cite{Yuriy2001}. That way we obtain a significant anharmonicity, see \cref{fig:cpbEnergies}(b) where the distance between the green and the yellow lines is significantly different from the distance between the blue and yellow line at, e.g., $n_g=1/2$.

The SCPB is, however, quite sensitive to small fluctuations of the gate voltage $V_g$, since this changes $n_g$ and the energy dispersion is steep around the working point $n_g=n+1/2$, as seen in \cref{fig:cpbEnergies}(b) for $E_J/E_C = 1.0$. This means that the qubit works only in this sweet spot as it is otherwise very sensitive to charge noise. This reduces the decoherence time of the system. The transmon qubit attempts to fix this problem.

\subsubsection{Transmon qubit}\label{sec:transmon}

\begin{figure}
	\centering
	\includegraphics[scale=0.75]{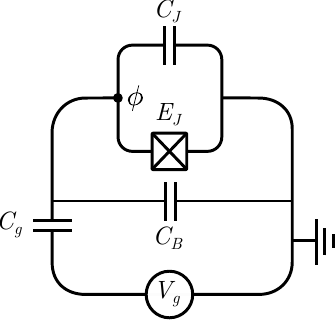}
	\caption{Circuit diagram of the transmon qubit consisting of a Josephson junction with energy $E_J$ and parasitic capacitance $C_J$ in series with a capacitor with capacitance $C_g$. The system is shunted by a large capacitance, $C_B$. The gate voltage is denoted $V_g$ and the system is connected to ground in the right corner. There is only one active node. The diagram should be compared to \cref{fig:scpb}.}
	\label{fig:transmon}
\end{figure}

The transmission-line shunted plasma oscillation qubit, or transmon qubit for short, was proposed in 2007 as an attempt to increase the coherence time in charge qubits \cite{Koch2007,Schreier2008}. It exploits the fact that the charge dispersion reduces exponentially in $E_J/E_C$, while the anharmonicity decreases only algebraically in $E_J/E_C$ following a power law. The setup resembles that of the single Cooper pair box, the difference being a large shunting capacitance, $C_B$, between the two superconducting islands, followed by a similar increase in the gate capacitance $C_g$. The circuit diagram is seen in \cref{fig:transmon}.

Capacitors in parallel add to one effective capacitor, hence the effective capacitance can be seen as the sum of the capacitance of the three capacitors $C_\Sigma = C_J + C_B + C_g$. With this elementary knowledge, the Hamiltonian of the transmon becomes identical to that of the single Copper pair box from \cref{eq:chargQubitH} with the exception that
\begin{equation}
	E_C = \frac{e^2}{2(C_J + C_B + C_g)},
\end{equation}
where we change the effective capacitance in \cref{eq:ECSCB}. This gives much more freedom in choosing the ratio $E_J/E_C$, and we can thus solve the Hamiltonian for the energy dispersion for larger $E_J/E_C$. The result is seen in \cref{fig:cpbEnergies}(c) and (d).

From these results, we observe that the energy dispersion becomes flatter for larger ratios of $E_J/E_C$, which means that the qubit becomes increasingly insensitive to charge noise. A completely flat dispersion would lead to no charge noise sensitivity at all. However, we also notice that the anharmonicity decreases for larger ratios. This is a result of the before-mentioned fact that the charge dispersion decreases exponentially in $E_J/E_C$ while the anharmonicity has a slower rate of change given by a power law.
Therefore we cannot just increase the shunting capacitance until all charge noise disappears as we still need a working qubit. We are thus left with some effective values for the transmon which are usually somewhere in the range $E_J/E_C\in [50,100]$.

Even though the transmon has a ratio $E_J/E_C$ close to that of the phase regime ($E_{C}\ll E_{J}$), it is still classified as a charge qubit due to the structural similarity to the single Cooper pair box qubit and the fact that the eigenstates still have reasonably well defined charge \cite{Koch2007}. Due to that and the fact that capacitors in parallel add, we often just put a Josephson junction and a parasitic capacitance in place of the transmon in larger circuits for simplicity. We further notice that if the ratio $E_J/E_C$ is very large, the bias voltage becomes irrelevant and can be omitted as well.

When implementing the transmon qubit on an actual chip various architectures are used, including the Xmon which is developed for nearest-neighbor capacitive coupling of qubits \cite{Barends2013,Barends2014,Kelly2015,Barends2015,Barends2016}, the three-dimensional (3D) transmon where the Josephson junction is coupled to a three-dimensional cavity \cite{Paik2011}, or the gatemon which is based on a semiconductor nanowire and controlled by an electrostatic gate \cite{Larsen2015,Casparis2016}. In general, there are many shapes of the transmon and these can often be tailormade to the specific experiment, see, e.g., Ref. \cite{Andersen2020}.
Common for these different architectures is that they can be treated theoretically equivalently to the basic transmon setup discussed above, and they are therefore often referred to as transmonlike qubits when the architecture is irrelevant from a theoretical point of view.

Recently a dual to the transmon qubit called a quasicharge qubit, or blochnium, has been proposed, where the shunting capacitance is replaced by a large shunting inductance. This large inductance makes the qubit very robust against flux noise, which could open up for exploring high-impedance circuits \cite{Pechenezhskiy2020}.

\subsection{Flux qubits}

In general, flux qubits are implemented in a looped superconducting circuit interrupted by one or more Josephson junctions. A current is induced in these circuits using the fact that fluxoid quantization means that only an integer number of magnetic flux quanta is allowed to penetrate the loop. As a response to the external flux, currents flow in superconducting materials to enhance or diminish the total flux such that an integer number of flux quanta is achieved in total.

A superposition of clockwise and counterclockwise currents is obtained by setting the external magnetic field at half a magnetic flux quantum. Changing to node flux space, this superposition of currents can be seen as a superposition of the ground states in a double-well potential. In the double-well potential, small tunneling occurs between the two sides of the well, which couples the two wave functions, making an avoided crossing, and thus a closely spaced two-level system, but with a very large gap to the remaining states. We now elaborate on some concrete realizations of these general ideas.

\subsubsection{$C$-shunted flux qubit}\label{sec:fluxqubit}
 
\begin{figure}
    \centering
    \includegraphics[width=.8\columnwidth]{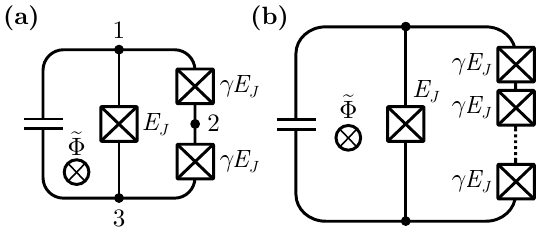}
    \caption{Circuit diagrams of different flux qubits. \textbf{(a)} $C$-shunted flux qubit. Two Josephson junctions in series are placed in parallel with a third Josephson junction. Both the parasitic and shunting capacitances are included in the capacitance. \textbf{(b)} Fluxonium qubit. An array of $N$ Josephson junctions are placed in parallel with another Josephson junction, effectively creating an inductor in parallel with the Josephson junction.}
    \label{fig:fluxQubits}
\end{figure}
 
The idea behind the C-shunted flux qubit (CSFQ) is the same as for the transmon. However, here the capacitive shunting is over a flux qubit, sometimes called a persistent-current qubit (PCQ) \cite{Orlando1999,Mooij1999}. As with the transmon qubit, the capacitive shunting improves the coherence of the qubit \cite{You2007,Yan2016}. We therefore consider the flux qubit without going into details of the shunting, see \cref{sec:transmon}. The coherence of the flux qubit can further be improved by placing it in a 3D \cite{Stern2014} or coplanar \cite{Orgiazzi2016} resonator.

The flux qubit consists of two Josephson junctions in series, with energy $\gamma E_J$, which are then placed in parallel with a third Josephson junction, with Josephson energy $E_J$. Here $\gamma$ is the ratio of the geometrical size of the Josephson junctions. 
To a good approximation, all capacitances (both parasitic and shunting) can be collected into one, as seen in \cref{fig:fluxQubits}(a), when assuming $\gamma > 1$.
When this is the case, the node in between the two Josephson junctions becomes a passive node.

Using the same trigonometric tricks as for the dc SQUID (see \cref{sec:dcSquid}), we can write the potential energy of the three Josephson junctions as
\begin{equation}
    U = -E_J \left[ 2\gamma\cos \left(\frac{\psi_+}{2} - \psi_2\right)\cos \frac{\psi_-}{2} +  \cos (\psi_- + \tilde\Phi)\right].
\end{equation}
Here we introduce the change of coordinates $\psi_\pm = \phi_1 \pm \phi_3$  and $\psi_2 = \phi_2$ where $n=2$ is the middle coordinate in between the two Josephson junctions. This coordinate transformation turns out to diagonalize the capacitance matrix as well as leaving only $\psi_-$ with a nonzero eigenvalue. Thus, the two remaining node fluxes are superfluous and from the constraints obtained from the Euler-Lagrange equations, we find that $\psi_2 = \psi_+/2$, which yields the potential energy
\begin{equation}
    U = -E_J \left[ 2\gamma\cos \frac{\psi_-}{2} +  \cos (\psi_- + \tilde\Phi)\right].
\end{equation}
This no longer has the usual sinusoidal form, and its final form depends on the external flux $\tilde \Phi$ and the junction ratio $\gamma$. The most common configuration for an external flux is $\tilde \Phi = (1+2l)\Phi_0/2$, where $l\in \mathbb{Z}$. These points are often called the flux degeneracy points and correspond to one half of the superconducting flux quantum threading the qubit loop. In this configuration the qubit frequency is most robust against flux noise, leaving the qubit with optimal coherence times.

As mentioned above we assume $\gamma > 1$, which led us to eliminate a degree of freedom. This can be seen as an approximation in which a particle that starts in two dimensions, but is rather forced to move along just one dimensions, and is sometimes called the quasi-one-dimensional (1D) approximation. This approximation fails if $\gamma < 1$. If $1 < \gamma < 2$, the potential takes the form of a double well, which has been investigated as the PCQ \cite{Orlando1999,Mooij1999}. If, on the other hand, $\gamma > 2$, the potential takes the form of a single well, very similar to the transmon qubit, which is why the CSFQ has been investigated in this regime \cite{You2007,Yan2016}. In both cases, if the anharmonicity is sufficiently large, the quantized potential can be truncated to the lower levels.

\subsubsection{Fluxonium}\label{sec:fluxonium}

\begin{figure}
    \centering
    \includegraphics[width=.9\columnwidth]{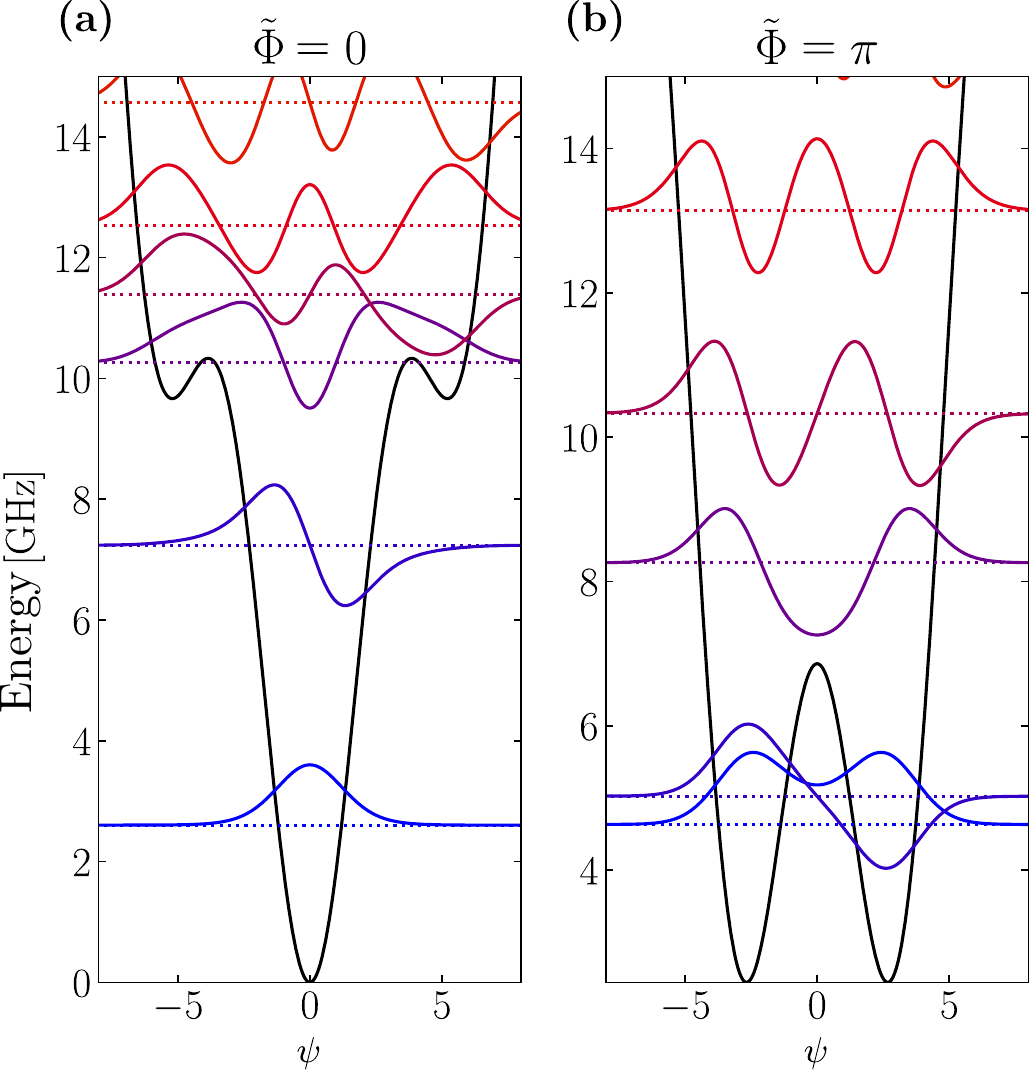}
    \caption{Spectrum of the fluxonium qubit at the two different flux-biasing points. For this plot the parameters are set to $E_C/h=\SI{1}{\giga\hertz}$, $E_J/h=\SI{3.43}{\giga\hertz}$ and $E_L/h=\SI{0.58}{\giga\hertz}$.}
    \label{fig:fluxonium_spectrum}
\end{figure}

The fluxonium qubit is the natural extension of the flux qubit. Instead of two Josephson junctions in parallel with another Josephson junction, the fluxonium features an array of up to $N=100$ Josephson junctions \cite{Manucharyan2009,Pop2014,Nguyen2019}, sometimes referred to as a \emph{superinductance} \cite{Bell2012,Masluk2012}. The circuit diagram can be seen in \cref{fig:fluxQubits}(b). Using the same quasi-1D approximation as in \cref{sec:fluxqubit} repeatedly, we arrive at a potential 
\begin{equation}
    U = -E_J \left[ N\gamma\cos \frac{\psi}{N} +  \cos (\psi + \tilde\Phi)\right],
\end{equation}
where $\psi$ is the sum of all node fluxes in between the array of Josephson junctions on the left side of \cref{fig:fluxQubits}(b). When the number of Josephson junctions $N$ becomes large the argument in the first cosine, $\psi/N$, becomes small such that the cosine can be approximated by a second-order approximation which yields
\begin{equation}
    U =   \frac{1}{2}E_L \psi^2 -  E_J\cos (\psi + \tilde\Phi),
\end{equation}
where $E_L = E_J\gamma/N$ is the resulting superinductance of the array of Josephson junctions. This has the same effective form as a rf SQUID \cite{Friedman2002}. However, the superinductance of the fluxonium qubit is much larger than the geometric inductance of the rf SQUID. This is because the superinductance is produced by the kinetic inductance of the array of Josephson junctions. It is therefore not limited, contrary to the geometrical inductance where the loop impedance cannot exceed $\alpha R_Q$. Here $\alpha$ is the fine-structure constant and $R_Q$ is the resistance quantum \cite{Bell2012}. Recent implementations of superinductors are based on nanowires of disordered granular aluminum or Nb alloys \cite{Niepce2019,Hazard2019,Wang2019Superinductor,Grunhaupt2019}.

When the external flux bias is $\tilde\Phi = 0$ the potential has minimum at $\psi=0$. For small fluctuations of $\psi$, the potential is approximately harmonic and the lowest-lying states are close to simple harmonic oscillator states. At higher energies, the anharmonic cosine term of the potential comes into play as seen in \cref{fig:fluxonium_spectrum}(a). This ensures the anharmonicity necessary for using the two lowest-lying states as the qubit subspace. However, the fluxonium qubit is most often operated at $\tilde\Phi = \pi$, similarly to the flux qubit. In this regime, the potential exhibits a double-well structure, and it is possible to achieve a much larger anharmonicity than in the $\tilde{\Phi}=0$ case, see \cref{fig:fluxonium_spectrum}(b).

In experiments, fluxonium qubits have reached impressive lifetimes of 100-$\SI{400}{\micro\second}$ \cite{Pop2014,Nguyen2019}, while recent experiments yields lifetimes in the 1-$\si{\milli\s}$ regime \cite{Somoroff2021}. This is done while maintaining a large anharmonicity suitable for fast gate operations. It puts fluxonium among the top qubit candidates for near future quantum-computing applications. In addition, the success of the fluxonium qubit proves that long coherence times can be achieved even in a more complicated system with a large number of spurious modes \cite{Giovanni2015}. This should encourage quantum engineers to further explore circuit design utilizing large superinductance.

A circuit element related to the fluxonium and the flux qubit is the superconducting nonlinear asymmetric inductive element (SNAIL), which has the same architecture as the fluxonium qubit in \cref{fig:fluxQubits}(b) but fewer Josephson junctions in the array than the fluxonium, i.e., $N\geq 2$ but less than for the fluxonium. For some particular choices of $\gamma$ and $\tilde\Phi$ it is possible to cancel any fourth-order term, $\phi^4$ while keeping a substantial cubic term, $\phi^3$ \cite{Frattini2017}. This can be used for amplifying three-wave-mixing \cite{Frattini2018,Sivak2019}.

\subsubsection{0-$\pi$ qubit}\label{sec:zeroPi}

\begin{figure}
    \centering
    \includegraphics[scale=.7]{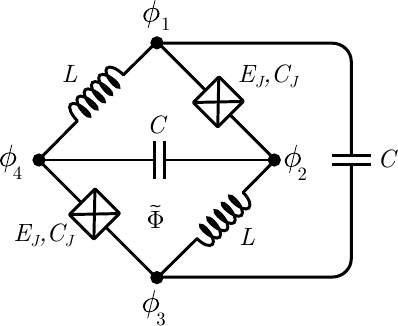}
    \caption{Circuit diagram of the 0-$\pi$ qubit. Four nodes are connected to each other by two large superinductors (drawn here as regular inductors) $L$, two Josephson junctions $E_J$ with parasitic capacitance $C_J$, and two shunting capacitors $C$.}
    \label{fig:zeropi}
\end{figure}

A new type of qubit is the 0-$\pi$ qubit. It has been proposed more recently than the above qubits, but it shows promising tendencies in topological protection from noise \cite{Doucot2002,Ioffe2002,Kitaev2006,Brooks2013,Groszkowski2018, gyenis2019experimental,Dempster2014}.

The $0\text{-}\pi$ qubit consists of four nodes that are all connected by two large superinductors, two Josephson junctions, and two large shunting capacitors, as shown in \cref{fig:zeropi}. We denote the shunting capacitors as $C$, the superinductors as $L$, and the Josephson junctions as $E_J$ and assume they have parasitic capacitances of $C_J$. The superinductors are usually made as an array of Josephson junctions (see \cref{sec:fluxonium}). However, here we draw them as regular inductors as this is their effective form. An external flux, $\tilde \Phi$, goes through the qubit. It is advantageous to choose the spanning tree such that only the Josephson junctions lie in the set of closure branches.

\begin{figure}
    \centering
    \includegraphics[width=\columnwidth]{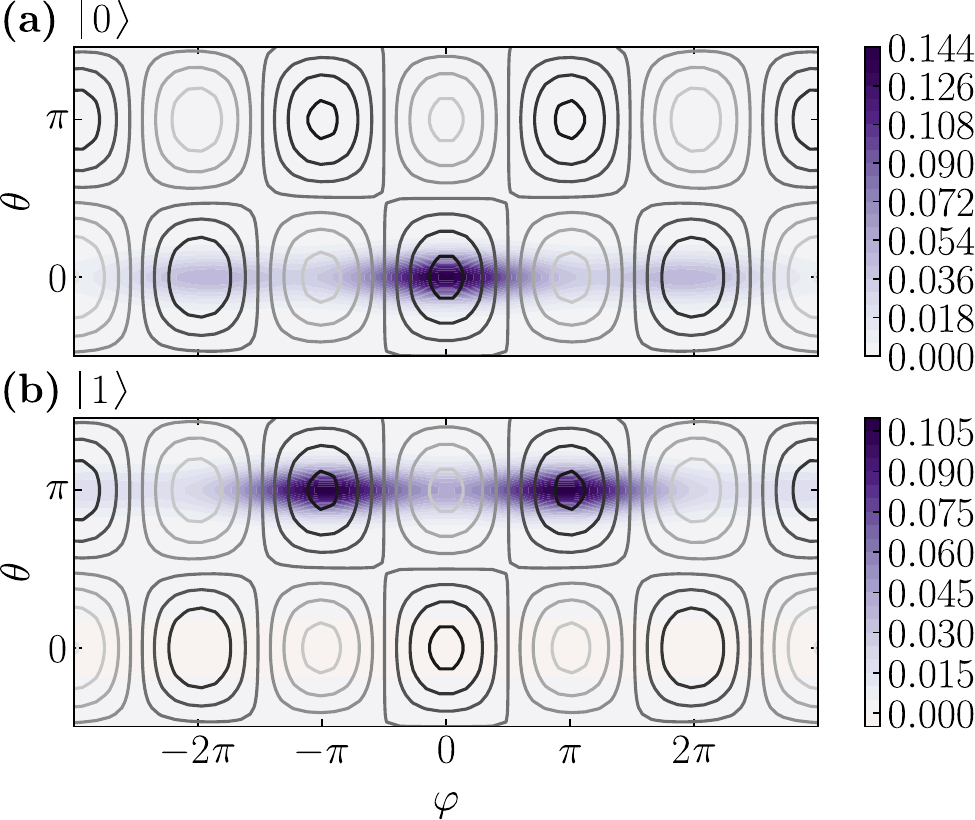}
    \caption{Wave functions of \textbf{(a)} the ground state and \textbf{(b)} first excited state for the $0$-$\pi$ qubit. The contour lines indicate the qubit potential. For this plot the parameters are set to $E_{C,\theta}=0.1$, $E_{C,\varphi}=10$, $E_J=10$ and $E_L=1$. In the figure $\tilde\Phi = 0$.  Changing the external flux would translate the potential along the $\theta$ axis.}
    \label{fig:zero_pi_states}
\end{figure}

The node fluxes of the circuit are denoted $(\phi_1,\phi_2, \phi_3, \phi_4)$, and the normal modes of the circuit can be written using the transformation
\begin{equation}
    \mat{\varphi \\ \theta \\ \zeta \\ \Sigma} = 
    \frac{1}{2}\mat{-1 & 1 & -1 & 1 \\ -1 & 1 & 1 & -1 \\ 1 & 1 & -1 & -1 \\ 1 & 1 & 1 & 1}\mat{\phi_1 \\ \phi_2 \\ \phi_3 \\ \phi_4}.
\end{equation}
Here $\Sigma$ is the CM coordinate, which has no influence on the dynamics of the system and can be discarded. This basis transformation diagonalizes the capacitance matrix $\vec C = 2 \, \diag{C_J, C_J+C, C}$. The Hamiltonian then takes the form
\begin{equation}
    \begin{split}
    \H &= 4E_{C\varphi}n_\varphi^2+4E_{C\theta}n_\theta^2+4E_{C\zeta}n_\zeta^2 +\frac{E_L}{2}(\varphi^2 + \zeta^2) \\
    &\phantom{=}-E_J\left[\cos (\theta+\varphi) + \cos (\theta-\varphi- \tilde\Phi)\right] ,
    \label{eq:zeropi_full}
\end{split}
\end{equation}
where $n_\varphi, n_\zeta, n_\theta$ are the canonical momenta, $E_{C\varphi}^{-1} = 16C_J$, $E_{C\theta}^{-1} = 16(C+C_J)$ and $E_{C\zeta}^{-1} = 16C$ are the charging energies of each mode, while $E_L = 2/L$ is the effective inductive energy. Note that the $\zeta$ mode completely decouples from the rest of the system and can thus be ignored. By transforming the $\theta$ variable $\theta \rightarrow \theta+\frac{\tilde\Phi}{2}$ we can rewrite the Hamiltonian into the simpler form
\begin{equation}
\begin{split}
    \H &= 4E_{C\varphi}n_\varphi^2+4E_{C\theta}n_\theta^2\\
    &\phantom{=}-2E_J\cos\theta\cos\left(\varphi+\frac{\tilde\Phi}{2}\right)+\frac{E_L}{2}\varphi^2.
    \label{eq:zeropi_simple}
\end{split}
\end{equation}
The circuit is engineered such that $C\gg C_J$, and we can thus think of the system as a heavy particle moving along the $\theta$ axis and a lighter particle moving along the $\varphi$ axis. In the basis of the computational states $\ket{0}$ and $\ket{1}$, which are chosen as the ground and first excited state, respectively, the $\theta$ variable is well localized around either $0$ or $\pi$, as shown in \cref{fig:zero_pi_states}. This is the reason for the naming of the qubit. Setting $\theta=0$ or $\pi$ in \cref{eq:zeropi_simple} we see that the potential along the $\varphi$ axis is similar to that of fluxonium biased by a flux of either $0$ or $\pi$. As a result the two states have vanishing matrix elements $\bra{0}\theta^n\ket{1},\bra{0}\varphi^n\ket{1}\simeq 0$. This makes the qubit highly resistant to noise-induced relaxation.

In recent experiments \cite{gyenis2019experimental} with the $0$-$\pi$ qubit, relaxation times above $\SI{1}{\milli\second}$ have been achieved, making it an exciting candidate for future research. As with fluxonium, the $0$-$\pi$ qubit proves that it is not only the most simple qubits, such as the charge and flux qubit families, that can achieve long coherence times. Researchers should make note of this when developing new circuit designs to tap into the potential that more complicated components, such as the superinductances used in fluxonium and the $0$-$\pi$ qubit, bring to the table. 

\subsection{Tunable couplers}\label{sec:dynamicalCouplers}

In \cref{sec:coupling} we have presented some simple static couplings of qubits. Here we present some tunable couplers from the literature. By tunable, we mean couplers where the interaction strength can be changed \emph{in situ}, without changing the circuit layout. We consider both capacitive and inductive coupling and finally $XXZ$ coupling. The list of couplers presented here is of course not exhaustive as there are other types of couplers in the literature, see, e.g., Refs. \cite{McKay2016,Rasmussen2020b,Chen2021}.

\subsubsection{Tunable capacitive coupler}

\begin{figure}
    \centering
    \includegraphics[scale=.8]{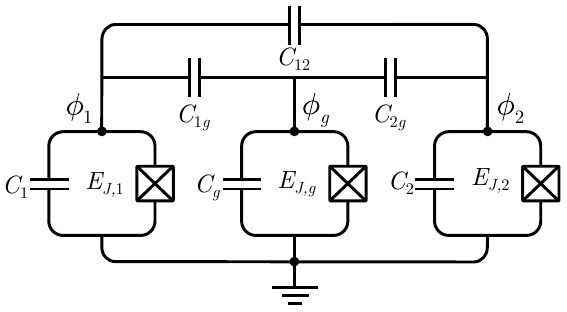}
    \caption{Circuit diagram implementing a tunable capacitive coupler. Two transmonlike qubits are connected via another tunable transmonlike qubit, and directly to each other.}
    \label{fig:tunableCoupler}
\end{figure}

Here we present a tunable capacitive coupling between two modes \cite{Neill2017,Yan2018,Sung2020,Li2020}.
Consider the circuit in \cref{fig:tunableCoupler} where two transmonlike qubits, subscript 1 and 2, are connected capacitively to each other and a mediating transmonlike qubit, subscript $g$. If we require the Josephson junctions of the qubits to be dc SQUIDs we can tune the frequency of the qubits. Writing down the Hamiltonian of this circuit following the approach in \cref{sec:MethodofNodes} is straightforward
\begin{equation}
    \H = \frac{1}{2}\vec q^T C^{-1} \vec q - \sum_j E_{J,j}\cos\phi_j,
\end{equation}
where the sum is over all three modes, i.e., $1,2,g$. The capacitance matrix is
\begin{equation}
    \vec C = \mat{C_1 + C_{1g} + C_{12} & -C_{1g} & -C_{12} \\
                -C_{1g} & C_g + C_{1g} + C_{2g} & -C_{2g} \\
                -C_{12} & -C_{2g} & C_2 + C_{2g} + C_{12}},
\end{equation}
which is invertible. We leave this inversion to the reader and note that assuming the qubit-coupler capacitances are smaller than the mode capacitances but larger than the qubit-qubit capacitance, i.e., $C_n \gg C_{ng} \gg C_{12}$, it can be simplified significantly, see e.g. Ref. \cite{Yan2018}. The diagonal terms of $\vec C^{-1}$ contribute to the frequencies of the modes, while the three off-diagonal terms contribute to the coupling. Quantizing the Hamiltonian, the interacting part takes the form
\begin{equation}
    \hat \H_{\text{int}} = \sum_{i>j} (\vec C^{-1})_{(i,j)} \hat n_i \hat n_j,
\end{equation}
where $\hat n_i$ is the Cooper pair number operator of the $i$th mode, and $i,j\in \{1,2,g\}$. Mapping to annihilation and creation operators connected to the harmonic degrees of freedom yields
\begin{equation}\label{eq:HintTunableCoupler}
    \hat \H_{\text{int}} = \sum_{i>j} g_{ij} \left( \hat b_i^\dagger \hat b_j + \hat b_i \hat b_j^\dagger - \hat b_i^\dagger \hat b_j^\dagger  - \hat b_i \hat b_j \right),
\end{equation}
where the coupling strength is given as
\begin{equation}
    g_{ij} = (\vec C^{-1})_{(i,j)}\frac{1}{2\sqrt{\zeta_i\zeta_j}},
\end{equation}
and the impedances are given in \cref{eq:impedance}. Note that we have to keep the nonconserving terms in \cref{eq:HintTunableCoupler} as these can be significant in the dispersive regime, i.e., when the coupler frequency is larger than the difference in qubit frequencies; $|\Delta_j| = |\omega_j - \omega_g| \gg g$.

To see this we perform a Schrieffer-Wolff transformation similar to the one performed in \cref{sec:resonators}. However, this time as we have three modes and include the nonconserving terms. We thus take
\begin{equation}
    \hat S = \sum_{j=1,2} \left[ \frac{g_{jg}}{\Delta_j}(\hat b_g^\dagger \hat b_j - \hat b_g \hat b_j^\dagger) - \frac{g_{jg}}{\Sigma_j}(\hat b_g^\dagger \hat b_j^\dagger  - \hat b_g \hat b_j) \right],
\end{equation}
where $\Sigma_j = \omega_j + \omega_g$. Assuming a small anharmonicity $\alpha_j \ll \Delta_j$, we can expand the transformation to second order (note that $g_{12}$ is considered a second-order small quantity on its own). We find the full Hamiltonian to be 
\begin{align}
    \hat \H_\text{disp} &= e^{\hat S} \hat\H e^{-\hat S} \\
    &= \sum_{j=1,2} \left[\tilde \omega_j \hat b^\dagger _j \hat b_j + \frac{\alpha_j}{2} \hat b_j^\dagger \hat b_j^\dagger \hat b_j \hat b_j\right] + \tilde g_{12} (\hat b_1^\dagger \hat b_2 + \hat b_1\hat b_2^\dagger), \nonumber
\end{align}
where
\begin{align}
    \tilde \omega_j &= \omega_j + g^2_{jg} \left( \frac{1}{\Delta_j} - \frac{1}{\Sigma_j}\right), \\
    \tilde g_{12} &= g_{12} + \frac{g_{1g}g_{2g}}{2} \left( \frac{1}{\Delta_1} + \frac{1}{\Delta_2} - \frac{1}{\Sigma_1} - \frac{1}{\Sigma_2}\right).
\end{align}
In the dispersive regime $|\Delta_j| \simeq |\Sigma_j|$ the nonconserving terms contribute to the coupling. The total effective coupling $\tilde g_{12}$ depends on $g_{jg}$ as well as $\Delta_j$ and $\Sigma_j$, all of which depend on the coupler frequency $\omega_g$, which can be tuned. Thus $\tilde g_{12}$ is tunable as it is implicitly a function of $\omega_g$.

Note that instead of the tunable transmon coupler, one could also have employed a tunable harmonic oscillator or cavity for coupling the two qubits as used in \cref{sec:resonators}. The analysis is largely the same.

\subsubsection{Delft coupler}\label{sec:delftCoupler}

The Delft coupler \cite{Kounalakis2018} introduces tunable nonlinear couplings between two qubits in a center-of-mass basis. As with the above coupler, it is based on capacitors. The following example is a simplification of the Supplementary Material of Ref. \cite{Kounalakis2018}.

\begin{figure}
    \centering
    \includegraphics[scale=0.75]{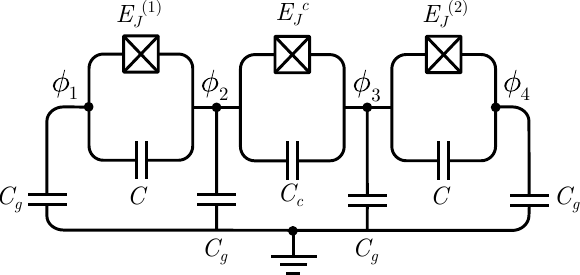}
    \caption{Circuit diagram of the Delft coupler. Two transmonlike qubits (1 and 2) are coupled via another transmonlike coupler (c).}
    \label{fig:DelftCoupler}
\end{figure}

Consider the circuit diagram in \cref{fig:DelftCoupler}. Following the approach in \cref{sec:MethodofNodes} we find the following capacitance matrix
\begin{equation}
    \vec C = \mat{
        C + C_{g} & -C & 0 & 0 \\ 
        -C & C + C_{g} + C_c & -C_c & 0 \\
        0 & -C_c & C + C_{g} + C_c & -C \\
        0 & 0 & -C & C + C_{g}
    },
\end{equation}
where we define the flux vector as $\vec\phi = (\phi_1, \phi_2, \phi_3, \phi_4)^T$. This yields the following circuit Lagrangian 
\begin{equation}
\begin{aligned}
    \L &= \frac{1}{2} \vec{\dot\phi}^T \vec C \vec{\dot\phi} + E_J^{(1)} \cos(\phi_1 - \phi_2) \\ 
    &\phantom{=}+ E_J^{c} \cos(\phi_2 - \phi_3) + E_J^{(2)} \cos(\phi_3 - \phi_4).
\end{aligned}
\end{equation}
We now change into a CM basis (see \cref{sec:normalModes}) of the capacitive subsystem using the following transformations
\begin{subequations}
\begin{align}
    \psi_\text{CM} &= \frac{1}{2}(\phi_1 + \phi_2 + \phi_3 + \phi_4), \\
    \psi_1 &= \frac{1}{\sqrt{2}}(\phi_1 - \phi_2), \\
    \psi_2 &= \frac{1}{\sqrt{2}}(\phi_4 - \phi_3), \\
    \psi_S &= \frac{1}{2}(\phi_1 + \phi_2 - \phi_3 - \phi_4). 
\end{align}
\end{subequations}
This decouples the center-of-mass coordinate, $\psi_\text{CM}$, from the remaining coordinates (note that this is due to the identical grounding capacitances $C_g$) as the transformed capacitance matrix takes the form
\begin{equation}\label{eq:KDelft}
    \vec K = \frac{1}{2}\mat{
        2C_g & 0 & 0 & 0 \\
        0 & 4C + 2C_g + C_c & -C_c & -\sqrt{2}C_c \\
        0 & -C_c & 4C + 2C_g + C_c & \sqrt{2}C_c \\
        0 & -\sqrt{2} C_c & \sqrt{2} C_c & 2C_g + 2C_c 
    },
\end{equation}
where we choose the basis such that $\vec \psi= ( \psi_\text{CM}, \psi_1, \psi_2, \psi_S)^T$. Doing a Legendre transformation and quantizing, we find the Hamiltonian
\begin{equation}
\begin{aligned}
    \hat\H &= \frac{1}{2} \vec{\hat p}^T \vec K^{-1} \vec{\hat p} - E_J^{(1)} \cos(\sqrt{2}\hat\psi_1) - E_J^{(2)} \cos(\sqrt{2}\hat\psi_2) \\
    &\phantom{=}- E_J^{c} \cos\left( \frac{\hat\psi_1 - \hat\psi_2}{\sqrt{2}} - \hat\psi_S \right),
\end{aligned}
\end{equation}
where $\vec{\hat p}$ is the vector of conjugate momentum of the $\vec{\hat \psi}$ vector. Expanding the cosines and changing into annihilation and creation operators [\cref{eq:stepOperators}], the noninteracting part of the Hamiltonian takes the form
\begin{equation}
    \hat \H_0 = \sum_{i = \{S,1,2\}} \left[\omega \hat b_i^\dagger \hat b_i + \frac{\alpha}{2}\hat b_i^\dagger \hat b_i^\dagger \hat b_i \hat b_i\right],
\end{equation}
where we define $\omega = 4\sqrt{E_C E_J} + \alpha $ and $\alpha = -\zeta^2(E_J + 3E_J^{(1)})/8$, with the effective capacitive energies being the usual $E_C^{(i)} = (\vec K ^{-1})_{(i,i)}/8$, which turn out to be the same for the 1 and 2 mode. Thus, we denote it $E_C = E_C^{(1)} = E_C^{(2)}$. We also define the effective Josephson energy $E_J = E_J^{(1)} + E^c_J/4$ and assume that the 1 and 2 modes are resonant, i.e., $E_J^{(1)} = E_J^{(2)}$. Lastly, we define the impedance as given in \cref{eq:impedance}.
We do not include the center-of-mass coordinate as it does not influence the dynamics of the system. Note how the 1 and 2 modes are affected by both their 'own' Josephson junction and the coupling Josephson junction.

Assuming that the so-called sloshing mode, $\psi_S$, is detuned from the remaining two modes, we can remove couplings to it, using the rotating-wave approximation from \cref{sec:rwa}. After this approximation, the interaction part of the Hamiltonian takes the form
\begin{equation}\label{eq:delftCoupling}
\begin{aligned}
    \hat \H_I &= J\hat b_1 \hat b_2^\dagger + \frac{V}{2} \hat b_1^\dagger  \hat b_1 \hat b_2^\dagger \hat b_2 + \frac{V}{4} \hat b_1^\dagger  \hat b_1^\dagger  \hat b_2\hat b_2 \\
    & \phantom{=} + \frac{V}{2} ( \hat b_1 \hat n_1 \hat b_2^\dagger + \hat b_2 \hat n_2 \hat b_1^\dagger) + \hc,
\end{aligned}
\end{equation}
where we use the assumption that the 1 and 2 modes are resonant. The swapping coupling strength is given by 
\begin{equation}
J = \frac{1}{2\zeta}(\vec K^{-1})_{(1,2)} - \frac{\zeta E_J^c}{4} - \frac{V}{2},
\end{equation}
where the nonlinear coupling factor is given as
\begin{equation}
    V = - \frac{E_J^c \zeta^2}{16}.
\end{equation}
The first nonlinear term in \cref{eq:delftCoupling} is sometimes called the cross-Kerr coupling term with coupling strength $V/2$, while the second nonlinear term tunnels a pair of excitations from one mode to the other with coupling strength $V/4$. Therefore this term does not contribute to the Hamiltonian if truncated to a two-level model, but it may result in corrections to the model. Thus truncating to a two-level model the Hamiltonian becomes
\begin{equation}
    \hat \H = \hat \H_0 + J(\sigma^+_1 \sigma_2^- + \sigma^-_1 \sigma_2^+) + \frac{V}{4} \sigma_1^z \sigma_2^z,
\end{equation}
where we have both transverse ($\sigma^+ \sigma^- + \sigma^- \sigma^+$) and longitudinal coupling ($\sigma_z\sigma_z$) between the 1 and 2 modes. Both $J$ and $V$ depend on the Josephson energy of the coupler, which can be tuned using the external flux, thus making the coupling tunable.

\subsubsection{Gmon coupler}\label{eq:Gmon}

\begin{figure}
    \centering
    \includegraphics[width=.65\columnwidth]{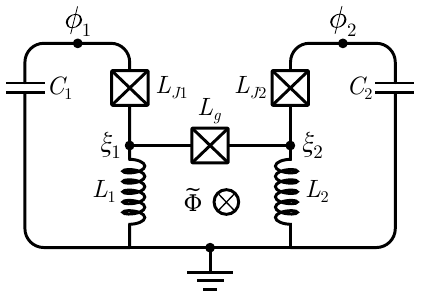}
    \caption{Circuit diagram of the gmon coupler. Two Josephson junctions, $L_{Ji}$, in parallel with a capacitor, $C_i$, and in series with a linear inductor, $L_{i}$, are coupled via a Josephson junction, $L_g$. The inductors lead to mutual inductance between the two loops. An external flux through the middle loop allows for tuning of the coupling.}
    \label{fig:gmon}
\end{figure}

The gmon coupler introduces tunable swapping couplings between two transmonlike qubits by exploiting mutual inductance \cite{Chen2014,Geller2015,Roushan2016,Neill2017}.

Consider the circuit diagram in \cref{fig:gmon} of the gmon. A Josephson junction in parallel with a capacitor and in series with a linear inductor is coupled to a similar setup via another Josephson junction. An external flux through the coupling loop makes it possible to tune the inductance of the coupling Josephson junction, such that $L_\text{eff} = L_g/\cos\delta$ [see \cref{eq:JJ_energy}]. Here we define the dc phase difference across the Josephson junction, $\delta = \tilde\Phi + \bar\xi_2 - \bar\xi_1$, where the bar indicates the equilibrium position of the coordinates.

The $\xi$ coordinates are passive nodes as they are only coupled to inductors and not any capacitors. We can therefore remove the $\xi$ coordinates from the Hamiltonian. To do this we must first determine the voltage of the $\phi$ coordinates. This can be done using Kirchhoff's voltage law, \cref{eq:KVL}, which yields
\begin{equation}
    V_i = (L_{Ji} + L_i) \dot I_i \pm M(\dot I_2 - \dot I_1),
\end{equation}
where $M$ is the mutual inductance between the right and left loop and we have plus for $i=1$ and minus for $i=2$. In order to simplify this expression we define $L_{qi} = L_{Ji} + L_i - M$, which is the inductance in the $i$th loop.

To determine the mutual inductance $M$ we consider a current $I_1$ in the left qubit. A small fraction of this current flows through the coupler Josephson junction. This fraction is
\begin{equation}
    I_{g} = \frac{L_1}{L_1+L_2+L_\text{eff}}I_1,
\end{equation}
where we use the effective inductance in place of $L_g$.
This current generates a flux in the right qubit $\phi_2 = L_2I_g$. This means that we can express the mutual inductance as
\begin{equation}
    M = \frac{\phi_2}{I_1} = \frac{L_1L_2}{L_1+L_2+L_\text{eff}}.
\end{equation}
With the mutual inductance determined we are ready to find the Hamiltonian of the circuit in \cref{fig:gmon}. It is as follows:
\begin{equation}
\begin{aligned}
    \H = \sum_{i=1,2}&\left[\frac{q_i^2}{2C_i} + \frac{\xi^2_i}{2L_i} - \frac{1}{L_{Ji}}\cos(\phi_i - \xi_i) \right. \\
    &- \left. \frac{1}{L_g}\cos(\xi_1 - \xi_2 +\tilde\Phi)\right],
\end{aligned}
\end{equation}
where $q_i$ is the conjugate momentum of the $i$th flux.
Since the $\xi$ coordinates are passive nodes we want to remove them from the Hamiltonian. We do this by minimizing $\xi_i$ for a fixed $\phi_i$. This is equivalent to solving Lagrange's equations, \cref{eq:Lagrang}, for $\xi_i$. This is straightforward but cumbersome work as we end up with transcendental equations for $\xi_i$. We, therefore, skip straight to the resulting Hamiltonian, details can be found in Ref. \cite{Geller2015}. The Hamiltonian in the harmonic and weak coupling limit, $L_q \gg M$, becomes
\begin{equation}
\begin{aligned}
    \H = \sum_{i=1,2}&\left[\frac{q_i^2}{2C_i} + \frac{\phi^2_i}{2L_{qi}}\right] + \Gamma \phi_1\phi_2,
\end{aligned}
\end{equation}
where we do not include the anharmonic corrections, see Ref. \cite{Geller2015}. The coupling is given as
\begin{equation}
\begin{aligned}
    \Gamma &= -\frac{M}{L_{q1}L_{q2}} \\
    &= - \frac{L_1L_2}{(L_1+L_{j1})(L_2+L_{J2})(L_\text{eff} + L_1+L_2)},
\end{aligned}
\end{equation}
and changing into annihilation and creation operators yields a coupling strength of 
\begin{equation}
    g = \frac{1}{2}\Gamma \sqrt{\zeta_1\zeta_2},
\end{equation}
where the impedances are found in \cref{eq:impedance}. This coupling strength is tunable via the parameter $L_\text{eff}$.

\subsection{$XXZ$ coupling and qutrits}\label{sec:coupledQutrits}

\begin{figure}
  \centering
    \includegraphics[width=.8\columnwidth]{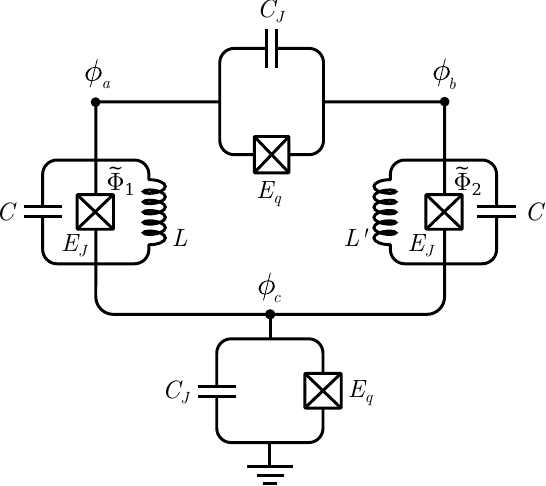}
  \caption{Circuit diagram of two coupled modes. The three circuit nodes, $\phi_a, \phi_b, \phi_c$, are indicated by dots.}
\label{fig:4qb_circuit}
\end{figure}

In this example, we present a system with two modes coupled via an effective Heisenberg $XXZ$ coupling. At the end of the example, we truncate the modes to the three lowest levels also known as qutrits. The circuit diagram is shown in \cref{fig:4qb_circuit}. The idea is to mix the nodes $ \phi_a$, $\phi_b$, and $\phi_c$ such that we obtain two low-dimensional degrees of freedom (after truncation) with the desired coupling and a decoupled third degree of freedom which can be seen as a center-of-mass coordinate.

We include driving lines to each of the three nodes, which enables us to control the mode energies dynamically by the ac Stark shift arising from detuned driving as explained in \cref{sec:driving}. We note that the inductances in \cref{fig:4qb_circuit} may be physically arranged in a manner that may allow for a mutual inductance as an additional manner of coupling. This can be analyzed as above in the gmon case, however, here we ignore mutual inductance for simplicity.

We start from the circuit in \cref{fig:4qb_circuit}, which yields the following Hamiltonian using the method of nodes as presented in \cref{sec:MethodofNodes}
\begin{equation}\label{eq:HqutritStart}
\begin{aligned}
        \H = &\frac{1}{2}\vec q^T \vec C \vec q +\frac{1}{2L}\left( \phi_a - \phi_c \right)^2 + \frac{1}{2L'}\left( \phi_b - \phi_c \right)^2 \\
    & - E_{q}\left[ \cos(\phi_a - \phi_b) + \cos\phi_c \right] \\
    &- E_J\left[ \cos(\phi_a - \phi_c + \tilde\Phi_1) + \cos(\phi_b - \phi_c + \tilde\Phi_2)  \right],
\end{aligned}
\end{equation}
where the capacitance matrix is 
\begin{equation}
    \vec C = \mat {C + C_J & -C_J & -C \\
                    -C_J & C+ C_J & -C \\
                    -C & -C & 2C + C_J},
\end{equation}
and we define the vector of conjugate momenta $\vec q^T = (q_a, q_b, q_c)$. We now follow \cref{sec:normalModes} and transform into a CM system of the capacitive subsystem using the following transformation
\begin{equation}\label{eq:CMbasis}
	\begin{aligned}
		\psi_1&= \frac{1}{\sqrt{2}}(\phi_a - \phi_b), \\
		\psi_2 &= \frac{1}{\sqrt{6}}(\phi_a + \phi_b - 2\phi_c), \\
		\psi_\text{CM} &= \frac{1}{\sqrt{3}}(\phi_a + \phi_b + \phi_c).
	\end{aligned}
\end{equation}
From this the transformation matrix $\vec{\mathcal{V}}$ can be constructed such that \cref{eq:basischange} is satisfied. With this transformation, the capacitance matrix takes the form
\begin{equation}
    \vec{K} = \mat{
    	2C_J + C & 0  & 0 \\
    	0 & \frac{2}{3}C_J + 3C  & -\frac{\sqrt{2}}{3}C_J \\
    	0 & -\frac{\sqrt{2}}{3}C_J  & \frac{1}{3}C_J
    },
\end{equation}
when the basis is chosen such that $\vec p^T = (p_1, p_2, p_\text{CM})$. Note that the CM mode is not decoupled from the second mode, as we do not transform into the normal modes of the system.
Assuming that $\vec K$ is invertible its inverse becomes
\begin{align}
    \vec{K}^{-1} = \mat{
    	\frac{1}{2C_J + C} & 0  & 0 \\
    	0 & \frac{1}{3C}  & \frac{\sqrt{2}}{3C} \\
    	0 & \frac{\sqrt{2}}{3C}  & \frac{3}{C_J} + \frac{2}{3C}
    }.
\end{align}
We notice that the diagonal terms for $\psi_1$ and $\psi_2$ are unequal, which becomes important when we later introduce the annihilation and creation operators related to the harmonic part of the full Hamiltonian. 

Returning to the potential part of the Hamiltonian in \cref{eq:HqutritStart} we rewrite it in the CM basis in \cref{eq:CMbasis} and apply the standard procedure of rewriting using trigonometric identities as in \cref{sec:dcSquid}, and requiring  $\tilde\Phi_1 = - \tilde\Phi_2 = \tilde \Phi$. Finally, we expand the cosine terms to fourth order, assuming that we are in the transmon regime. The potential part of the Lagrangian takes the form
\begin{equation} 
\begin{aligned}
    U(\vec \psi) \simeq & E_1\psi_1^2 - \frac{1}{24}\left( 4E_{q} + \frac{E_J}{2}\cos\tilde\Phi \right)\psi_1^4 \\
    &+E_2\psi_2^2 - \frac{1}{24}\left( \frac{4}{9}E_{q} + \frac{9}{2}E_J\cos\tilde\Phi \right)\psi_2^4 \\
    &- g\psi_1\psi_2 - \frac{3E_J}{8}\cos\tilde\Phi\psi_1^2\psi_2^2 ,
\end{aligned}
\end{equation}
where we neglect all energy and coupling terms involving $\psi_\text{CM}$ as the $\psi_\text{CM}$-degree of freedom will typically have an energy spectrum far from the rest. We also remove all nonenergy-conserving terms, see \cref{sec:rwa}. The effective energies and coupling strength are defined by
\begin{subequations}\label{eq:enegiesQutrit}
\begin{align}
    E_1 &= E_L + E_{L'} + \frac{E_J}{2}\cos\tilde\Phi + E_q, \\
    E_2 &= 3\left(E_L + E_{L'} + \frac{E_J}{2}\cos\tilde\Phi + \frac{1}{9}E_{q}\right), \\
    g &=  2\sqrt{3}(E_L - E_{L'}),
\end{align}
\end{subequations}
where $E_L = 1/4L$ and $E_{L'}= 1/4L'$ are the effective inductive energies. Note the asymmetry between the 1 and 2 modes. Ignoring the CM mode we quantize the Hamiltonian and change into annihilation and creation operators using \cref{eq:stepOperators}, with the impedances $\zeta_i = \sqrt{4E_{C,i}/E_i}$, where $E_{C,i}$ is the usual effective capacitive energy. The Hamiltonian takes the form
\begin{equation} \label{eq:H_b}
    \begin{aligned}
    	\hat\H &= \sum_{i=1}^2 \left[ 4\sqrt{E_{C,i}E_i}\hat b_i^\dagger \hat b_i + \frac{\alpha_i}{12}( \hat b_i^\dagger + \hat b_i )^4 \right] \\
    	&\phantom{=}+\frac{1}{2}g\sqrt{\zeta_1\zeta_2}( \hat b_1^\dagger + \hat b_1 )( \hat b_2^\dagger + \hat b_2 ) \\
        &\phantom{=}- \frac{3E_J}{32}\cos\tilde\Phi\zeta_1\zeta_2( \hat b_1^\dagger + \hat b_1 )^2( \hat b_2^\dagger + \hat b_2 )^2,
    \end{aligned}
\end{equation}
where we define the anharmonicities
\begin{subequations}\label{eq:anharmonicitiesQutrit}
\begin{align}
    \alpha_1 &= -\frac{\zeta_1^2}{8}\left( 4E_{q} + \frac{E_J}{2}\cos\tilde\Phi \right), \\
    \alpha_2 &= -\frac{\zeta_2^2}{8}\left( \frac{4}{9}E_{q} + \frac{9}{2}E_J\cos\tilde\Phi \right).
\end{align}
\end{subequations}

\begin{figure}
  \centering
  \includegraphics[width=0.4\columnwidth]{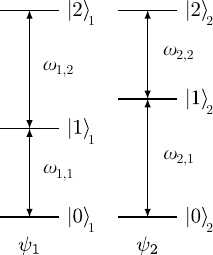}
  \caption{Energy diagram of the system of two qutrits described by the Hamiltonian in \cref{eq:H0qutrit} resulting from the circuit in \cref{fig:4qb_circuit}. The qutrits are connected with a Heisenberg $XXZ$ coupling seen in \cref{eq:Hxqutrit,eq:Hzqutrit}.}
\label{fig:4qb_diagram}
\end{figure}

We now wish to truncate the two modes to qutrits, following the procedure presented in \cref{sec:qutrit}. Note that we can choose any other amount of levels to truncate to as well. We choose the zero-point energy to be at the $\ket{0}_i$ state for both qutrits. This is contrary to the qubit where we usually choose the zero-point energy to lie in between the two states. The diagonal part of the Hamiltonian becomes
\begin{equation}\label{eq:H0qutrit}
    \hat \H_0 = \sum_{i=1}^2 \left( \omega_{i,1}\ketbra{1}_i + (\omega_{i,1}+\omega_{i,2})\ketbra{2}_i \right),
\end{equation}
where $\ketbra{1}_i$ and $\ketbra{2}_i$ are the projection operators of the $i$th qutrit on to the first and second excited state, respectively. The energies are given as
\begin{subequations}
\begin{align}
    \omega_{i,1} &= 4\sqrt{E_{C,i}E_i} +\alpha_i,\\
    \omega_{i,2} &= \omega_{i,1} +\alpha_i, 
\end{align}
\end{subequations}
from which we see the effect of the anharmonicity. An energy diagram of the two qutrits is shown in \cref{fig:4qb_diagram}.
The transverse interaction part, which swaps excitation between the two qutrits, is
\begin{equation}\label{eq:Hxqutrit}
\begin{aligned}
    \H_X &= J_{X} (\ketbra{01}{10} + 2\ketbra{12}{21})\\
    &\phantom{=}+ 2J_{Z}\ketbra{02}{20} + \hc,
\end{aligned}
\end{equation}
where we define the shorthand notation $\ket{nl} = \ket{n}_1\ket{l}_2$ used in the projection operators. The coupling constants are given as
\begin{subequations}
\begin{align}
    J_X &= \frac{1}{2}g\sqrt{\zeta_1\zeta_2}, \\
    J_Z &= -\frac{3E_J}{32}\cos\tilde\Phi\zeta_1\zeta_2.
\end{align}
\end{subequations}
Note that there is also a $\ketbra{11}{20}$ term, however, this is suppressed due to the anharmonicity and thus we can remove it using the RWA from \cref{sec:rwa}.
Finally, the longitudinal interaction part of the Hamiltonian is
\begin{equation}\label{eq:Hzqutrit}
    \hat\H_Z = J_Z S_1^zS_2^z,
\end{equation}
where $S^z = \ketbra{0} + 3\ketbra{1} + 5\ketbra{2}$, which is a generalization of the qubit $\sigma^z_1\sigma^z_2$ longitudinal coupling.

\subsubsection{External driving of the modes}

We wish to control the two qutrit modes using external microwave driving following the procedure presented in \cref{sec:quditDriving}. We drive the three original modes $\phi_a$, $\phi_b$, and $\phi_c$ of the system. This gives the following three additional terms for the Lagrangian
\begin{equation}\label{eq:qutritDriving}
\begin{aligned}
    \L_\text{ext} &= \frac{C_a}{2}\left( \dot{\phi}_a - V_a(t) \right)^2 + \frac{C_a}{2}\left( \dot{\phi}_b - V_b(t) \right)^2 \\
    &\phantom{=}+ \frac{C_c}{2}\left( \dot{\phi}_c - V_c(t) \right)^2,
\end{aligned}
\end{equation}
where $V_i(t)$ is the external microwave driving.
 
We want to effectively couple the $\psi_1$ and $\psi_2$ modes to (independent) external fields. Because these are specific linear combinations of $\phi_a$, $\phi_b$, and $\phi_c$ via \cref{eq:CMbasis}, the coupling fields must also be linear combinations. We, therefore, change the basis and expand the parenthesis in \cref{eq:qutritDriving}. Considering only the terms that contribute to the external driving of the modes, we find
\begin{equation}
\begin{aligned}
    \L_\text{ext} &= -\frac{C_a}{\sqrt{2}}[V_a(t) - V_b(t)]\dot{\psi}_1\\ 
    &\phantom{=}- \frac{C_a}{\sqrt{6}}\left[V_a(t) + V_b(t) - \frac{2C_c}{C_a}V_c(t)\right]\dot{\psi}_2 \\
    & \phantom{=}- \frac{C_a}{\sqrt{3}}\left[V_a(t) + V_b(t) + \frac{C_c}{C_a}V_c(t)\right]\dot{\psi}_\text{CM}.
\end{aligned}
\end{equation}
The remaining terms are either irrelevant offset terms or simple corrections to the energies of the modes.
We want the external drivings in the first and second term to be equal to a simple sinusoidal driving, while we want the driving in the last term, regarding the CM mode, to be zero. This yields the following equations
\begin{subequations}
\begin{align}
		V_a(t) - V_b(t) &= \sqrt{2}A_1\cos(\omega^\text{ext}_1t), \\
		V_a(t) + V_b(t) - \frac{2C_c}{C_a}V_c(t) &= \sqrt{6}A_2\cos(\omega^\text{ext}_2t), \\
		V_a(t) + V_b(t) + \frac{C_c}{C_a}V_c(t) &= 0,
\end{align}
\end{subequations}
where $\omega_i^\text{ext}$ and $A_i$ is the external driving frequency and amplitude of the $i$th qubit. Note that we do not include a phase in the driving for simplicity. The $\sqrt{2}$ and $\sqrt{6}$ factors are chosen to simplify the result.
If we choose $C_a = 2C_c$, we are left with three equations with three unknowns. These equations can be solved by
\begin{subequations}
	\begin{align}
		V_a(t) &= \frac{1}{\sqrt{2}}A_1\cos(\omega^\text{ext}_1t) + \frac{1}{\sqrt{6}}A_2\cos(\omega^\text{ext}_2t), \\
		V_b(t) &= -\frac{1}{\sqrt{2}}A_1\cos(\omega^\text{ext}_1t) + \frac{1}{\sqrt{6}}A_2\cos(\omega^\text{ext}_2t), \\
		V_c(t) &= -\frac{4}{\sqrt{6}} A_2\cos(\omega^\text{ext}_2t).
	\end{align}
\end{subequations}
Expanding and collecting the terms leads to a total kinetic energy
\begin{equation}
	T = \frac{1}{2}\dot{\vec{\psi}}^T\vec{\tilde K}\dot{\vec{\psi}} - \frac{C_a}{2}\sum_{i=1}^2 A_i\cos(\omega^\text{ext}_it)\dot{\psi_i},
\end{equation}
where $\vec{\tilde K} = \vec{K} + \vec{K}_\text{ext}$ is the adjusted capacitance matrix in the CM frame and $\vec K_\text{ext} = \vec{\mathcal{V}}^T \vec C_\text{ext} \vec{\mathcal{V}}$ is the contribution from the coupling to the external nodes with $\vec C_\text{ext} = \diag{C_a,C_a,2C_a}$.
Performing a Legendre transformation and changing into annihilation and creation operators, the driving term takes the form
\begin{equation}
    \hat\H_\text{ext} = i\sum_{i=1}^2 \frac{C_a}{2\sqrt{2\zeta_i}} A_i\cos(\omega_i^\text{ext}t) \sum_{j=1}^3(\vec{\tilde K}^{-1})_{(j,i)}(\hat b_j^\dagger - \hat b_j).
\end{equation}
We may now define the Rabi frequency of the driving as
\begin{equation}
    \Omega_i = C_a\sum_{j=1}^3(\vec{\tilde K}^{-1})_{(j,i)}/ 2\sqrt{2\zeta_i}.    
\end{equation}
Truncating to the lowest three states the driving term takes the form
\begin{equation}
    \hat\H_\text{ext} = i\sum_{i=1}^2 \Omega_i \cos(\omega^\text{ext}_it) \left( \ketbra{0}{1}_i + \sqrt{2} \ketbra{1}{2}_i \right) + \hc
\end{equation}
Depending on which transition we want to drive we must match the driving frequencies with the transition energy, e.g., $\omega^\text{ext}_1 = \omega_{1,2}$ if we want to drive the $\ket{1} \leftrightarrow \ket{2}$ transition of the first qutrit, see \cref{fig:4qb_diagram}.

Such a system can, besides arbitrary one-qutrit gates and generalized controlled-\textsc{not} gates, implement both single-qubit and two-qubit nonadiabatic holonomic gates \cite{Pachos:2012:ITQ:2331123,HQC}.

\subsection{Multibody interactions}\label{sec:multibodyInteractions}

\begin{figure}
    \centering
	\includegraphics[width=.7\columnwidth]{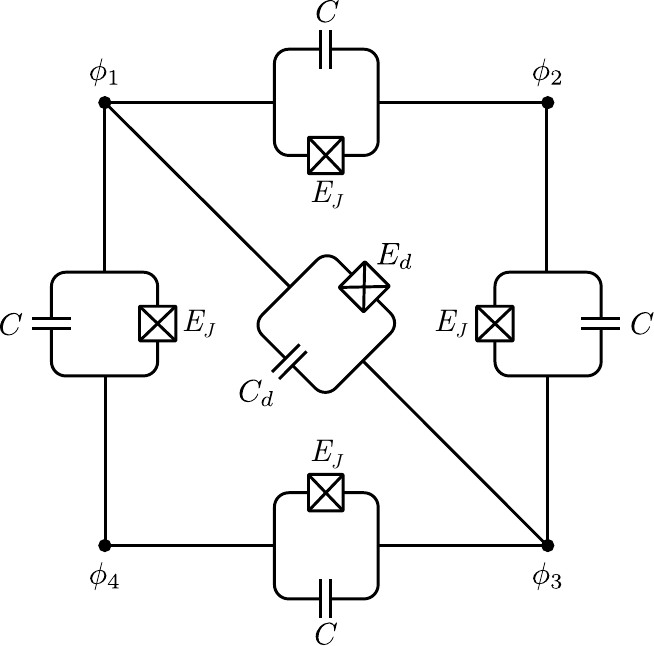}
	\caption{An example of a simple circuit with four nodes. The nodes are connected in a square with Josephson junctions and a capacitor, and two of the nodes are connected to the opposite corner.}
	\label{fig:square}
\end{figure}

The smallest example of multibody interaction must consist of four nodes, as we can always decouple the CM mode leaving three true degrees of freedom.
Consider therefore the circuit in \cref{fig:square} inspired by Ref. \cite{Pedersen2019}. If we approximate the Josephson junctions as linear inductors, we quickly realize that the capacitive and inductive matrices can be diagonalized easily. We therefore first consider the capacitive subgraph, which, following the method in \cref{sec:MethodofNodes} yields a capacitance matrix 
\begin{equation}
\vec{C} = \mat{2C + C_{d} & -C & -C_{d} & -C \\ -C & 2C & -C & 0 \\ -C_{d} & -C & 2C + C_{d} & -C \\ -C & 0 & -C & 2C}.
\end{equation}
To begin with we set the diagonal capacitance $C_d = C$, which yields the eigenmodes
\begin{equation}\label{eq:eigenmodes}
\begin{aligned}
&\vec{v}_\text{CM} = \frac{1}{2}\mat{1 \\ 1 \\ 1 \\ 1} ,\quad 
\vec{v}_{1} = \frac{1}{\sqrt{2}}\mat{1 \\ 0 \\ -1 \\ 0} ,\\
&\vec{v}_{2} = \frac{1}{\sqrt{2}}\mat{0 \\ 1 \\ 0 \\ -1} ,\quad 
\vec{v}_{3} = \frac{1}{2}\mat{1 \\ -1 \\ 1 \\ -1},
\end{aligned}
\end{equation}
with eigenvalues $ \lambda_\text{CM} = 0 $ (as we do not ground any node), $ \lambda_{1} = \lambda_3 = 4C $, and $ \lambda_{2} = 2C $. Note how the choice of identical capacitances ensures that the $ v_{i} $ are independent of the capacitances. 

These modes correspond to charge oscillating across the diagonal between nodes $ 1 $ and $ 3 $ ($ \vec{v}_{1} $), oscillation across the sides of the square between nodes $ 2 $ and $ 4 $ ($ \vec{v}_{2} $), and finally an oscillation involving the entire circuit between the nodes $ 1 $ and $ 3 $, and the nodes $ 2 $ and $ 4 $ ($ \vec{v}_{3} $). We can see that changing the capacitance of the diagonal branch does not disturb the eigenmodes. The fact that this is the only branch where the capacitance can be changed is an intuitive result when we consider the modes in terms of oscillating charge. We can think of the mode $ \vec{v}_{1} $ as the only one involving the diagonal branch.

Changing the diagonal capacitance to $ C_{d} \neq C $, we change only the first mode, and the diagonalized capacitance matrix takes the form
\begin{equation}
\vec{K} = \mat{0 & 0 & 0 & 0 \\ 0 & 2(C + C_{d}) & 0 & 0 \\ 0 & 0 & 2C & 0 \\ 0 & 0 & 0 & 4C}.
\end{equation}
From this point on we remove the center-of-mass coordinate. Consider now the inductive subgraph of the circuit in \cref{fig:square}. It yields the potential energy
\begin{equation}
\begin{aligned}
U &=- E_J\Bigg[\cos\left(\frac{\psi_{1} - \psi_{2}}{\sqrt{2}} + \psi_{3}\right) \\
    &\phantom{=- E_J} + \cos\left(\frac{\psi_{1} + \psi_{2}}{\sqrt{2}} - \psi_{3}\right)\\
	&\phantom{=- E_J} + \cos\left(\frac{\psi_{1} - \psi_{2}}{\sqrt{2}} - \psi_{3}\right) \\
	&\phantom{=- E_J} + \cos\left(\frac{\psi_{1} + \psi_{2}}{\sqrt{2}} + \psi_{3}\right)\Bigg]\\
	& \phantom{=}- E_{d}\cos(\sqrt{2}\psi_{1}),
\end{aligned}
\end{equation}
where we change to the diagonal basis. With some trigonometric identities, this can be reduced to
\begin{equation}
\begin{aligned}
U = &- E_{d}\cos(\sqrt{2}\psi_{1}) \\ &- 4E_J\cos\left(\frac{\psi_{1}}{\sqrt{2}}\right)\cos\left(\frac{\psi_{2}}{\sqrt{2}}\right)\cos(\psi_{3}).
\end{aligned}
\end{equation}
From this, we see that the diagonal Josephson junction, with $E_d$, does not contribute to the coupling between the three modes. We also see that the remaining four Josephson junctions with $E_J$ lead to three-body interaction between the $\psi_1$, $\psi_2$, and $\psi_3$ modes. 
Such a three-body interaction is a sixth-order effect, $\psi_1^2\psi_2^2\psi_3^2$, and one would therefore need to keep all terms to sixth order when expanding the cosine. This expansion to sixth order also leads to corrections to the frequencies and two-body couplings between the three modes.
Note that the diagonal branch can be removed from the system without changing the dynamics of the system.

By introducing external fluxes, we can tune the triple-cosine term to involve odd terms as well. The cosine terms themselves result only in products of even powers of the $ \psi_{i} $'s, but with flux threading the circuit a cosine term can be turned into a sine term, making the contributions from the corresponding mode completely odd. This opens up the possibility for further multibody couplings achieved through the normal modes, including couplings that do not require an expansion to sixth order.
Multibody couplings are useful, e.g., in gauge theories \cite{Marcos2013,Hauke2013,Marcos2014,Mezzacapo2015} or quantum annealing \cite{Lloyd2016,Chancellor2017,Kafri2017,Schondorf2018,Lechner2015,Leib2016,Perdomo2008,PerdomoOrtiz2012}.

\subsubsection{External coupling to eigenmodes}

If we wish to couple an eigenmode circuit into a larger configuration, we need to couple the eigenmodes to the external degrees of freedom. Such external degrees of freedom can be used to control or measure the system. While a nontransformed node flux can be controlled by coupling a single control line to the corresponding node (see \cref{sec:driving}), we must employ several control lines to couple the eigenmodes as these are generally linear combinations of the flux node variables as we have seen several times in the previous sections.
For concreteness, we consider the circuit in \cref{fig:square} transformed to its eigenmodes. We now want to capacitively couple the $\psi_1$ degree of freedom to an external control line without coupling to the two remaining degrees of freedom.

We therefore couple each node in the (nontransformed) circuit via identical capacitors of capacitance  $C_{\textup{ext}}$ to an external driving voltage $ V_{i}(t) $. This results in the following additional terms in the Lagrangian
\begin{equation}
\L_{\textup{ext}} = \frac{C_{\textup{ext}}}{2}\sum_{i=1}^4\left(\dot{\phi}_{i} - V_i\right)^2,
\end{equation}
similarly to the previous example.
Writing the Lagrangian in terms of the eigenmodes, expanding the parenthesis, and throwing away constant terms, we have
\begin{equation}
\begin{aligned}
\L_{\textup{ext}} 
&= \frac{C_{\textup{ext}}}{2}\Big[\dot{\psi}_\text{CM}^2 + \dot{\psi}_{1}^2 + \dot{\psi}_{2}^2 + \dot{\psi}_{3}^2\\
	&\phantom{=} - \dot{\psi}_\text{CM}\left(V_{1} + V_{2} + V_{3} + V_{4}\right)\\
	&\phantom{=} + \sqrt{2} \dot{\psi}_{1}\left(-V_{1} + V_{3}\right)\\
	&\phantom{=} + \sqrt{2}\dot{\psi}_{2}\left(-V_{2} + V_{4}\right)\\
	&\phantom{=} + \dot{\psi}_{3}\left(-V_{1} + V_{2} - V_{3} + V_{4}\right) \Big].
\end{aligned}
\end{equation}
If we now want to only couple to $\psi_{1}$, we choose $ V_{1} = -V_{3} $ and $ V_{2} = V_{4} = 0 $. This yields
\begin{align}
\L_{\textup{ext}} = \frac{C_{\textup{ext}}}{2}\Big[\dot{\psi}_\text{CM}^2 + \dot{\psi}_{1}^2 + \dot{\psi}_{2}^2 + \dot{\psi}_{3}^2 - 2\sqrt{2}\dot{\psi}_{1}V_{1}\Big]
\end{align}
The four first terms contribute to the diagonal of $ \vec{K} $ and can be viewed as corrections to the energy of the modes, and the last term is exactly an interaction term between $ \psi_{1} $ and the external $ V_{1}(t) $. Note that we could have just coupled to the 1 and 3 nodes to get the same result, which we could have guessed from the form of $\vec v_1$ in \cref{eq:eigenmodes}. Also, note that the center-of-mass mode obtains a nonzero eigenvalue, because all nodes are coupled to the ground.

\section{Summary and outlook}\label{sec:summary}

\begin{figure}
    \centering
    \includegraphics[scale=0.5]{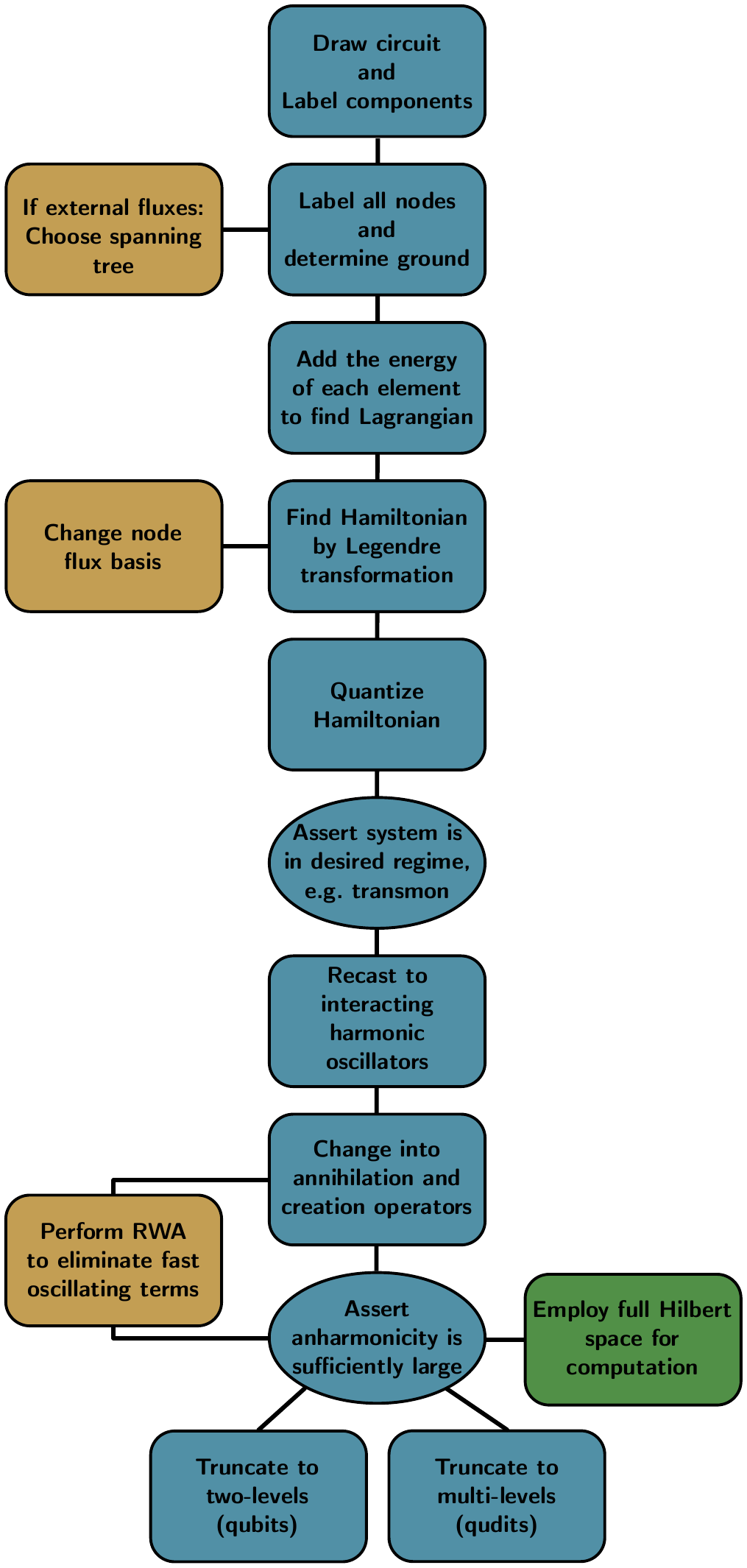}
    \caption{Overview of the methods presented in this tutorial. Blue blocks indicate the essential methods, while yellow blocks indicate optional steps. Green boxes are beyond the scope of this tutorial. Round blocks are assertions that must be satisfied before advancing in the flowchart.}
    \label{fig:flowchart}
\end{figure}

In this tutorial, we have presented various methods used when analyzing superconducting electrical circuits. We have summarized the methods in \cref{fig:flowchart}. 

An analysis usually starts by determining over which components possible external flux should be added, either using Kirchhoff's law directly, as in \cref{sec:KirchhoffsDirectly} or via constructing a spanning tree as described in \cref{sec:MethodofNodes}. The Lagrangian can then be constructed by determining the capacitor (kinetic) energy and subtracting the inductive (potential) energy as in \cref{sec:lagrangianApproach}. The Hamiltonian is found using a Legendre transformation in \cref{sec:ObtainHamiltonian}. One can then optionally change basis, e.g., into normal modes as in \cref{sec:normalModes,sec:changeOfBasis}.
The Hamiltonian can then be quantized using the canonical quantization in \cref{sec:quantize}. Asserting that the system is only weakly anharmonic it can be rewritten into interacting harmonic oscillators perturbed by the anharmonicity following the approach in \cref{sec:recasting}. After changing to annihilation and creation operators, the rotating-wave approximation can be applied if needed as in \cref{sec:rwa}. If the anharmonicity is large, the system can be truncated into qubits or qudits using either the methods in \cref{sec:truncation} or the more advanced techniques in \cref{sec:exactTrunc}. 
Note that this final truncation of the Hilbert space is not strictly necessary to perform computations using the superconducting circuit as other approaches work with the full Hilbert space of the oscillators. These approaches are beyond the scope of this tutorial and more information can be found in Refs. \cite{Vlastakis2013,Mirrahimi2014,Ofek2016,Rosenblum2018,Gao2019,Gertler2021,Joshi2021,Cai2021,Puri2017,Puri2019,Grimm2020,Braunstein2005,Lau2016}.

Besides the essential steps mentioned above, we have also discussed control of the modes via microwave driving in \cref{sec:driving} and used to perform single-qubit gates. Simple coupling of modes is discussed in \cref{sec:coupling} and this enables the implementation of two-qubit gates. In the same section, we also discussed coupling to linear resonators and inductive coupling via mutual inductance. Finally, we have discussed how to include noise when calculating the dynamics of the system using the Bloch-Redfield model and master equation in \cref{sec:noise}. We have illustrated the methods with concrete examples throughout the tutorial to aid the reader, and finally in \cref{sec:examples} we discussed a number of key examples of contemporary qubit designs and a number of couplers that allow the qubits to interact.

The methods presented here are by no means exhaustive in regards to circuit analysis. Classical electrical circuit analysis has been performed for decades by both physicists and engineers, and much more information on this subject can be found in the existing literature. The methods presented here should therefore not be seen as a limit to what can be done with superconducting circuits, but merely as a starting point for researchers new to the field of superconducting electrical circuit analysis. 

So where to go from here? If you want to explore controlling and measuring superconducting circuits we recommend Ref. \cite{Girvin2014} which discusses the coupling to microwave resonators in greater detail. For more information on the method in \cref{sec:MethodofNodes} and the more advanced methods in \cref{subsec:GraphMethod}, see Refs. \cite{Vool2017} and \cite{Burkard2004}, respectively. Both of these references also discuss dissipation in more detail. For each of the examples, we reference the original research which should be consulted, and finally for an overview of the field see Ref. \cite{Krantz2019} which reviews recent state-of-the-art concepts.

\begin{acknowledgements}
We wish to thank our numerous experimental and theoretical collaborators and colleagues for discussions and collaborations on superconducting circuits over the past few years, in particular D. Petrosyan, M. Kjaergaard, W. Oliver, S. Gustavsson, and C.~K. Andersen. Thank you also to all the students that have challenged us to provide clear and introductory explanations in this exciting field. They are the real reason that we undertook the task of writing this tutorial. This work is supported by the Danish Council for Independent Research and the Carlsberg Foundation.
\end{acknowledgements}

\appendix

\section{Fundamental graph theory of electrical networks}\label{subsec:Fundamental graph theory}

\begin{figure*}
    \centering
    \includegraphics[width=.9\textwidth]{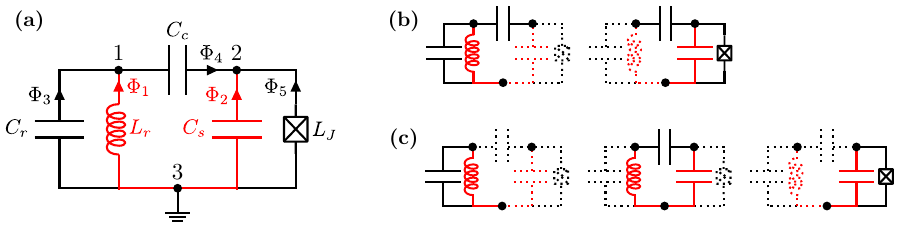}
    \caption{\textbf{(a)} Example transmon-resonator circuit. The chosen spanning tree (red) consists of the resonator $L_r$ inductance and the shunting capacitance $C_s$. \textbf{(b)} Fundamental cutsets of the circuit (in solid) with respect to the chosen spanning tree. \textbf{(c)} Fundamental loops of the circuit (in solid) with respect to the chosen spanning tree.}
    \label{fig:example_tr_circuit}
\end{figure*}

In this appendix, we present some fundamental definitions from graph theory. The reason for this is that graph theory is the natural language of electromagnetic circuits where each circuit element can be represented as an edge on a graph. We introduce these definitions as a supplement to the discussion in the main text. The first three definitions are directly related to the main text, while the remaining definitions provide an alternative way of stating Kirchhoff's laws.
We describe the quantities important to circuit analysis using the example circuit shown on \cref{fig:example_tr_circuit}(a). The example circuit consists of a transmon qubit capacitively coupled to a resonator, which is a very common setup \cite{Koch2007,Kounalakis2018}. For more material on graph theory see, e.g., Refs. \cite{Peikari1974,Bondy1976}.

\begin{definition}[Graph]\label{def:graph}
    A \textbf{graph} $\mathcal{G} = (\mathcal{N}, \mathcal{B})$ is a set of \textbf{nodes} $\mathcal{N} = \{n_1, \dots, n_{N}\}$ where $N$ is the number of nodes, and a set of branches (sometimes called edges) $\mathcal{B} = \{b_1, \dots, b_B\}$ where each branch connects a pair of nodes and $B$ is the number of branches. The number of nodes is called the \textbf{order} of the graph and is denoted $|\mathcal{G}| = N$. We allow multiple branches to connect the same pair of nodes. Sometimes this is called a \textbf{multigraph} in order to distinguish it from \textbf{simple graphs} where only one branch can connect the same pair of nodes.
\end{definition}

Using this definition, we can consider each circuit as a graph where each component corresponds to a branch. 
The first step of any circuit analysis is to label every branch of the graph. These can be labeled in different ways, usually via the element or the flux through the current. These are equivalent, and often both are used as they complement each other. 

Using the components as the labels, the set of branches in \cref{fig:example_tr_circuit}(a) becomes $\mathcal{B} = \{L_r, C_s, C_r, C_c, L_J\}$. The order of the graph is $|\mathcal{G}|=3$ and the nodes can be labeled arbitrarily, here we label them 1,2, and 3.
The number of branches is $B=5$, and we can thus write the flux over all branches as a vector with five elements
\begin{equation}
    \vec{\Phi} = \mat{\Phi_1 & \Phi_2 & \Phi_3 & \Phi_4 & \Phi_5}^T,
\end{equation}
where the order of the fluxes corresponds to the number of the branches in $\mathcal{B}$.
Note that we have indicated the direction of every branch in \cref{fig:example_tr_circuit}(a) using arrows. We define positive branch currents $I_b>0$ as the case where current flows through a branch in the direction of the arrow. Using the passive sign convention the voltage over a branch is then given by $V_b = V_b^\text{start}-V_b^\text{end}$, which ensures that the power $P_b = I_bV_b$ is positive if energy is being stored or dissipated in the branch element. Strictly speaking, this makes our graph a \emph{directed} graph, but since all electrical network graphs are directed graphs, we are simply going to call them graphs. We are also going to assume that our graph is connected, meaning that there exists a path between every pair of nodes. 

\begin{definition}[Subgraph]\label{def:sub_graph}
    A graph $\mathcal{H} = (\mathcal{N}_\mathcal{H}, \mathcal{B}_\mathcal{H})$ is called a \textbf{subgraph} of $\mathcal{G} = (\mathcal{N}_\mathcal{G}, \mathcal{B}_\mathcal{G})$, written $\mathcal{H} \subseteq \mathcal{G}$, if $\mathcal{N}_\mathcal{H} \subseteq \mathcal{N}_\mathcal{G}$ and $\mathcal{B}_\mathcal{H} \subseteq \mathcal{B}_\mathcal{G}$. If $\mathcal{H}$ is a subgraph of $\mathcal{G}$ but $\mathcal{H} \neq \mathcal{G}$ it is called a proper subgraph.
\end{definition}

In the electrical circuit setting, the notion of subgraphs is often used to describe the capacitive and inductive subgraphs of the circuit. In the case of the example in \cref{fig:example_tr_circuit}(a) the capacitive subgraph is defined to be the set of branches $\mathcal{B}_C = \{C_s, C_r,C_c\}$, while the inductive subgraph is defined by $\mathcal{B}_L = \{L_r, L_J\}$. Note that the set of nodes are identical for the capacitive and inductive subgraph as well as the full (super)graph, i.e., $\mathcal{N}_C = \mathcal{N}_L = \mathcal{N}$. Also, even though we assume our graph to be connected, its subgraphs are not necessarily connected.

The next step in the analysis is to specify a subgraph called a \emph{spanning tree} for our graph.
\begin{definition}[Spanning tree]\label{def:spanning_tree}
    A \textbf{spanning tree} of a graph $\mathcal{G}$ is a connected subgraph $\mathcal{T}$ that contains the same nodes as $\mathcal{G}$ (i.e., $\mathcal{N}_\mathcal{T} = \mathcal{N}_\mathcal{G}$) and contains no loops. 
    
    The branches of the spanning tree are called \textbf{twigs} and branches of the complement of the spanning tree are called \textbf{links} (or \textbf{chords}). Note that there are $B_\mathcal{G} - (N_\mathcal{G} - 1)$ links.
\end{definition}

The spanning tree connects every pair of nodes through exactly one path. For our example, we choose branches 1 and 2 as our spanning tree as shown in red on \cref{fig:example_tr_circuit}. The linear inductor, together with the $C_s$ shunting capacitor, constitutes the twigs of the tree, while the remaining capacitors ($C_r, C_c$) and Josephson junction ($L_J$) are links. Note that we are free to choose our spanning tree differently as long as it obeys the definition. We could, e.g., have chosen the inductive subgraph, as mentioned above, however, we can not choose the capacitive subgraph as it includes a loop. This freedom in choosing the spanning tree corresponds to a gauge freedom in the equations of motion.

Choosing a spanning tree also allows us to define fundamental cutsets and fundamental loops, which are useful when deriving the equations of motion for a circuit. The following definitions used in the main text but can be used for an alternative statement of Kirchhoff's laws. We start with the fundamental cutsets. 

\begin{definition}[Cut]\label{def:cut}
    Given a graph $\mathcal{G} = (\mathcal{N}, \mathcal{B})$ a \textbf{cut} is a partitioning of nodes $\mathcal{N}$ into two disjoint sets $\mathcal{N}_A$ and $\mathcal{N}_B$. 
    With every cut, we can associate a \textbf{cutset}, which is the set of branches that have endpoints in both $\mathcal{N}_A$ and $\mathcal{N}_B$.
\end{definition}

Note that removing a single twig cuts the spanning tree, $\mathcal{T}$, into two disjoint subgraphs with nodes $\mathcal{N}_A$ and $\mathcal{N}_B$. Such a cut is called a \emph{fundamental cut}, and the branches that must be removed to complete the same cut on the full graph is called a \emph{fundamental cutset}. More formally:

\begin{definition}[Fundamental cut]\label{def:fundamental-cut}
    Given a graph $\mathcal{G}$ and a spanning tree $\mathcal{T}$ we define a \textbf{fundamental cut} or \textbf{f-cut} as a cut whose \textbf{cutset} contains only one twig.
\end{definition}

In practice, the fundamental cutsets can be found by removing one twig from the spanning tree. This creates two disjoint subgraphs of the spanning tree with nodes $\mathcal{N}_A$ and $\mathcal{N}_B$. Now remove the links of the full graph with endpoints in both partitions. The cutset is then the set of all the removed links and the single twig. We thus end up with a unique cutset with one twig and any number of links. This can be done for every twig, and the number of fundamental cutsets is thus equal to the number of twigs $|\mathcal{T}| = N-1$. The fundamental cutsets of our example graph can be seen in \cref{fig:example_tr_circuit}(b).

We now turn our attention to the loops. By taking the spanning tree and adding a single link from the full graph we form a unique loop. Such a loop contains exactly one link and one or more twigs. We call these loops the \emph{fundamental loops} of the $\mathcal{G}$ with respect to the spanning tree $\mathcal{T}$.

\begin{definition}[Fundamental loop]\label{def:fundamental-loop}
    Given a graph $\mathcal{G}$ and a spanning tree $\mathcal{T}$, we define a \textbf{fundamental loop} or \textbf{f-loop} as a loop consisting of exactly one link and one or more twigs.
\end{definition}

The number of fundamental loops that can be formed is equal to the number of links. The fundamental loops of our example graph can be seen in \cref{fig:example_tr_circuit}(c).

As we shall see in the following section the fundamental loops and cuts allow us to write Kirchhoff's laws in a compact and useful way.

\subsection{Circuit matrices}\label{subsec:Circuit matrices}

Using the notion of f-loops and f-cuts, we define two characteristic matrices for the network graphs, which can be used to write Kirchhoff's laws more compactly.

For every loop, we can define the orientation, i.e., clockwise or anti-clockwise. For an f-loop, we let the orientation be determined by the orientation of its link. We can then define the fundamental loop matrix.
\begin{definition}[Fundamental loop matrix]\label{def:f_loop_matrix}
    Given a graph $\mathcal{G}=(\mathcal{N},\mathcal{B})$, with spanning tree $\mathcal{T}$, we define the \textbf{fundamental loop matrix}, or \textbf{f-loop matrix}, $ \vec{F}^{(L)} $ as
    \begin{equation}\label{eq:f_loop_matrix_defintion}
    	\vec F^{(L)}_{ij} = 
        \begin{cases}
        +1 & \text{if $b_j\in f_i$ and $l_i, b_j$ same orientation}\\
        -1 & \text{if $b_j\in f_i$ and $l_i, b_j$ opposite orientation}\\
        0 & \text{if $b_j\notin f_i$}
        \end{cases},
    \end{equation}
    where $l_i$ is the link in the $i$th f-loop, $f_i$, with $1\leq i \leq |\mathcal{G}\backslash \mathcal{T}| = B-(N-1)$ and $b_j$ is the $j$th branch in $\mathcal{B}$ with $1\leq j \leq B$.
\end{definition}
In other words, we iterate through the branches and the set of f-loops. If the given branch is in the given f-loop, the matrix entry becomes $\pm1$, with a plus if the branch has the same orientation as the f-loop (which is determined by the link of the f-loop). If the branch is not in the given f-loop, the matrix entry is 0.

Consider our example circuit and its fundamental loops from \cref{fig:example_tr_circuit}(c). The first fundamental loop consists of the link $\Phi_3$ and the twig $\Phi_1$. The orientation of the loop (determined by $\Phi_3$) is clockwise, which means that the $\bm{F}^{(L)}_{11} = -1 $, since the twig $\Phi_1$ points in the anti-clockwise direction. The only other nonzero entry in the first row is $\bm{F}^{(L)}_{13} = 1 $, corresponding to the link $\Phi_3$ oriented in the clockwise direction. Following the same method for the other two f-loops, we find
\begin{equation}\label{eq:FLexample}
    \vec{F}^{(L)} = \mat{
    -1 & 0 & 1 & 0 & 0\\
    1 & -1 & 0 & 1 & 0\\
    0 & -1 & 0 & 0 & 1
    },
\end{equation}
where the columns correspond to the branches in their respective order and the rows correspond to the loops in the same order as in \cref{fig:example_tr_circuit}(c).

As with the loops, we can also choose an orientation for the cutsets. If a cut is oriented from $\mathcal{N}_A$ to $\mathcal{N}_B$, we say that a branch in the cutset has positive orientation if it begins in $\mathcal{N}_A$ and ends in $\mathcal{N}_B$. We choose to orient every f-cutset such that its twig in an f-cutset has positive orientation. We can then define the \emph{fundamental cutset matrix}.

\begin{definition}[Fundamental cut matrix]\label{def:f_cut_matrix}
    Given a connected graph $\mathcal{G}=(\mathcal{N},\mathcal{B})$, with spanning tree $\mathcal{T}$, we define the \textbf{fundamental cut matrix}, or \textbf{f-cut matrix}, $ \vec{F}^{(C)} $ as
    \begin{equation}\label{eq:f-cut_matrix_definition}
    	\vec{F}^{(C)}_{ij} = 
        \begin{cases}
        +1 & \text{if $b_j\in c_i$ and $t_i, b_j$ same orientation}\\
        -1 & \text{if $b_j\in c_i$ and $t_i, b_j$ opposite orientation}\\
        0 & \text{if $b_j\notin c_i$}
        \end{cases},
    \end{equation}
    where $t_i$ is the twig of the $i$th cutset, $c_i$, with $1\leq i \leq |\mathcal{T}| = N-1$ and $b_j$ is the $j$th branch in $\mathcal{B}$ with $1\leq j \leq B$.
\end{definition}
In other words, we iterate through the branches and the set of cutsets. If the given branch is in the given cutset, the matrix entry becomes $\pm1$, with a plus if the branch has the same orientation as the cutset (which is determined by the orientation of the twig of the cutset). If the branch is not in the given f-cutset, the matrix entry is 0.

As an example take the first cutset from \cref{fig:example_tr_circuit}(b). The twig $\Phi_1$ and link $\Phi_3$ both points towards the same node and thus have positive orientation. The final link $\Phi_4$ points away from the node and has negative orientation. Thus the first row of the cutset matrix becomes $[1, 0, 1, -1, 0]$. By analyzing the other cutset, in the same manner, we find the fundamental cutset matrix
\begin{equation}\label{eq:FCexample}
    \vec{F}^{(C)} = \mat{
    1 & 0 & 1 & -1 & 0\\
    0 & 1 & 0 & 1 & 1\\
    },
\end{equation}
where the columns correspond to the branches in their respective order, and the rows correspond to the cutsets in the same order as in \cref{fig:example_tr_circuit}(b).

All branches of the graph are either twigs or links. Every f-cutset contains only one twig, and every f-loop contains only one link. Additionally, for every partition of nodes defined by an f-cut, every f-loop must begin and end in the same partition. Thus every f-cutset and f-loop share either 0 or exactly 2 branches. Now consider the elements
\begin{equation}\label{eq:f-cut-and-loop-matrix-product-proof-eq_1}
	\left(\vec{F}^{(L)}(\vec{F}^{(C)})^T\right)_{ij} = \sum_{k}\vec{F}^{(L)}_{ik}\vec{F}^{(C)}_{jk}.
\end{equation}
Evidently, the $(i,j)$th element depends only on the $ i $th f-loop and the $j$th f-cut. If the f-cutset and f-loop share no branches, all the terms are zero, and in the case where they share exactly two branches, we get two nonzero terms with opposite signs. We thus have
\begin{equation}\label{eq:f-cut-and-loop-matrix-product}
    \vec{F}^{(L)}(\vec{F}^{(C)})^T = \vec{0}.
\end{equation}
Multiplying \cref{eq:FCexample,eq:FLexample} we see that this is exactly the case for the example graph, as it should be.

\section{Method of electrical network graph theory}\label{subsec:GraphMethod}

In this section, we present a more mathematical stringent method for obtaining the Hamiltonian of an electrical superconducting circuit. This method is based upon Ref. \cite{Burkard2004} and uses electrical network graph theory \cite{Peikari1974}. This method is a more advanced alternative to the method presented in \cref{sec:EOM}, however, the resulting equations of motion are the same.

The first step is to label and order all the circuit components (branches) of the network graph and choose a spanning tree for the graph. Without loss of generality, we order the components such that the first $|\mathcal T|$ branches are the twigs, and we then write the fluxes and currents through all components as vectors
\begin{equation}\label{eq:fluxAndCurrentsVectors}
\vec \Phi = 
\mat{
\vec \Phi_t \\ \vec \Phi_l
}, \quad
\vec I = 
\mat{
\vec I_t \\ \vec I_l
},
\end{equation}
where $\vec \Phi_t$ ($\vec I_t$) are the fluxes (currents) of all the twigs and $\vec \Phi_l$ ($\vec I_l$) are the fluxes (currents) of all the links. For the example circuit in \cref{fig:example_tr_circuit} we have $\vec\Phi_t = (\Phi_1, \Phi_2)^T$ and $\vec\Phi_l = (\Phi_3, \Phi_4, \Phi_5)^T$ and likewise for the current vector.

After all components have been labeled and a tree has been selected, we construct the fundamental matrices of the graph $\vec{F}^{(L)}$ and $\vec{F}^{(C)}$ following \cref{def:f_loop_matrix,def:f_cut_matrix}, respectively. In the following, we show how these matrices may be used to set up the equations of motion and reduce the number of free coordinates.

\subsection{Kirchhoff's laws}\label{subsec:Kirchhoffs laws}

Using \cref{eq:fluxAndCurrentsVectors} and the f-matrices, we reformulate Kirchhoff's laws as stated in \cref{eq:KirchhoffsLaws1}.

\subsubsection{Kirchhoff's current law}\label{subsubsec:Kirchhoffs current law}

Kirchhoff's current law states that no charge may accumulate at a node. Mathematically we may write this as
\begin{equation}\label{eq:KCL_raw}
	\sum_{b \text{ incident on } n} s_{n,b}I_b = 0, \quad \text{for every node $n$},
\end{equation}
where we have $ s_{n,b}=+1 $ if the branch $ b $ ends at node $ n $ and $ s_{n,b}=-1 $ if $ b $ begins at $ n $. This is equivalent to the definition in \cref{eq:KCL}, but with currents instead of charges, i.e., \cref{eq:KCL_raw} is the time derivative of \cref{eq:KCL}.
Recall that a cutset is the set of branches between two partitions of nodes. Thus if no charge has accumulated at a single node, the total current from one partition of nodes to another must be zero. We can write this using the f-cut matrix as
\begin{equation}\label{eq:KCL_using_f-cut_matrix}
	\vec{F}^{(C)}\vec{I} = \vec{0}.
\end{equation}
If we calculate this matrix product for the example circuit using \cref{eq:FCexample} we find
\begin{align*}
    \vec{F}^{(C)}\vec{I} &= \mat{
    1 & 0 & 1 & -1 & 0\\
    0 & 1 & 0 & 1 & 1\\
    } \mat{I_1 \\ I_2 \\ I_3 \\ I_4 \\ I_5} \\
    &= \mat{I_1 + I_3 - I_4 \\ I_2 + I_4 + I_5} = \mat{0 \\ 0},
\end{align*}
which is equivalent to applying Kirchhoff's current law directly to nodes 1 and 2 in \cref{fig:example_tr_circuit}.

\subsubsection{Kirchhoff's voltage law}\label{subsubsec:Kirchhoffs voltage law}

Kirchhoff's voltage law states that if we choose some oriented loop of branches $l$, the algebraic sum of voltages around the loop must equal the electromotive force induced by external magnetic flux, $\tilde  {\Phi}_l$, through the face enclosed by the loop, i.e.,
\begin{equation}\label{eq:KVL_raw}
	 \sum_{b\in l} s_{l, b} V_b = \dot{\tilde  {\Phi}}_l,  \quad \text{for all loops $l$},
\end{equation}
where $ s_{l, b} = +1 $ if $ b $ is oriented along $ l $, and $ s_{l, b} = -1 $ if $ b $ is oriented against $ l $. The external flux through the loop $l$ is denotes $\tilde \Phi_l$. This is equivalent to the definition in \cref{eq:KVL}, but with voltages instead of fluxes, i.e., \cref{eq:KVL_raw} is the time derivative of \cref{eq:KVL}.
Thus, the f-loops of the graph define a set of equations and using \cref{eq:flux_definition} we may write Kirchhoff's voltage law as
\begin{equation}\label{eq:KVL_using_f-loops}
	\vec{F}^{(L)}\vec{\Phi} =  \vec{\tilde \Phi},
\end{equation}
where $ \vec{\tilde{\Phi}} = (\tilde\Phi_1, \dots, \tilde\Phi_{B-N+1})^T$ is the vector external fluxes through the fundamental loops.

For the example circuit, we calculate the matrix product using \cref{eq:FLexample} and find
\begin{align*}
    \vec{F}^{(L)}\vec{\Phi} &= \mat{
    -1 & 0 & 1 & 0 & 0\\
    1 & -1 & 0 & 1 & 0\\
    0 & -1 & 0 & 0 & 1
    } \mat{\Phi_1 \\ \Phi_2 \\ \Phi_3 \\ \Phi_4 \\ \Phi_5} \\
    &= \mat{-\Phi_1 + \Phi_3 \\ \Phi_1 - \Phi_2 + \Phi_4 \\ -\Phi_2 + \Phi_5} = \mat{\tilde\Phi_1 \\ \tilde\Phi_2 \\ \tilde\Phi_3},
\end{align*}
where each row is equivalent to applying Kirchhoff's voltage law directly to the corresponding loop. We assume external fluxes of $ \vec{\tilde\Phi} = (\tilde\Phi_1, \tilde\Phi_2, \tilde\Phi_3)^T$ through the loops.
 
\subsubsection{Reducing the number of coordinates}\label{subsubsec:Reducing the number}

Using Kirchhoff's voltage law, we can reduce the number of free coordinates. We only need to specify the fluxes of the spanning tree to calculate the remaining fluxes. In order to do so, we write our f-cut matrix as 
\begin{equation}\label{eq:dividedFC}
    \vec{F}^{(C)} = 
    \mat{
        \mathbb{1} & \vec{F}
    },
\end{equation}
where $\vec{F}$ is a $|\mathcal{T}|\times|\mathcal{G}\backslash\mathcal{T}| = (N-1) \times (B - N + 1)$ matrix and the identity is a $(N-1) \times (N-1)$ matrix. Note that our specific ordering of the circuit components (twigs first, then links) allows for the simple block structure of \cref{eq:dividedFC}. This structure is clearly seen in the example in \cref{eq:FCexample}, from which it is evident that 
\begin{equation}\label{eq:Fexample}
    \vec F = \mat{1 & -1 & 0 \\ 0 & 1 & 1},
\end{equation}
for the example circuit in \cref{fig:example_tr_circuit}.

Reordering the components shuffles the rows and columns of the fundamental cut matrix, and the following derivations can easily be generalized. 
From \cref{eq:f-cut-and-loop-matrix-product} and \cref{def:f_loop_matrix} we find that we can write the f-loop matrix in a similar manner
\begin{equation}
    \vec{F}^{(L)}=   
    \mat{
        -\vec{F}^T & \mathbb{1}
    },
\end{equation}
where $\vec{F}$ is the same matrix as in \cref{eq:dividedFC}, meaning that the identity is now $(B - N+1) \times (B - N+1)$. This structure is again seen in the example in \cref{eq:FLexample} where the transpose of \cref{eq:Fexample} occurs.
We can then rewrite Kirchhoff's voltage law in \cref{eq:KVL_using_f-loops} and isolate the fluxes of the links
\begin{equation}\label{eq:voltage_coordinates_reduced}
	\vec{\Phi}_l = \vec{\tilde{\Phi}} + \vec{F}^T\vec{\Phi}_t ,
\end{equation}
and use this to write our flux vector in \cref{eq:fluxAndCurrentsVectors} in terms of the twig and external fluxes
\begin{equation}\label{eq:voltage_vector_reduced_expression}
	\vec{\Phi} = \mat{
    \vec{\Phi}_t \\ \vec{F}^T\vec{\Phi}_t+\vec{\tilde{\Phi}}
    }
    = (\vec{F}^{(C)})^T\vec{\Phi}_t+	\mat{
    \vec{0} \\ \vec{\tilde{\Phi}}
    },
\end{equation}
meaning that we have eliminated the fluxes of the links.

Using \cref{eq:voltage_vector_reduced_expression} on the example circuit in \cref{fig:example_tr_circuit} we can write the fluxes as
\begin{equation}
    \vec \Phi = \mat{\Phi_1 \\ \Phi_2 \\ \Phi_1 + \tilde\Phi_1 \\ \Phi_1 - \Phi_2 + \tilde\Phi_2 \\ \Phi_2 + \tilde\Phi_3},
\end{equation}
which means that we have eliminated the three fluxes on the links.

\subsection{Equations of motion}\label{subsec:Equations of motion}

In this section, we use Kirchhoff's current law, to set up the equations of motion for the system. For this purpose, it is convenient to introduce the species-specific vectors $\vec I_S$ and  $\vec \Phi_S$
\begin{subequations}\label{eq:SpeciesVectors}
    \begin{align}
    (\vec I_S)_i &= 
    \begin{cases}
    \vec I_i & \text{ if the $i$th element is of species $S$,}\\
    0 & \text{ otherwise, }
    \end{cases},\label{eq:SpeciesCurrentVector}\\
    (\vec \Phi_S)_i &= 
    \begin{cases}
    \vec \Phi_i & \text{ if the $i$th element is of species $S$,}\\
    0 & \text{ otherwise, }
    \end{cases}
\end{align}
\end{subequations}
where the species subscript, $S$, indicates the element species, i.e., capacitor, inductor, etc. This can be understood as the current and flux vectors with everything but $S$ species removed. We use $C$ for capacitors, $L$ for linear inductors, and $J$ for Josephson junctions.
For the example circuit this yields
\begin{subequations}
\begin{align}
    \vec I_C &= (0, I_2, I_3, I_4, 0)^T, \label{eq:capacitorCurrent}\\
    \vec I_L &= (I_1, 0, 0, 0, 0)^T, \label{eq:inductorCurrent}\\
    \vec I_J &= (0, 0, 0, 0, I_5)^T, \label{eq:JJCurrent}
\end{align}
\end{subequations}
and likewise for the fluxes.

The first step of the analysis is to express the current of every branch in terms of the tree fluxes $\vec \Phi_t$. The current flowing through a capacitor with capacitance $C$ is given by \cref{eq:currentCapacitor}, and we can thus write the current flowing through all capacitors as
\begin{equation}\label{eq:capacitor_currents}
    \vec I_C = \vec D_C \ddot{\vec{\Phi}},
\end{equation}
where $\vec D_C$ is a diagonal matrix with the circuit capacitances on the diagonal. In this context, all other circuit components are counted as having zero capacitance. For the example circuit the capacitance matrix becomes $\vec D_C = \diag{0,C_s, C_r, C_c, 0}$, which multiplied to $\ddot{\vec{\Phi}} = (\ddot\Phi_1, \ddot\Phi_2, \ddot\Phi_3, \ddot\Phi_4, \ddot\Phi_5)^T$ yields \cref{eq:capacitorCurrent}.

The flux stored in the linear inductors is related to the currents through
\begin{equation}\label{eq:inductor_fluxes}
    \vec L\vec I = \vec{\Phi}_L,
\end{equation}
where $\vec L$ is a symmetric matrix with diagonal elements $\vec L_{ii} = L_i$ where $L_i$ is the inductance of the $i$th element. For all other components than linear inductors, we set $L_i=0$. The off-diagonal elements are the mutual inductances $\vec L_{ij} = M_{ij} = k_{ij}\sqrt{L_iL_j}$ between the $i$th and $j$th inductor, with $-1< k_{ij} < 1$ being the coupling coefficient. If a positive current in one inductor results in a positive magnetic flux contribution through another, we have $k_{ij}>0$. If the contribution is negative, we instead have $k_{ij}<0$. The numerical value of $k_{ij}$ depends on the placement of the inductors relative to each other.

In the example circuit, there is only one inductor and thus no mutual inductance, which means that $\vec L = (L_r, 0, 0, 0, 0)^T$, which multiplied to $\vec \Phi$ gives \cref{eq:inductorCurrent}.

Note that all the rows and columns belonging to components not on the inductor subgraph are zero. By removing these zero rows and columns, we get a $N_L\times N_L$ matrix $\vec L'$, where $N_L$ is the number of inductors. We can then rewrite \cref{eq:inductor_fluxes} as
\begin{equation}\label{eq:inductor_fluxes_reduces}
    \vec L'\vec I'_L = \vec{\Phi}'_L,
\end{equation}
where $\vec I_L'$ and $\vec \Phi_L'$ are the corresponding vectors found by removing all the noninductor entries of the full-size vectors $\vec I$ and $\vec \Phi$. In our example this becomes a single equation $L_r \Phi_1 = I_1$

The magnetic field energy stored in the inductors is 
\begin{equation}
    0 \leq E_L = \frac{1}{2}\vec I_L'^T\vec L' \vec I_L' ,
\end{equation}
which means that $\vec L'$ must be positive semi-definite. We further assume $\vec L'$ is positive definite, meaning that $0<\vec I_L'^T\vec L' \vec I_L'$ for $\vec I'_L \neq 0$. This assumption is also physically sensible since any current through the inductors must store at least some magnetic field energy in a realistic configuration. It also ensures that the symmetric $\vec L'$ matrix is invertible, and we can write 
\begin{equation}\label{eq:inductor_fluxes_reduces_rewritten}
    \vec I'_L = \vec L'^{-1} \vec{\Phi}'_L.
\end{equation}
We can expand the matrix $\vec L'^{-1}$ to work on the full flux vector by  inserting zeros on the noninductor columns and rows. Similarly, we also build the corresponding full inductor current vector $\vec I_L$. The resulting equation can be written 
\begin{equation}\label{eq:inductor_currents}
    \vec I_L = \vec L^+\vec{\Phi},
\end{equation}
where $\vec L^+$ is the matrix found by expanding $\vec L'^{-1}$ with the zero-columns and rows of the noninductor components. Formally, $ \vec L^+$ is the Moore-Penrose pseudo-inverse \cite{penrose_1955} of the original full inductance matrix $\vec L$. 

For our example we can easily invert $\vec L' = \mat{L_r}$ in order to find the psuedo-inverse $\vec L^+ = \diag{1/L_r, 0, 0, 0, 0}$, which fulfill \cref{eq:inductor_currents}.

Now we only need to include the current through the Josephson junctions, which follows from the Josephson relation
\begin{equation}\label{eq:jj_currents}
    \vec I_J = \vec D_J \textbf{sin} \vec \Phi,
\end{equation}
where $\vec D_J$ is a diagonal matrix with the Josephson critical currents on the diagonal, see \cref{eq:josephson_current_phase} for the case of a single Josephson junction. As with $\vec L$ and $\vec C$, all other components than Josephson junction are counted as having zero critical currents. The vector $\textbf{sin} \vec \Phi = (\sin\Phi_1, \dots ,\sin \Phi_B)^T$ is understood as the vector of sines of the branch fluxes.

We have only one Josephson junction in the example circuit in \cref{fig:example_tr_circuit} which means that $\textbf{sin} \vec \Phi = (0, 0, 0, 0, \sin\Phi_5)^T$ and $\vec D_J = \diag{0, 0, 0, 0, I_c}$, where $I_c = 1/L_J$ in our notation, see \cref{subsubsec:Josephson Junction}. Multiplying these two gives \cref{eq:JJCurrent}.

Thus, the current through each branch can be written as a function of the branch flux and its derivatives as seen in \cref{eq:capacitor_currents,eq:inductor_currents,eq:jj_currents}, and Kirchhoff's current law thus gives a set of coupled second-order differential equations
\begin{equation}\label{eq:circuit_EQM_generalized_with_matrices}
\begin{aligned}
	\vec 0 &= \vec F^{(C)} \vec{I} = \vec F^{(C)} \left[\vec I_C+ \vec I_L+\vec I_J\right]\\
	&=\vec M \ddot{\vec \Phi}_t + \dot{\vec Q}_0 + \vec  K \vec \Phi_t +\vec I_0\\&\phantom{=}+
    \vec F^{(C)}\vec D_J\textbf{sin}\left((\vec{F}^{(C)})^T\vec{\Phi}_t+	\mat{
    \vec{0} \\ \vec{\tilde{\Phi}}
    }\right),
\end{aligned}
\end{equation}
where we define the \enquote{mass} and \enquote{spring constant} matrices (analogous to in \cref{sec:normalModes})
\begin{subequations}
    \begin{align}
    \vec{M} &= \vec F^{(C)}\vec{D_C}(\vec F^{(C)})^T,\label{eq:capacitance_or_mass_matrix_advanced} \\ 
    \vec{K} &= \vec F^{(C)}\vec{L^+}(\vec F^{(C)})^T,
    \label{eq:inductance_or_spring_matrix_advanced}
\end{align}
\end{subequations}
and the offset charges and flux induced currents
\begin{subequations}
    \begin{align}
    \vec Q_0 &= \vec F^{(C)}\vec D_C \mat{\vec{0} \\ \dot{\vec{\tilde{\Phi}}} }, \label{eq:flux_induced_offset_charges}\\
    \vec I_0 &= \vec F^{(C)}\vec L^+ \mat{\vec{0} \\ \vec{\tilde{\Phi}} }.
\end{align}
\end{subequations}
Note that these matrices are different from the capacitive and inductive matrices presented in \cref{sec:lagrangianApproach}. 

Consider again the example circuit in \cref{fig:example_tr_circuit}. The \enquote{mass} and \enquote{spring constant} matrices are in this case
\begin{subequations}
\begin{align}
    \vec M &= \mat{C_c + C_r & -C_c \\ -C_c & C_c + C_s}, \\
    \vec K &= \mat{1/L_r & 0 \\ 0 & 0}.
\end{align}
\end{subequations}
Note how these are identical to how we constructed the capacitance matrix and the inductor matrix in \cref{sec:lagrangianMatrices}, respectively. Thus we have derived how to formulate the capacitance and inductive matrices from the main text.

The offset charges and flux induces currents are
\begin{subequations}
\begin{align}
    \vec Q_0 &= \mat{C_r\dot{\vec{\tilde\Phi}}_1 - C_c\dot{\vec{\tilde\Phi}}_2  \\ C_c\dot{\vec{\tilde\Phi}}_2}, \\
    \vec I_0 &= \mat{0 \\ 0}.
\end{align}
\end{subequations}
The offset charges $\vec Q_0$ disappear if we assume the external fluxes to be time-independent. The offset flux induced currents are zero since no linear inductors are links, meaning that we have chosen no external fluxes over the linear inductors.

The final term of \cref{eq:circuit_EQM_generalized_with_matrices} reduces to
\begin{equation}
    \vec F^{(C)}\vec D_J\textbf{sin}\vec \Phi = \mat{0 \\ I_c  \sin(\Phi_2 + \tilde\Phi_3)},
\end{equation}
where we can move the external flux into the offset charges by choosing a spanning tree over the Josephson junction instead.

\subsection{Voltage and current sources}

Until now, we have assumed that external fluxes are our only control parameters, but we can also add current and voltage sources. Voltage sources can be added in series with existing components without introducing new constraints on the branch fluxes. This effectively transforms the external flux vector
\begin{equation}
    \tilde{\vec \Phi}(t) \rightarrow \tilde{\vec \Phi}(t) - \int_{-\infty}^t\vec V_V(t')\dd{t'},
\end{equation}
where $(\vec {V}_V)_i$ is the voltage generated by the source on the $i$th branch, or $0$ if the $i$th branch is not a voltage source, i.e., defined analogously to \cref{eq:SpeciesVectors}. 

Similarly, we can add a current source in parallel with an existing element without introducing additional constraints on the free currents. This modifies $\vec I_0$ according to
\begin{equation}
    \vec I_0 \rightarrow \vec I_0 + \vec F^{(C)} \vec I_B,
\end{equation}
where $\vec I_B$ is the bias current vector with zeros on all entries except those belonging to a branch with a current source, where instead it has the applied current, i.e., as in \cref{eq:SpeciesCurrentVector}.

\subsection{Lagrangian and Hamiltonian}

One can show, using \cref{eq:Lagrang}, that a Lagrangian fulfilling the equations of motion in \cref{eq:circuit_EQM_generalized_with_matrices} is
\begin{equation}\label{eq:lagrangian_from_matrices}
\begin{split}
	\L &= \frac{1}{2}\dot{\vec{\Phi}}_t^T \vec M
    \dot{\vec{\Phi}}_t + \vec{Q}_0\cdot\dot{\vec{\Phi}}_t\\
    &\phantom{=}-\frac{1}{2}\vec{\Phi}_t^T\vec K
    \vec{\Phi}_t- \vec{I}_0\cdot\vec{\Phi}_t\\
    &\phantom{=}+\vec J_C \cdot \textbf{cos}\left((\vec F^{(C)})^T  
    \vec{\Phi}_t+\mat{\vec{0} \\ \vec{\tilde{\Phi}} }\right),
\end{split}
\end{equation}
where we define the critical current vector
\begin{equation}
    (\vec{J}_C)_i = (\vec D_J)_{ii}. \label{eq:JJ_critical_current_vector}
\end{equation}
The conjugate momenta of the twig branches are then given by 
\begin{equation}
	\vec{Q}_t = \pdv{\L}{\dot{\vec{\Phi}}_t} = \vec{M} \dot{\vec{\Phi}}_t+\vec Q_0,	\label{eq:canonical_momenta}
\end{equation}
and the Hamiltonian can be found performing a Legendre transformation
\begin{equation}\label{eq:HamiltonianDissipationFree}
	\begin{split}
    	\H &= \vec Q_t\cdot \dot{\Phi}_t-\L\\
    	&=
    	\frac{1}{2}\left(\vec Q_t-\vec Q_0\right)^T\vec M^{-1}\left(\vec Q_t-\vec Q_0\right)\\
    	&\phantom{=}+\frac{1}{2}\vec{\Phi}_t^T\vec K\vec{\Phi}_t+\vec{I}_0\cdot\vec{\Phi}_t\\
    	&\phantom{=}-\vec J_C \cdot \textbf{cos}\left((\vec F^{(C)})^T  
    \vec{\Phi}_t+\mat{\vec{0} \\ \vec{\tilde{\Phi}} }\right).
    \end{split}
\end{equation} 
This Hamiltonian can easily be quantized using the approach presented in \cref{sec:quantize}, where this time the canonical variables are the branch fluxes $\Phi_b$ and $Q_b$ of the twigs, with the commutator relation in \cref{eq:commutatorPhiQ}.

\section{Exact truncation}\label{sec:exactTrunc}

In the main text, the truncated cosine contribution from Josephson junctions is expanded to fourth order. This expansion introduces some errors and limits the qubits to operate in a given regime. However, the behavior of the exact Hamiltonian should, in the same regime, be nearly the same with only negligible differences. In this appendix, we introduce a method for truncating cosine terms without expanding.
This can be done numerically by writing the node fluxes in terms of creation and annihilation operators, which would be represented by finite matrices in numerical calculations, and calculating the matrix cosine function of these. However, that requires much more difficult computations than if we were able to write the truncated cosine in matrix form directly. We, therefore, present a method for doing so here. The method presented here employs the displacement operator, which is often used in quantum optics studies of optical phase space \cite{Gerry2005}.

Consider the standard cosine term of a Josephson junction bridging two nodes with node fluxes $ \phi_{1} $ and $ \phi_{2} $. Written in terms of the creation and annihilation operators, this becomes
\begin{equation}\label{eq:cosinOperator}
\cos (\hat\phi_{1} - \hat\phi_{2}) = \cos\left(\sqrt{\frac{\zeta_1}{2}}(\hat b_{1}^{\dagger} + \hat b_{1}) - \sqrt{\frac{\zeta_2}{2}}(\hat b_{2}^{\dagger} + \hat b_{2})\right),
\end{equation}
where $\zeta_i$ is the impedance [\cref{eq:impedance}] of the two nodes. We want to find the matrix elements of the cosine operator in \cref{eq:cosinOperator} for the lowest levels of the two anharmonic oscillator modes. The trick is to rewrite the cosine in terms of exponentials that contain only one mode. We write the cosine in terms of two complex exponentials as usual and note that each exponential can be written as a product because the operators of different modes commute
\begin{equation}\label{eq:cosineDisplacement}
\begin{aligned}
\cos&\left(\sqrt{\frac{\zeta_1}{2}}(\hat b_{1}^{\dagger} + \hat b_{1}) - \sqrt{\frac{\zeta_2}{2}}(\hat b_{2}^{\dagger} + \hat b_{2})\right) \\
&= \frac{1}{2}\bigg[e^{i\sqrt{\frac{\zeta_1}{2}}(\hat b_{1}^{\dagger} + \hat b_{1})}e^{-i\sqrt{\frac{\zeta_2}{2}}(\hat b_{2}^{\dagger} + \hat b_{2})}\\
&\phantom{=}+ e^{-i\sqrt{\frac{\zeta_1}{2}}(\hat b_{1}^{\dagger} + \hat b_{1})}e^{i\sqrt{\frac{\zeta_2}{2}}(\hat b_{2}^{\dagger} + \hat b_{2})}\bigg].
\end{aligned}
\end{equation}
From this expression it is clear that we need to find the matrix representation of the general operator $ \exp[ik(\hat b^{\dagger} + \hat b)] $ for some real number $ k=\sqrt{\zeta/2} $. 
To find the desired matrix representation we consider the displacement operator
\begin{equation}
\hat D(\xi) = e^{\xi \hat b^{\dagger} - \xi^{*}\hat b},
\end{equation}
where $ \xi $ is complex number. This operator is unitary and satisfies $ \hat D(\xi)^{\dagger} = \hat D(-\xi) $, as well as the following commutation relation
\begin{equation}
[\hat D(\xi), \hat b^{\dagger}] = - \xi^{*}\hat D(\xi).
\end{equation} 
The operator creates coherent states by ''displacing`` the vacuum state
\begin{equation}
\hat D(\xi)\ket{0} = \ket{\xi},
\end{equation}
where $ \ket{\xi} $ is the coherent state defined by $ \hat b\ket{\xi} = \xi\ket{\xi} $. A coherent state can be written in terms of Fock-states as
\begin{align}
\ket{\xi} = e^{-\frac{\vert\xi\vert^2}{2}}\sum_{n = 0}^{\infty}\frac{\xi^{n}}{\sqrt{n!}}\ket{n}.
\end{align}
We have
\begin{align}
e^{ik(\hat b^{\dagger} + \hat b)} = \hat D(ik).
\end{align}
Using the above commutation relation, we can derive the effect of the displacement operator on any other Fock-state. With that, we can calculate its matrix elements as desired. For ease let us write the commutation relation and coherent state for $ \xi = ik $
\begin{subequations}
	\begin{align}
	\hat D(ik)\hat b^{\dagger} &= (\hat b^{\dagger} + ik)\hat D(ik),\\
	\ket{ik} &= e^{-\frac{k^2}{2}}\sum_{n = 0}^{\infty}\frac{(ik)^{n}}{\sqrt{n!}}\ket{n}.
	\end{align}
\end{subequations}
For truncation to the two lowest levels we need only calculate three matrix elements
\begin{subequations}
	\begin{align}
	\mel{0}{\hat D(ik)}{0} &= e^{-\frac{k^2}{2}}\sum_{n = 0}^{\infty}\frac{(ik)^{n}}{\sqrt{n!}}\braket{0}{n} = e^{-\frac{k^2}{2}},\\
	\mel{1}{\hat D(ik)}{0} &= e^{-\frac{k^2}{2}}\sum_{n = 0}^{\infty}\frac{(ik)^{n}}{\sqrt{n!}}\braket{1}{n} = ike^{-\frac{k^2}{2}},\\
	\begin{split}
	\mel{1}{\hat D(ik)}{1} &= \mel{1}{\hat D(ik)\hat b^{\dagger}}{0}\\
	&= \mel{1}{(\hat b^{\dagger} + ik)\hat D(ik)}{0}\\
	&= e^{-\frac{k^2}{2}}\sum_{n = 0}^{\infty}\frac{(ik)^{n}}{\sqrt{n!}}\left(\braket{0}{n} + ik\braket{1}{n}\right)\\
	&= (1 - k^2)e^{-\frac{k^2}{2}}.
	\end{split}
	\end{align}
\end{subequations}
In general we would find that arbitrary matrix elements can be expressed in terms of Laguerre polynomials.
From $\hat D(\xi)^{\dagger} = \hat D(-\xi) $ we have $ \mel{0}{\hat D(ik)}{1} = (\mel{1}{\hat D(ik)^{\dagger}}{0})^{\dagger} = ike^{-\frac{k^2}{2}} $. With this we conclude that the matrix representation of $e^{ik(\hat b^{\dagger} + \hat b)}$ is
\begin{align}
\begin{split}
M_{2}\left[e^{ik(\hat b^{\dagger} + \hat b)}\right] &= \mat{e^{-\frac{k^2}{2}} & ike^{-\frac{k^2}{2}} \\ ike^{-\frac{k^2}{2}} & (1 - k^2)e^{-\frac{k^2}{2}}}\\
	&= \left(1 - \frac{k^2}{2} + \frac{k^2}{2}\sigma^{z} + ik\sigma^{x}\right)e^{-\frac{k^2}{2}},
\end{split}
\end{align}
which is exactly what we need to truncate cosine-operators.

Consider again the standard Josephson junction term from \cref{eq:cosineDisplacement}. We can now perform exact truncation of it to its lowest two levels using the above identity. We find 
\begin{align}
\begin{split}
M_{2}\left[\cos(\hat\phi_{1} - \hat\phi_{2})\right] &= \frac{1}{2}M_{2}\bigg[e^{i\sqrt{\frac{\zeta_1}{2}}(\hat b_{1}^{\dagger} + \hat b_{1})}e^{-i\sqrt{\frac{\zeta_2}{2}}(\hat b_{2}^{\dagger} + \hat b_{2})}\\
&\phantom{=}+ e^{-i\sqrt{\frac{\zeta_1}{2}}(\hat b_{1}^{\dagger} + \hat b_{1})}e^{i\sqrt{\frac{\zeta_2}{2}}(\hat b_{2}^{\dagger} + \hat b_{2})}\bigg]\\
&= \Bigg[\left(\frac{\zeta_1}{4} - \frac{\zeta_1\zeta_2}{16}\right)\sigma_{1}^{z} + \left(\frac{\zeta_2}{4} - \frac{\zeta_1\zeta_2}{16}\right)\sigma_{2}^{z} \\
&\phantom{=}+ \frac{\zeta_1\zeta_2}{16}\sigma_{1}^{z}\sigma_{2}^{z} + \frac{\sqrt{\zeta_1\zeta_2}}{2}\sigma_{1}^{x}\sigma_{2}^{x}\Bigg]e^{-(\zeta_{1} + \zeta_{2})/4}
\end{split}
\end{align}
where we ignore constant terms. Hence, the cosine term has resulted in the usual contribution to the qubit energies, and transverse and longitudinal couplings. The difference, however, is that the coefficients of these terms are more accurate as the calculation did not involve any Taylor expansions. In particular, each coefficient has a factor of $ \exp\left(-(\zeta_{1} + \zeta_{2})/4\right) $, which contains contributions corresponding to the infinitely many possible virtual processes of exciting the modes to any higher lying-level and de-exciting again, which affect the dynamics of the two-level subspace.

\subsection{Exact four-level model}

Using the method of exact truncation, the cosine terms can also be truncated exactly to more than the two lowest levels. This can be useful for numerical studies of the higher levels' effect on the two-level dynamics. It can be advantageous to perform the truncation analytically to avoid having to do it numerically. Here we show an exact truncation to the four lowest levels. First, we define some new matrices
\begin{subequations}
	\begin{align}
	\vec{Z} &= \mat{1 & 0 & 0 & 0 \\ 0 & -1 & 0 & 0 \\ 0 & 0 & -3 & 0 \\ 0 & 0 & 0 & -5},\\
	\vec{A} &= \mat{0 & 0 & 0 & 0 \\ 0 & 0 & 0 & 0 \\ 0 & 0 & 1 & 0 \\ 0 & 0 & 0 & 3},\\
	\vec{B} &= \mat{0 & 0 & 0 & 0 \\ 0 & 0 & 0 & 0 \\ 0 & 0 & 0 & 0 \\ 0 & 0 & 0 & 1},
	\end{align}
\end{subequations}
and finally $ \vec{X}_{ij} $ and $ \vec{Y}_{ij} $ for $ i,j = 0,1,2,3 $ with $ i<j $, whose $ (a,b) $th entries are
\begin{subequations}
	\begin{align}
	(\vec{X}_{ij})_{ab} &= \begin{cases}
	0, & \textup{ for } ab \neq ij,ji\\
	1, & \textup{ for } ab = ij,ji
	\end{cases}\\
	(\vec{Y}_{ij})_{ab} &= \begin{cases}
	0, & \textup{ for } ab \neq ij,ji\\
	-i, & \textup{ for } ab = ij\\
	i, & \textup{ for } ab = ji
	\end{cases}
	\end{align}
\end{subequations}
Together with the identity, $ \vec{Z} $, $ \vec{A} $, and $ \vec{B} $ describe contributions to the energy levels. In particular $ \vec{Z} $ can be seen as the $4\times 4$ expansion of $ \sigma^{z} $, while $ \vec{A} $ describes the anharmonicity, and $ \vec{B} $ describes a similar anharmonic energy shift beyond the regular anharmonicity only relevant for the third excited state and higher. In terms of the usual bosonic number operator $ \hat n = \hat b^{\dagger}\hat b = \diag{0,1,2,3} $, we may say that $ \vec{Z} $ corresponds to $ \hat n $, $ \vec{A} $ corresponds to $ \hat n^2 $ and $ \vec{B} $ to $ \hat n^3 $. Alternatively, we may say that $ \vec{A} $ is proportional to $ \hat b^{\dagger}\hat b^{\dagger}\hat b\hat b $ and $ \vec{B} $ to $ \hat b^{\dagger}\hat b^{\dagger}\hat b^{\dagger}\hat b\hat b\hat b $, which shows how $ \vec{A} $ does not affect levels below the second excited one, while $ \vec{B} $ does not matter for levels below the third excited. The $ \vec{X}_{ij} $ and $ \vec{Y}_{ij} $ describe rotation or flipping between the $ i $th and $ j $th energy level, and are thus the generalizations of $ \sigma^{x} $ and $ \sigma^{y} $. In terms of these matrices, we can write
\begin{widetext}
	\begin{subequations}
		\begin{align}
		M_{4}[\hat b^{\dagger} - \hat b] &= \vec{Y}_{01} + \sqrt{2}\vec{Y}_{12} + \sqrt{3}\vec{Y}_{23}, \label{eq:truncFour1}\\
		M_{4}[(\hat b^{\dagger} - \hat b)^2] &= -2 + \vec{Z} + \sqrt{2}\vec{X}_{02} + \sqrt{6}\vec{X}_{13}, \label{eq:truncFour2}\\
		M_{4}[\hat b^{\dagger} + \hat b] &= \vec{X}_{01} + \sqrt{2}\vec{X}_{12} + \sqrt{3}\vec{X}_{23}, \label{eq:truncFour3}\\
		M_{4}[(\hat b^{\dagger} + \hat b)^2] &= 2 - \vec{Z} +   \sqrt{2}\vec{X}_{02} + \sqrt{6}\vec{X}_{13}, \label{eq:truncFour4}\\
		\begin{split}
		M_{4}\left[e^{i\sqrt{\zeta/2}(\hat b^{\dagger} + \hat b)}\right] &= \bigg[1 - \frac{\zeta}{4} + \frac{\zeta}{4}\vec{Z} + \frac{\zeta^2}{8}\vec{A} - \frac{\zeta^3}{48}\vec{B} + i\sqrt{\frac{\zeta}{2}}\vec{X}_{01} - \frac{\zeta}{\sqrt{8}}\vec{X}_{02} - i\frac{\zeta^{3/2}}{\sqrt{48}}\vec{X}_{03} \\
		&\phantom{=}+ i\left(\zeta - \frac{\zeta^2}{4}\right)\vec{X}_{12} - \frac{\sqrt{6}}{4}\left(\zeta - \frac{\zeta^2}{6}\right)\vec{X}_{13} + i\sqrt{\frac{3}{2}}\left(\zeta - \frac{\zeta^2}{2} + \frac{\zeta^3}{24}\right)\vec{X}_{23}\bigg]e^{-\zeta/4}. \label{eq:truncFour5}
		\end{split}
		\end{align}
	\end{subequations}
\end{widetext}
With the above method of truncating to four levels, we can find the exact anharmonicities as the coefficients of the standalone $ \vec{A} $-operators. Just as there are longitudinal couplings among spins, which change energy levels depending on the state of the system, there will be couplings involving $ \vec{Z} $ that change the anharmonicities. But just as we only look at standalone $Z$-operators when determining basic energy levels, we would not include the interaction contributions to the anharmonicity when calculating it. These contributions will, in general, also be smaller, as they originate from terms involving more node fluxes and therefore more $ \zeta $'s. If one wishes to find the anharmonicity without finding and reducing the complete four-level Hamiltonian, one can replace the exponentials $ e^{i\sqrt{\zeta/2}(\hat b^{\dagger} + \hat b)} $ with only $ \left(1 - \frac{\zeta}{4} + \frac{\zeta^2}{8}\vec{A}\right)e^{-\zeta/4} $, and then find the standalone $ \vec{A} $ matrices. We do not need to include the other terms from \cref{eq:truncFour5} as they will only contribute to interactions that are not interesting in this case.


\end{document}